\documentclass[amsmath,amssymb,aps,prd,nofootinbib]{revtex4-1}
\usepackage[utf8]{inputenc}
\usepackage{graphicx}
\usepackage{booktabs}
\usepackage[english]{babel}
\usepackage{lineno}
\usepackage{localdefinitions}

\usepackage{xcolor} 

\AtBeginDocument{
\heavyrulewidth=.08em
\lightrulewidth=.05em
\cmidrulewidth=.03em
\belowrulesep=.65ex
\belowbottomsep=0pt
\aboverulesep=.4ex
\abovetopsep=0pt
\cmidrulesep=\doublerulesep
\cmidrulekern=.5em
\defaultaddspace=.5em
}

\begin{document}

\preprint{1803.03594}

\title{Advancing LHC Probes of Dark Matter from\\the Inert 2-Higgs Doublet Model with the Mono-jet Signal}

\author{A.\ Belyaev}
\affiliation{Rutherford Appleton Laboratory, Didcot, United Kingdom}
\affiliation{University of Southampton, Southampton, United Kingdom}
\author{T.\ R.\ Fernandez Perez Tomei}
\affiliation{Universidade Estadual Paulista, São Paulo, Brazil}
\author{P.\ G.\ Mercadante}
\affiliation{Universidade Federal do ABC, São Paulo, Brazil}
\author{C.\ S.\ Moon}
\affiliation{Kyungpook National University, Daegu, Korea}
\author{S.\ Moretti}
\affiliation{Rutherford Appleton Laboratory, Didcot, United Kingdom}
\affiliation{University of Southampton, Southampton, United Kingdom}
\author{S.\ F.\ Novaes}
\affiliation{Universidade Estadual Paulista, São Paulo, Brazil}
\author{L.\ Panizzi}
\affiliation{Uppsala University, Uppsala, Sweden}
\affiliation{University of Southampton, Southampton, United Kingdom}
\author{F.\ Rojas}
\affiliation{Universidad Técnica Federico Santa María, Valparaíso, Chile}
\author{M.\ Thomas}
\affiliation{University of Southampton, Southampton, United Kingdom}



\date{\today}

\begin{abstract}
The inert 2-Higgs Doublet Model (i2HDM) is a well-motivated minimal consistent
Dark Matter (DM) model, but it is rather challenging to test at the Large Hadron
Collider (LHC) in the parameter space allowed by relic density and DM direct
detection constraints. This is especially true when considering the latest XENON
1T data on direct DM searches which we use here to present the best current
combined limit on the i2HDM parameter space. In this analysis, we present
prospects to advance the exploitation of DM mono-jet signatures from the i2HDM
at the LHC, by emphasising that a shape analysis of the missing transverse
momentum distribution allows one to sizably improve the LHC discovery potential.
As a key element of our analysis, we explore the validity of using an effective
vertex, $ggH$, for the coupling of the Higgs boson to gluons using a full
one-loop computation. We have found sizeable differences between the two
approaches, especially in the high missing transverse momentum region, and
incorporated the respective K-factors to obtain the correct kinematical
distributions. As a result, we delineate a realistic search strategy and present
the improved current and projected LHC sensitivity to the i2HDM parameter space.
\end{abstract}

\maketitle

\section{Introduction}

Despite several independent evidences of Dark Matter (DM) at the cosmological
scale, its nature remains unknown  since no experiment so far has been able to
claim its detection in the laboratory and probe its properties. Potentially, DM
can be probed in direct or indirect detection experiments as well as  be
produced at the Large Hadron Collider (LHC) or future machines, though the
latter can only detect DM \emph{candidates}, as any observed missing energy can
still be interpreted as generated by long-lived neutral particles. This combined
effort on advancing our  knowledge of DM properties is one of the key goals of
the astroparticle and high energy physics communities.

A convenient way to understand the potential of both collider and non-collider
experiments to probe DM is to explore simple, fully calculable, renormalisable
models  with viable DM candidates, which we refer to as Minimal Consistent Dark
Matter (MCDM) models. We do not know yet which theoretical scenario corresponds
to reality, but any model of this kind offers an excellent opportunity to gain
insight into the intricate interplay between collider and non-collider
constraints. MCDM models, which can be viewed as robust toy models, are
self-consistent and can be easily incorporated into larger theoretically-driven
scenarios of physics Beyond the Standard Model (BSM). Because of their
attractive features, MCDM models can be considered as the next step beyond DM
Effective Field Theory (EFT) (see e.g. \cite{Fox:2011pm,Rajaraman:2011wf,Goodman:2010ku,Bai:2010hh,Beltran:2010ww,Goodman:2010yf,Fox:2011fx,Shoemaker:2011vi,Fox:2012ru,Haisch:2012kf,Busoni:2013lha,Busoni2014a,Belyaev:2016pxe}) and simplified DM models (see e.g. \cite{Buchmueller:2013dya,Busoni:2014sya,Busoni:2014haa,Buchmueller:2014yoa,Buckley:2014fba,Abdallah:2015ter,Abdallah:2014hon,Abercrombie:2015wmb}).

The inert 2-Higgs Doublet Model (i2HDM), which has been initially suggested more
than 30 years ago in~\cite{Deshpande:1977rw}, is one of the most representative
MCDM models which has  become very attractive lately~\cite{Ma:2006km,Barbieri:2006dq,LopezHonorez:2006gr,Arina:2009um,Nezri:2009jd,Miao:2010rg,Gustafsson:2012aj,Arhrib:2012ia,Swiezewska:2012eh,Goudelis:2013uca,Arhrib:2013ela,Krawczyk:2013jta,Krawczyk:2013pea,Ilnicka:2015jba,Diaz:2015pyv,Modak:2015uda,Queiroz:2015utg,Garcia-Cely:2015khw,Hashemi:2016wup,Poulose:2016lvz,Alves:2016bib,Datta:2016nfz,Belyaev:2016lok}
in the light of intensive DM searches. In fact, besides providing a good DM
candidate, the i2HDM can also give rise to an  `improved
naturalness'~\cite{Barbieri:2006dq} since large radiative corrections from the
inert Higgs sector can `screen' the SM Higgs contribution to the Electro-Weak
(EW) parameter $\Delta T$.

It was shown in~\cite{Belyaev:2016lok}
that the LHC has limited sensitivity to probe the i2HDM with the mono-jet
signature using the cut-based analyses optimised for the low luminosity Run 2
data. To complement these studies, in the present paper, we explore the LHC
potential to probe DM via the mono-jet signature in the i2HDM scenario by
exploiting a larger amount of information from observables at the differential
level. More specifically, we will consider the shape of the missing transverse
momentum (\MET{}) distribution. New findings of this study include: a) updating
limits on the i2HDM parameter space following the recent XENON 1T results on DM
Direct Detection (DD) searches; b) exploration of the range of validity of the
effective $ggH$ vertex in the heavy top mass limit by considering the \MET{}
distribution and comparing its shape to the full one-loop result, which will
allow us to determine a realistic LHC potential for probing DM in different
kinematical regions; c) optimisation and improvement of the LHC sensitivity to
the DM mono-jet signal from the i2HDM defined by Higgs and $Z$-boson mediation
processes using a shape analysis of the \MET{} distribution; d) projection of
our results to the High Luminosity LHC (HL-LHC) phase.

The rest of the paper is organised as follows. In section 2 we discuss the i2HDM
parameter space together with the current status of theoretical and experimental
constraints. In section 3 we present the main results of the paper which include
the analysis of the validity of the effective $ggH$ ($H$ being the SM-like
Higgs) vertex approach, the exploration of several model benchmarks and finally
finding the LHC potential to probe the i2HDM at present and projected
luminosities via exploitation of the \MET{} shape in the mono-jet signature. In
section 4 we draw our conclusions.

\section{The i2HDM}

 \subsection{Parameter space}
The i2HDM \cite{Deshpande:1977rw,Ma:2006km,Barbieri:2006dq,LopezHonorez:2006gr} is an extension of the SM
with a second scalar doublet $\phi_2$ possessing the same quantum numbers as the SM Higgs doublet $\phi_1$
but with no couplings to fermions, thus providing its inert nature.
This construction is protected by a discrete $\mathcal{Z}_2$ symmetry under which $\phi_2$ is odd and all the other fields are even.
The  Lagrangian of the scalar sector is
  \begin{equation}
  \mathcal{L} =
  |D_{\mu}\phi_1|^2 + |D_{\mu}\phi_2|^2 -V(\phi_1,\phi_2)
  \textrm{,}
  \label{eq:lagrangian}
  \end{equation}
where   $V$ is the potential with all scalar interactions compatible with the $\mathcal{Z}_2$ symmetry:
\begin{eqnarray}
  V &=& -m_1^2 (\phi_1^{\dagger}\phi_1) - m_2^2 (\phi_2^{\dagger}\phi_2) + \lambda_1 (\phi_1^{\dagger}\phi_1)^2 + \lambda_2 (\phi_2^{\dagger}\phi_2)^2    \nonumber  \\
  &+&  \lambda_3(\phi_1^{\dagger}\phi_1)(\phi_2^{\dagger}\phi_2)
  + \lambda_4(\phi_2^{\dagger}\phi_1)(\phi_1^{\dagger}\phi_2)
  + \frac{\lambda_5}{2}\left[(\phi_1^{\dagger}\phi_2)^2 + (\phi_2^{\dagger}\phi_1)^2 \right]\,.\label{eq:potential}
\end{eqnarray}
In the unitary gauge, the doublets take the form
\begin{equation}
\phi_1=\frac{1}{\sqrt{2}}
\begin{pmatrix}
0\\
v+H
\end{pmatrix},
  \qquad
  \phi_2= \frac{1}{\sqrt{2}}
\begin{pmatrix}
 \sqrt{2}{h^+} \\
 h_1 + ih_2
\end{pmatrix},
\end{equation}
where we consider the parameter space in which only the first, SM-like doublet,  acquires a Vacuum Expectation Value (VEV), $v$.
In the notation $\langle\phi_i^0\rangle = v_i/\sqrt{2}$, this inert minimum corresponds to
$v_1 = v$, $v_2 = 0$.
After EW Symmetry Breaking (EWSB), the $\mathcal{Z}_2$ symmetry is still conserved by the vacuum state, which forbids direct coupling of any single inert field to the SM fields
and protects the lightest inert boson from decaying, hence providing the DM candidate in this scenario.
In contrast, the interactions of  {\it pair} of  inert scalars with the SM  gauge-bosons and  SM-like Higgs $H$ are allowed, thus
giving rise to various signatures at colliders and at DM detection experiments.

In addition to the SM-like scalar $H$, the model contains one inert charged $h^\pm$ and two further inert neutral $h_1, h_2$ scalars. The two neutral scalars of the i2HDM have opposite $CP$-parities, but it is impossible  to unambiguously determine which of them is $CP$-even and which one  is $CP$-odd since
the model has two $CP$-symmetries, $h_1 \to h_1, h_2 \to -h_2$ and $h_1 \to -h_1, h_2 \to h_2$, which get interchanged upon a change of basis $\phi_2 \to i \phi_2$. This makes the specification of the $CP$-properties of $h_1$ and $h_2$ a basis dependent statement.
Therefore, following Ref.~\cite{Belyaev:2016lok},  we denote the two neutral inert scalar masses as $M_{h_1} < M_{h_2}$, without specifying which is scalar or pseudoscalar, so that $h_1$ is the DM candidate.

The model can be conveniently  described by a five dimensional parameter space\cite{Belyaev:2016lok}
using the following   phenomenologically relevant variables:
\begin{equation}
\label{eq:model-parameters}
M_{h_1}\,,\quad M_{h_2} > M_{h_1}\,,\quad M_{h^+} > M_{h_1}\,, \quad \lambda_2 > 0\,,\quad \lambda_{345} > -2\sqrt{\lambda_1\lambda_2},
\end{equation}
where $M_{h_1},M_{h_2}$ and $M_{h^+} $ are the masses of the two neutral and  charged inert scalars, respectively, whereas   $\lambda_{345}=\lambda_3+\lambda_4+\lambda_5$
is the coupling which governs the Higgs-DM interaction vertex $H h_1 h_1$.
The masses of the physical scalars are expressed in terms of the parameters of the Lagrangian in Eqs.~(\ref{eq:lagrangian})--(\ref{eq:potential}) as follows:
\begin{eqnarray}
\label{masses}
\begin{array}{lcl}
\MH^2     &=& 2 \lambda_1 v^2 = 2m_1^2, \\[2pt]
M_{h^+}^2 &=& \frac{1}{2} \lambda_3 v^2 - m_2^2, \\[2pt]
M_{h_1}^2 &=& \frac{1}{2}(\lambda_3 +\lambda_4 - |\lambda_5|) v^2 - m_2^2, \\[2pt]
M_{h_2}^2 &=& \frac{1}{2}(\lambda_3 +\lambda_4 + |\lambda_5|) v^2 - m_2^2 \ >M_{h_1}^2.
\end{array}
\end{eqnarray}

\subsection{Theoretical and experimental constraints}

Constraints on the Higgs potential from requiring vacuum stability and a global minimum take the following  form\cite{Belyaev:2016lok}:
\begin{eqnarray}
\left\{
\begin{array}{lcr}
M_{h_1}^2 > 0 \text{ (the trivial one)} & \text{for} & |R|<1, \label{eq:scalar-pot1}\\
M_{h_1}^2 > (\lambda_{345}/2\sqrt{\lambda_1\lambda_2}-1) \sqrt{\lambda_1\lambda_2} v^2 = (R-1) \sqrt{\lambda_1\lambda_2} v^2 & \text{for} & R>1,~\label{eq:scalar-pot2}
\end{array}
\right.
\end{eqnarray}
where $R=\lambda_{345}/2\sqrt{\lambda_1\lambda_2}$ and  $\lambda_1 \approx 0.129$ is fixed as in the SM by the Higgs mass in Eq. (\ref{masses}).
The latter condition places an important upper bound on $\lambda_{345}$ for a given DM mass $M_{h_1}$.

\begin{figure}[htb]
\centering
\includegraphics[width=0.55\textwidth]{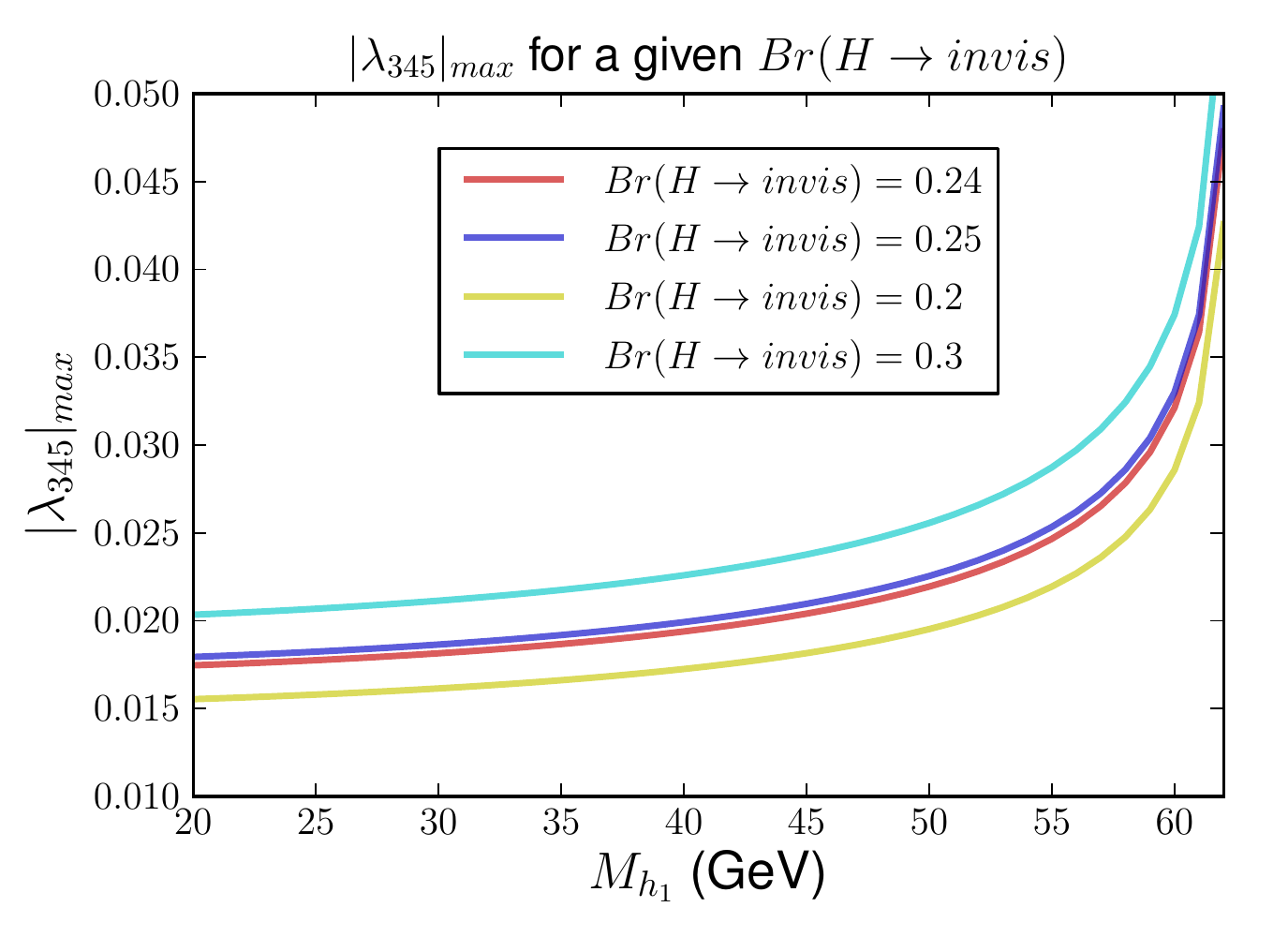}
\caption{\label{fig:l345max}The values of $|\lambda_{345}|_{max}$  as a function of $M_{h_1}$ for selected choices of  $Br(H\to~{invis})$.
This limit is found under the assumption that
$H\to h_1 h_1$ is the only invisible decay channel of the SM-like  Higgs boson.}
\end{figure}

The theoretical  upper limit on $\lambda_{345}$ for a given DM mass  comes from the vacuum stability constraint. Using Eq. (17) from  \cite{Belyaev:2016lok} and an upper limit on $\lambda_2$ (which  is about $4\pi/3$ for DM masses below 300 GeV) we find:
\begin{equation}
\lambda_{345}< 2\left(\frac{M_{h_1}^2}{v^2}+\sqrt{\lambda_1\lambda_2^{max}}\right)
\simeq 2\left(\frac{M_{h_1}^2}{v^2}+\sqrt{\lambda_1 \frac{4\pi}{3}}\right).
\label{l345limit}
\end{equation}
When $M_{h_1}<\MH/2$, $\lambda_{345}$ has a much stronger limit coming from the invisible Higgs boson decay measured at the LHC.
In this region, the limit on $|\lambda_{345}|$ can be written in the following form:
\begin{equation}
|\lambda_{345}| <
\left(
\frac{8\pi g_W^2 \Gamma_{\rm SM} \MH}{M_W^2 \left(\frac{1}{Br(H\to~{invis})}-1\right)\sqrt{1-4\frac{M_{h_1}^2}{\MH^2}}}
\right)^{1/2},\label{lam345-limit-from-inv}
\end{equation}
where ${Br(H\to~{invis})}$ is the experimental limit on the
Branching ratio ($Br$) for  invisible Higgs boson decays, $\Gamma_{\rm SM}$ the SM-like Higgs boson width and $g_W$ the SM weak coupling.
This formula is derived under the assumption that
$H\to h_1 h_1$ is the only invisible channel of the SM-like   Higgs boson.
In Fig.~\ref{fig:l345max} we present values of  $|\lambda_{345}|_{max}$
as  function of $M_{h_1}$ for several values of ${Br(H\to~{invis})}$
including $0.25$ and $0.24$ corresponding to the most up-to-date limits
on ${Br(H\to~{invis})}$ from ATLAS~\cite{Aad:2015txa} and CMS~\cite{Khachatryan:2016whc},  respectively. One should note that experimental limits on
 ${Br(H\to~{invis})}$ are actually placed for $H\to invis$ to {\it any channel}, thus
 also include  $H\to ZZ \to \ \mbox{neutrinos}$, which is however below the per mille level and  can thus be neglected in our study.

The $|\lambda_{345}|_{max}$ value  increases
when $M_{h_1}$ approaches $\MH/2$, ranging
from 0.024 at $M_{h_1}=50$ GeV to 0.053 at $M_{h_1}=62$ GeV.
At the same time the  $\Omega h^2<0.1$ constraint sets the lower limit  $M_{h_1}\gtrsim 40$~GeV since below it  there are no
effective annihilation and/or co-annihilation DM channels to bring DM relic density to a low enough level consistent with Planck constraints.
One should note that, when the decay $H\to h_2 h_2$ also takes place, and, when $h_1$ and $h_2$ are close in mass (below, say few GeV), this channel will also contribute to the invisible Higgs
decay.
In this case the limit on $\lambda_{345}$ can be easily modified, taking into account that
$\lambda_{Hh_2h_2} = \lambda_{345}+\frac{M_{h_2}^2-M_{h_1}^2}{v^2}$ and thus, for $M_{h_2}\simeq M_{h_1}$, one has
$\lambda_{Hh_2h_2}\simeq \lambda_{Hh_1h_1}=\lambda_{345}$.

\begin{figure}[tbh]
\begin{center}
\begin{tabular}{p{0.28\textwidth} p{0.4\textwidth} p{0.3\textwidth}}
\hspace*{1.9cm}theory constraints (a) &\hspace*{1.1cm}+LEP,EWPT,LHC(Higgs) (b) &\hspace*{-0.7cm}+relic density, LUX (c)
\end{tabular}
\hspace*{-0.3cm}
\includegraphics[trim={0.5cm   0     2.4cm     0.3cm},clip,height=0.18\textheight]{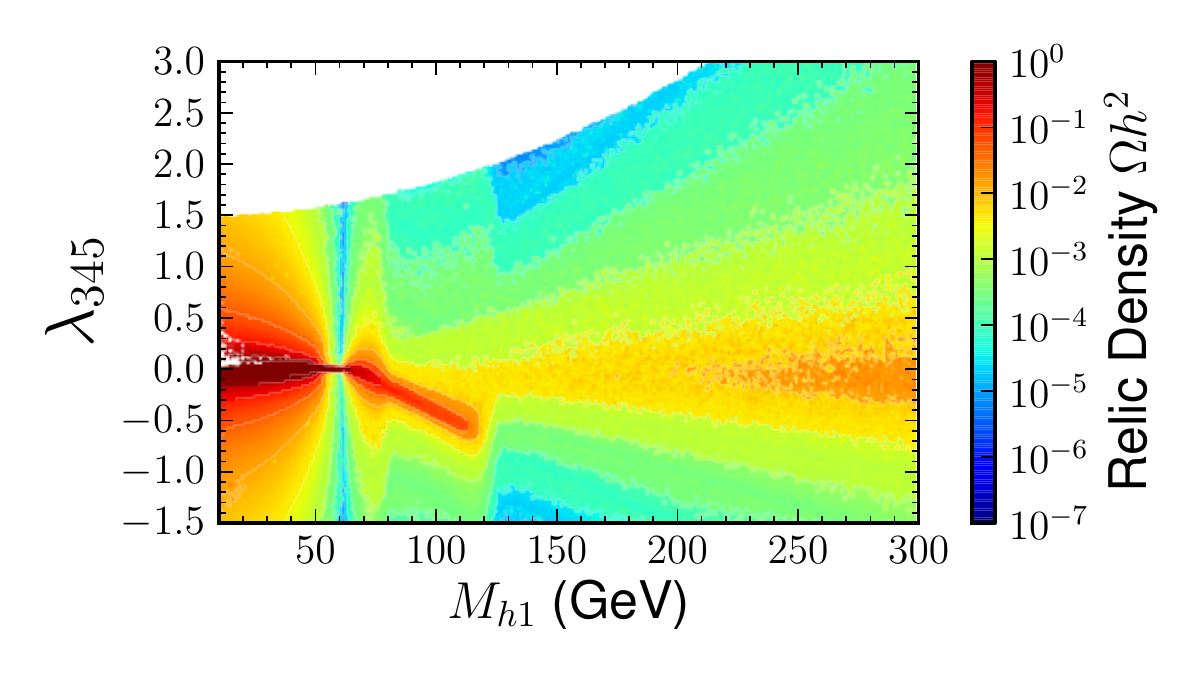}%
\includegraphics[trim={2.2cm 0   2.4cm   0.3cm},  clip,height=0.18\textheight]{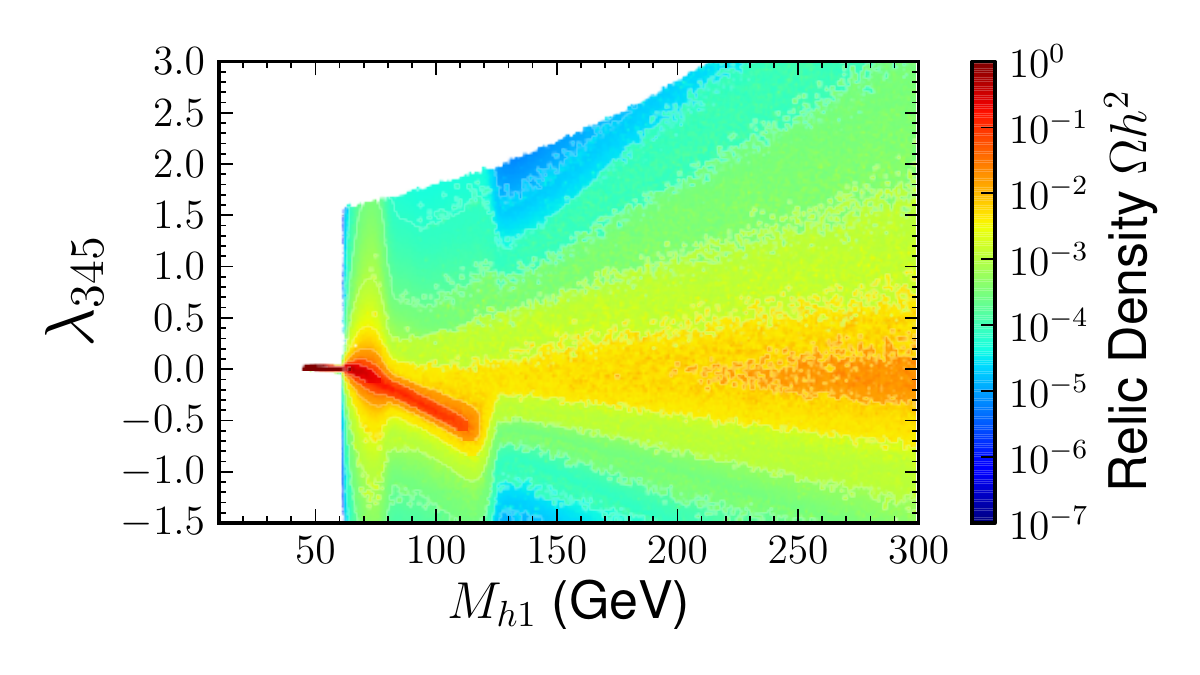}%
\includegraphics[trim={2.2cm 0   0      0.3cm },  clip,height=0.18\textheight]{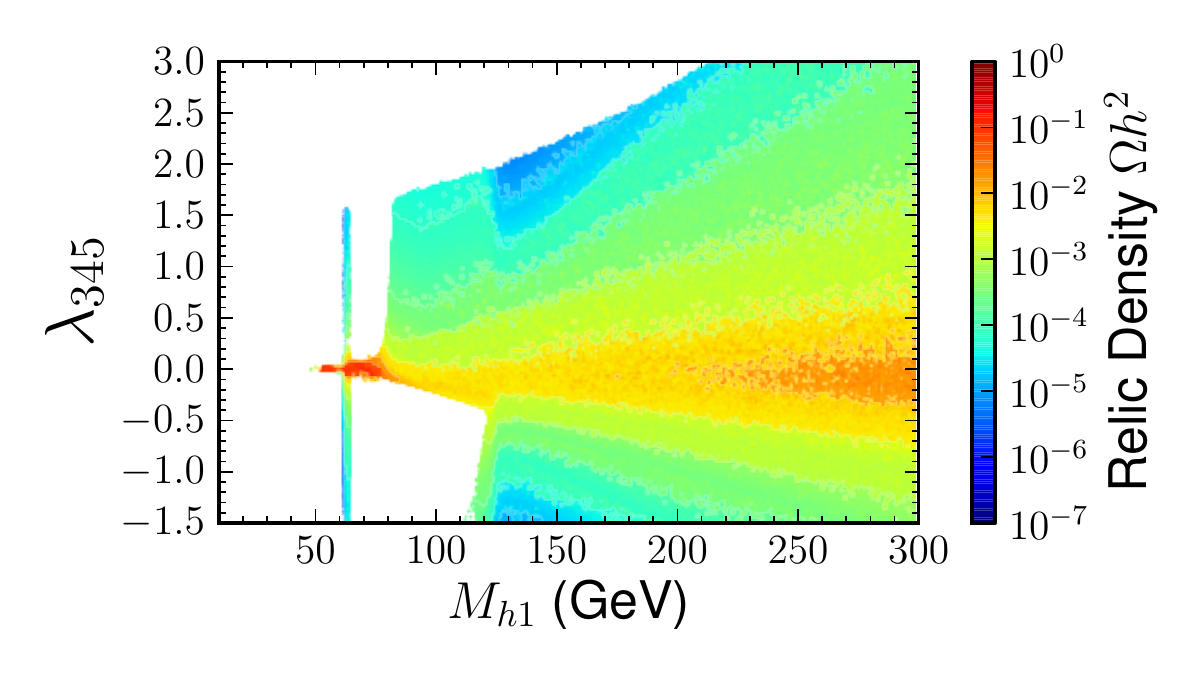}
\caption{Colour maps of DM relic abundance projected on the plane $(M_{h_1},\lambda_{345})$ from~\cite{Belyaev:2016lok}.
The three plots correspond to the surviving points after
the sequential application of the sets of constraints described in the text.\label{fig:scan-simplified}}
\end{center}
\end{figure}

The comprehensive analysis of the i2HDM parameter space performed in~\cite{Belyaev:2016lok} using an i2HDM implementation into the CalcHEP~\cite{CALCHEP} and micrOMEGAs~\cite{Belanger:2001fz,Belanger:2004yn} frameworks demonstrates an important  complementarity of
various constraints, which is presented  in Fig.~\ref{fig:scan-simplified} as an effect  of the sequential  application  of:
a) theoretical constraints from vacuum stability, perturbativity and unitarity (theory);
b) experimental constraints from colliders (LEP, LHC Higgs data, including those from EW Precision Test (EWPT) data);
c) the upper bound on the DM relic density at $\Omega_{\rm DM} h^2$ given by  Planck~\cite{Ade:2013zuv,Ade:2015xua} and constraints from DM DD searches at LUX~\cite{Akerib:2013tjd}.

From  Fig.~\ref{fig:scan-simplified}(a) and (b) one can see
the large effect   of the  invisible Higgs decay constraint on $\lambda_{345}$
(of the order of $10^{-2}$) in the   $M_{h_1}<\MH/2$
region, which is two orders of magnitude stronger than
the constraint on $\lambda_{345}$ from  vacuum stability.
The constraint from DM DD searches from LUX~\cite{Akerib:2013tjd}
further limits $\lambda_{345}$ as one can see from  Fig.~\ref{fig:scan-simplified}(c).
Let us recall first that we use the re-scaled DD Spin-Independent (SI) cross section, $\hat{\sigma}_{\rm SI}=  R_\Omega\times \sigma_{\rm SI}$, where the scaling factor
$R_\Omega = \Omega_{\rm DM}/\Omega^{\rm Planck}_{\rm DM}$ takes into account the case of $h_1$ representing only a part of the total DM budget, thus allowing for a convenient comparison
of the model predictions with the DM DD limits.
One can see that this constraint is not symmetric with respect to the sign of $\lambda_{345}$: the parameter space with
$\lambda_{345}<0$ receives stronger constraints.
The reason for  this is that the sign of $\lambda_{345}$ defines the sign of  the interference of DM annihilation into  EW gauge bosons via Higgs boson and  via the $h_1 h_1 VV$  quartic coupling.
For positive $\lambda_{345}$ the interference is positive and the relic density is correspondingly lower, so that the DM DD rates rescaled with relic density, $\hat{\sigma}_{\rm SI}$, are lower than for the case
of negative $\lambda_{345}$, when the corresponding interference is negative
and the relic density higher.
One should also note that  the combined constraints exclude
$M_{h_1}<45$~GeV for the whole parameter space of the i2HDM.

Since DM DD constraints play an important role, in the light
of recent results from the XENON1T experiment~\cite{Aprile:2017iyp},
we have performed a further comprehensive scan of the i2HDM parameter space
analogously to  Ref.~\cite{Belyaev:2016lok} and have found new constraints\footnote{For the XENON1T limit we have used digitised data from the \textsc{PhenoData} database~\cite{phenodata-xenon1t}.}.
\begin{figure}
\centering
\includegraphics[width=0.8\textwidth]{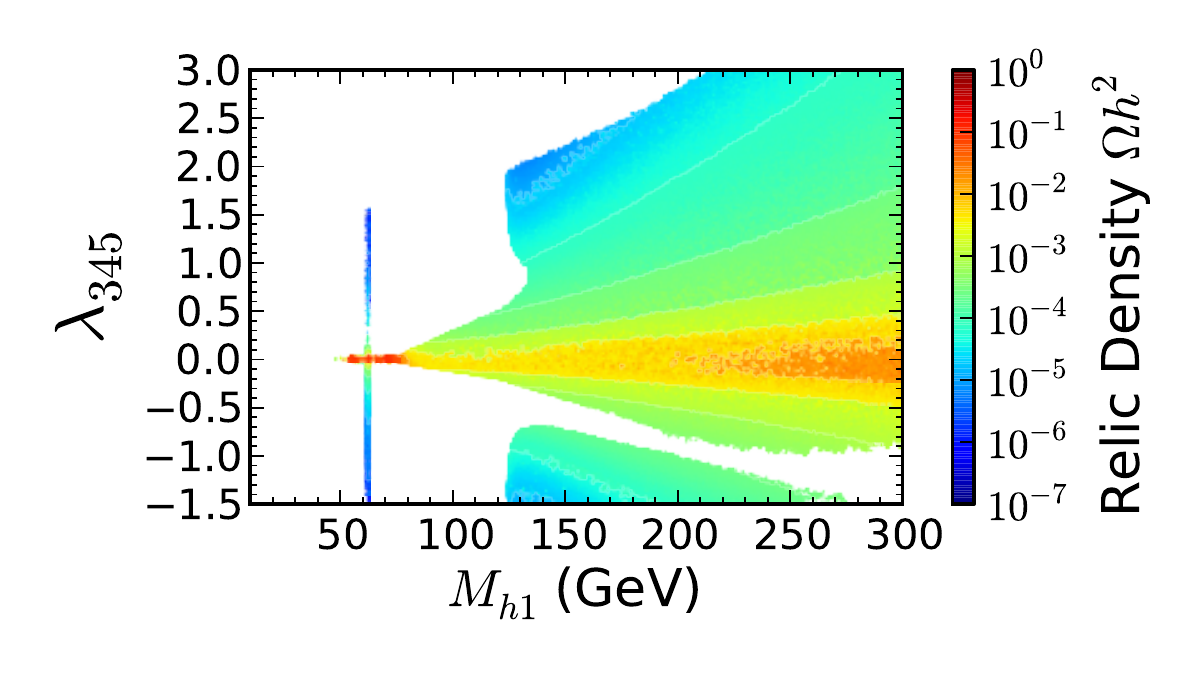}%
\caption{The new constraints on the i2HDM parameter space from XENON1T
searches for DM~\cite{Aprile:2017iyp}.
\label{fig:scan-simplified_xenon1t}}
\end{figure}
Our results are shown in Fig.~\ref{fig:scan-simplified_xenon1t}, where we present the i2HDM parameter space left after the application of theory, LEP,
 EWPT, LHC constraints as well as upper bounds on the relic density from Planck and DM DD limits from XENON1T. One can see a large effect of the XENON1T constraints on $\lambda_{345}$,
which improve LUX limits by more than one order of magnitude, chiefly,  over the
   $\MH/2<M_{h_1}<$ 125~GeV region.
In particular, in this region, $|\lambda_{345}|$ is limited to be always below about 0.05
which is crucial for one of the main signatures of  DM searches at the LHC which we discuss below.\footnote{One should note that in \cite{Ilnicka:2018def} authors have also analysed i2HDM parameter space using  XENON1T (2017)  constraints. However, the pattern of their surviving parameter space is quite different in some specific regions.  For example, for $M_{h1}$ just above  $M_{\rm H}/2$ we have found parameter space with $\lambda_{345}\simeq 1$, $M_{h1} \simeq M_{h2}$ which satisfy experimental limits and which have DM relic density below the PLANCK upper limit primarily because of the strong $h_1-h_2$ co-annihilation channel. We believe that this region  was missed in \cite{Ilnicka:2018def}.}

The asymmetric picture with respect to negative and positive values of
$\lambda_{345}$  is even more pronounced in case of these latest results
as one can clearly see the white funnel region excluded for $\lambda_{345}<0$.
The reason for this is again the negative interference between  DM annihilation into  EW gauge bosons via Higgs boson exchange and   $h_1 h_1 VV$  quartic couplings described above:
 in this funnel region this negative interference brings the DM relic density up,
which in turn increases the DM DD rates.

One should note that though constraints from DM DD and invisible Higgs decay
on  $|\lambda_{345}|$ dominate the one from vacuum stability,
the latter sets the most strict upper bound on $\lambda_{345}$
for $M_{h_1}\simeq \MH/2$. In this region the invisible Higgs decay
is suppressed by the phase space while DM DD rates rescaled by  relic density
are suppressed because  $\Omega h^2$ is driven to low values
in this  parameter space which is dominated by $h_1 h_1 \to H$ resonant annihilation.
Therefore the constraint from vacuum stability which becomes important in this region limits $\lambda_{345} \lesssim 1.6$ as follows from  Eq.~(\ref{l345limit}).

\section{Mono-jet signatures at the LHC}

\begin{figure}[htb]
\includegraphics[width=\textwidth]{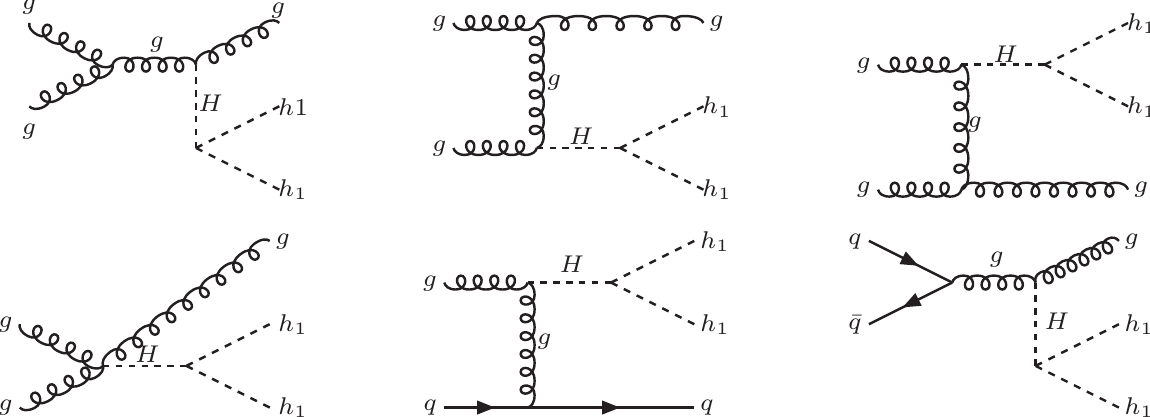}
\caption{Feynman diagrams for the $gg\to h_1 h_1+g$ process
contributing to the mono-jet signature.}
\label{fig:fd-mono-jet1}
\end{figure}

The i2HDM exhibits different collider signatures which can  potentially be accessible at the LHC. In this analysis we will focus on mono-jet final states, which arise from  $gg\to h_1 h_1+g$, $qg\to h_1 h_1+q$ and $q\bar{q}\to h_1 h_1+g$ processes, to which we will  refer cumulatively as the $h_1h_1j$ process. The corresponding Feynman diagrams are presented in Fig.~\ref{fig:fd-mono-jet1}.

For this signature, and for $M_{h_1}>\MH/2$, the relevant non-trivial parameter space is one dimensional and corresponds to the DM mass, $M_{h_1}$, since the production cross section is proportional to $(\lambda_{345})^2$.
For $M_{h_1}<\MH/2$, however, the situation can be different, for two reasons:
a)  only $H\to h_1 h_1$ takes place, so that the cross section is defined by the  production of the SM-like Higgs times  $Br(H\to h_1 h_1)$ which is a function of $\lambda_{345}$ and $M_{h_1}$;
b)  both  $H\to h_1 h_1$ and  $H\to h_2 h_2$ contribute to the invisible Higgs decay which then implies that both  $h_1 h_1j$ and $h_2 h_2j$ will contribute to the same signature
(for a few GeV mass difference between  ${h_2}$ and ${h_1}$,  $h_2\to h_1 f\bar{f}$ is invisible because of the soft fermions $f$ in the final state),  the cross section of which is defined by the  production of the SM-like Higgs state times  $(Br(H\to h_1 h_1)+ Br(H\to h_2 h_2)$ which is a function of $\lambda_{345}$, $M_{h_1}$ as well as $M_{h_2}$.

\begin{figure}[htb]
\centering
\includegraphics[width=0.8\textwidth]{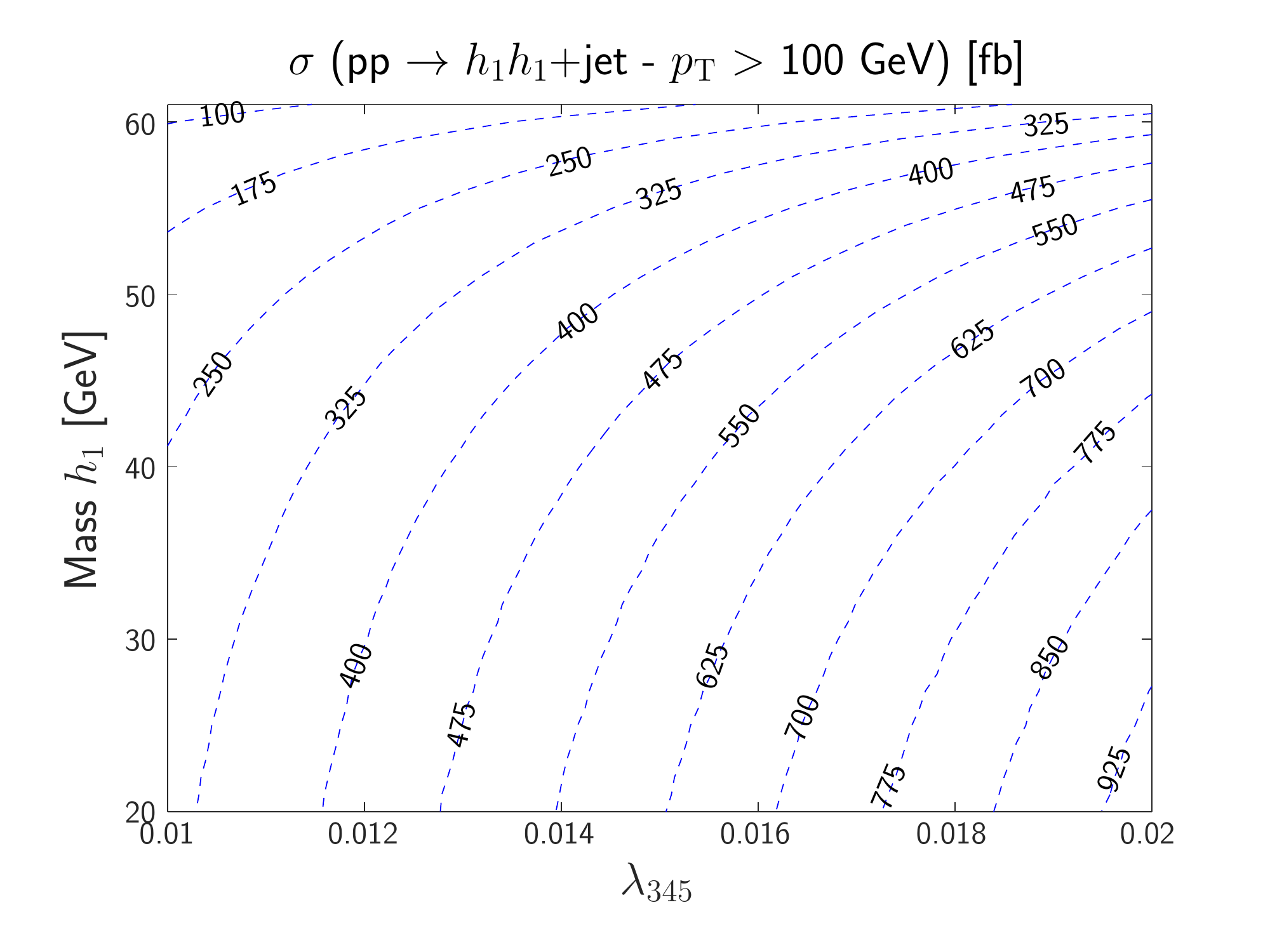}
\caption{Cross sections versus
DM mass $M_{h_1}$ and coupling constant $\lambda_{345}$ for the mono-jet process $h_1h_1j$ at the LHC@13 TeV. The mass of the $h_2$ particle is  set to  $M_{h_2} = \unit{200}{\GeV}$. Here, the cross section was evaluated for the initial cut on $p_\mathrm{T}^\mathrm{\rm jet} > \unit{100}{\GeV}$.}
\label{fig:csh_1h_1j}
\end{figure}

In Fig.~\ref{fig:csh_1h_1j} we present the cross sections  for the mono-jet process $h_1h_1j$ at the LHC@13 TeV in  the ($M_{h_1},\lambda_{345}$) plane. The mono-jet cross section was evaluated with the initial cut on $p_\mathrm{T}^\mathrm{\rm jet} > \unit{100}{\GeV}$, $\lambda_{345}$ has been chosen to be in the range [0.01, 0.02], $M_{h_1}$ has been chosen in the range \unit{[20, 60]}{\GeV} and $M_{h_2}$ has been fixed to $\unit{200}{\GeV}$.
We can see that, for this range of parameters, the cross section rate is between 100 and \unit{1000}{\fb}, which gives us a strong motivation to probe this signal at the LHC.
For this and the following parton level calculations and simulations  we have used the HEPMDB site \cite{HEPMDB}, the CalcHEP package~\cite{CALCHEP} and  the  NNPDF23LO (\verb|as_0130_qed|) Parton Distribution Function (PDF) set~\cite{Ball:2012cx} with both factorisation and
renormalisation  scale   set to the  transverse mass of the final state particles.

An important remark is that the mass of the top-quark in the loop which defines the $ggH$ coupling can be less than the energy scale of the $h_1 h_1 j$ process which is related to the jet transverse momentum,  $p_T^{\rm jet}$. Hence, in the region of high $p_T^{\rm jet}$, one should check the validity of the EFT approach based  on the heavy top-quark approximation
which is often used  for simplification.
This  is the subject of the next section.

\begin{figure}[htb]
\includegraphics[width=\textwidth]{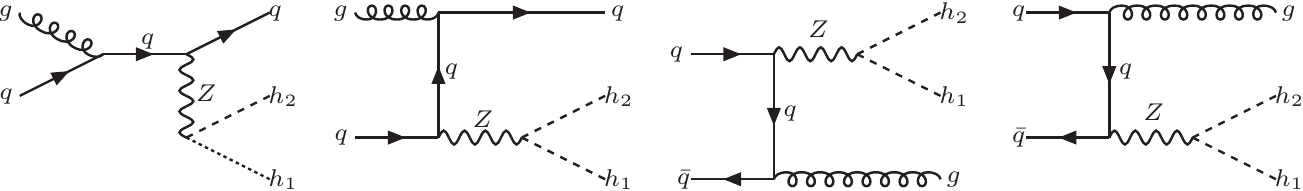}
\caption{Feynman diagrams for $q\bar{q}\to h_1 h_2+g$ ($gq\to h_1 h_2+q$) process.}
\label{fig:fd-mono-jet2}
\end{figure}
\begin{figure}[htb]
\centering
\includegraphics[width=0.8\textwidth]{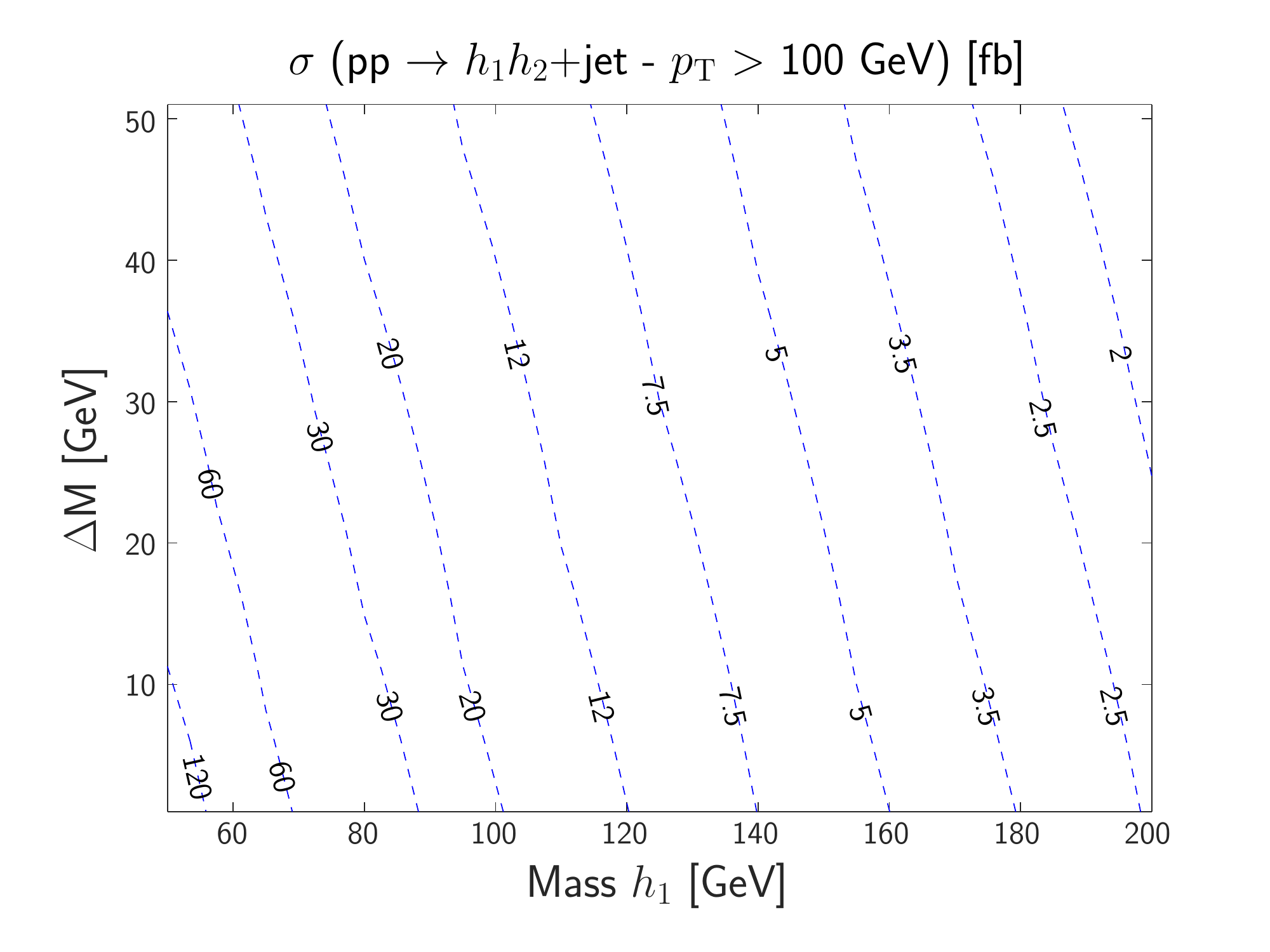}
\caption{Cross section versus DM mass $M_{h_1}$ and $\Delta M = M_{h_2} - M_{h_1}$ for the mono-jet process $h_1h_2j$ at the LHC@13 TeV. This process gives rise to a mono-jet signal if the mass difference  $\Delta M$  is small enough, such that the decay $h_2 \to h_1 + X$ gives rise only to \MET{} + soft undetected leptons or jets. Here, the cross section was evaluated for the initial cut  $p_\mathrm{T}^\mathrm{\rm jet} > \unit{100}{\GeV}$.}
\label{fig:csh_1h2j}
\end{figure}

There is one more process that potentially contributes to the mono-jet signature in the i2HDM, namely,
$q\bar{q}\to h_1 h_2+g$ ($gq\to h_1 h_2+q$), which we will refer to as the $h_1h_2j$ process. Feynman diagrams for this process are presented in
Fig.~\ref{fig:fd-mono-jet2}.
This process contributes to the mono-jet signature when the mass splitting between $h_1$ and $h_2$ is small, of the order of few GeV.
In this case, $h_2$ will decay to $h_1$ and  soft jets or  leptons from a  virtual $Z$ which escapes detection.
In spite of the fact that there is one  mediator for this process, {i.e.} the $Z$ boson,
one can  see that $t-$ and $s-$channel topologies with a light quark in the propagator
make this process different from simplified models with fermionic DM and a vector mediator which have been
studied so far in literature, so it is worth exploring it in detail.

The parameter space for this process is characterised by two variables, $M_{h_1}$ and $M_{h_2}$, which fix its cross section for a given collider energy.
It is also convenient to use $\Delta M = M_{h_2} - M_{h_1}$, the mass difference between the two particles, instead of $M_{h_2}$.
In Fig.~\ref{fig:csh_1h2j} we present the cross section for the $h_1h_2j$ process in the $(M_{h_1}, \Delta M)$ plane.
The cross section has been evaluated with an initial cut, $p_\mathrm{T}^\mathrm{\rm jet} > \unit{100}{\GeV}$.
One can see that, in this plane, the pattern of the cross section iso-levels takes a simple form. One can also note that in case of $M_{h_1}\simeq 50-60$~GeV and small $\Delta M$ the cross section is of the order of 100 fb, which could be in the region of the LHC sensitivity. {This cross section is comparable to that of the $h_1h_1j$ process for $\lambda_{345}$ = 0.01 and the same mass, which makes this kind of process important for the parameter space region where $\lambda_{345}$ is small.} 
It is important to stress that the cross section for the $h_1h_2j$ process is independent of $\lambda_{345}$, therefore this process would provide a probe of the i2HDM parameter space which is  complementary to the $h_1h_1j$ process\footnote{On a similar footing, we should finally mention that the mono-jet signature emerging from the
i2HDM also sees a component in which an $h_2$ pair is produced, when $\Delta M$ is very small. The production topologies of this process, henceforth $h_2h_2j$, are the same as those for the $h_1h_1j$ case, though the yield is generally smaller. We nonetheless include this channel in our simulations yet we will not dwell on it separately.}.

\subsection{Validity of the effective $ggH$ vertex approach}

The  SM $ggH$ vertex is dominantly generated by the top-quark
loop (with a  small bottom quark contribution). It is known that integrating out the top quark  is a good approximation for Higgs production processes when considering inclusive rates,  as long as the Higgs boson is not far off-shell or with high transverse momentum. The literature on this subject is vast and we refer the reader  to the corresponding  sections in Ref.~\cite{deFlorian:2016spz} and references therein. In case of our study, however,  the selection of large transverse momentum of the  jet (done to increase the signal-to-background ratio),
which is typically bigger then the top-quark mass, is likely to lead to the breakdown of the heavy top-quark  approximation.

\begin{figure}[htb]
\centering
\includegraphics[width=0.4\textwidth]{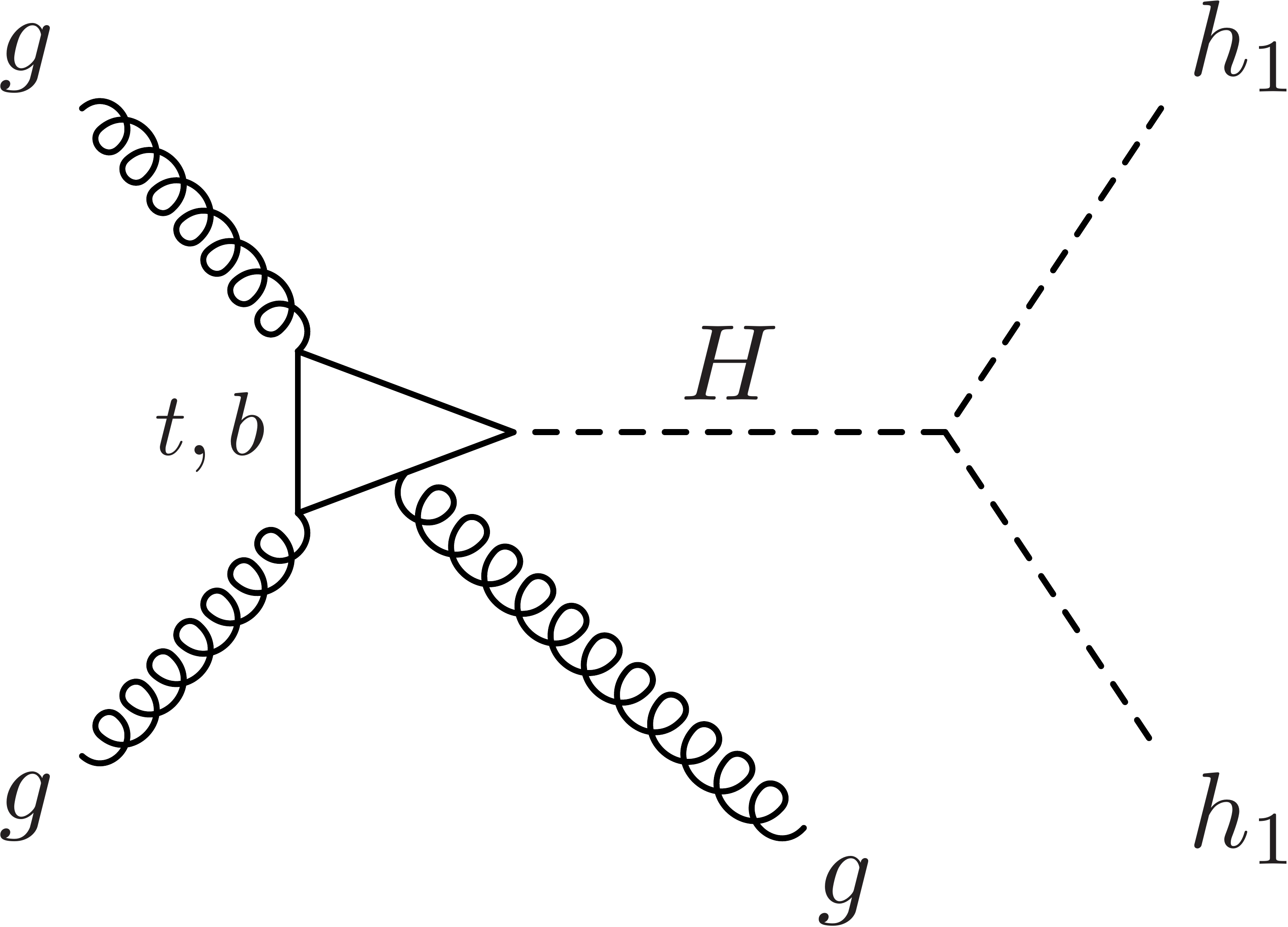}
\caption{\label{fig:topLoopExpanded} Representative Feynman diagram
for the  one-loop (top- and bottom-quark) induced $gg\to h_1 h_1 g$ process under study.}
\end{figure}

From one of the representative one-loop diagram presented in Fig.~\ref{fig:topLoopExpanded} for the $gg\to h_1 h_1 g$ process one can see  that a high $p_T$ jet  emitted from the top-quark loop can `resolve" the top-quark in the loop if the transverse momentum of the jet is large enough. This effect is crucial  since the mono-jet $p_T$ and $\MET$ distributions from the EFT approximation (which one could be tempted to use for the sake of simplicity) could be different from those  described by the exact loop calculation. This is even more crucial for us, due to the $\MET$ shape-analysis techniques which we use in our study. Therefore, we have compared the $\MET$ shapes for the events simulated using the EFT heavy top-quark approximation to those from the exact one-loop calculation.
For this purpose we have simulated the process of Higgs boson production in association with a jet and scanned over the mass of the Higgs boson, corresponding to the different invariant masses of the DM pair.

\begin{figure}[htb]
\centering
\includegraphics[width=0.725\textwidth]{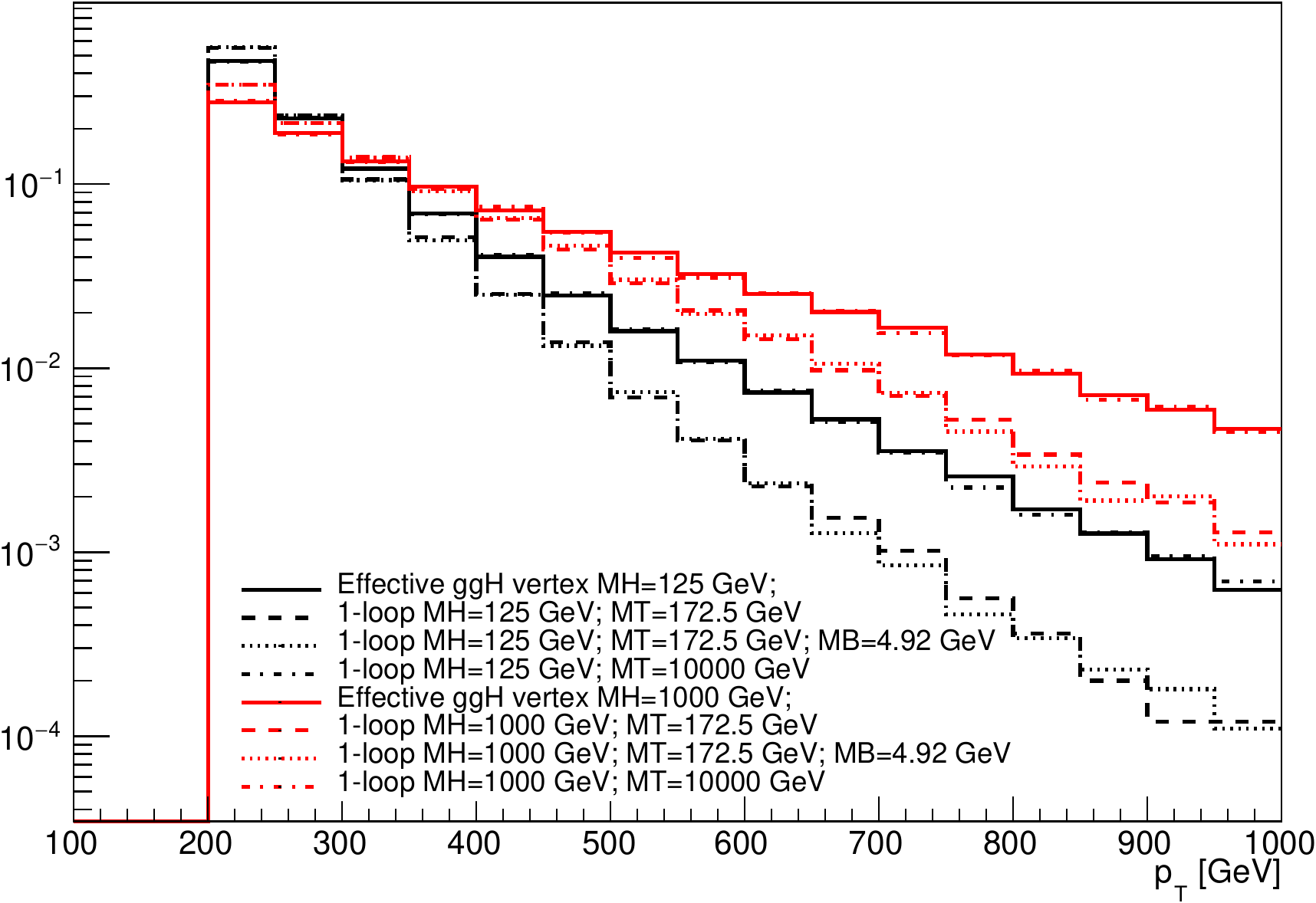}\\
\includegraphics[width=0.725\textwidth]{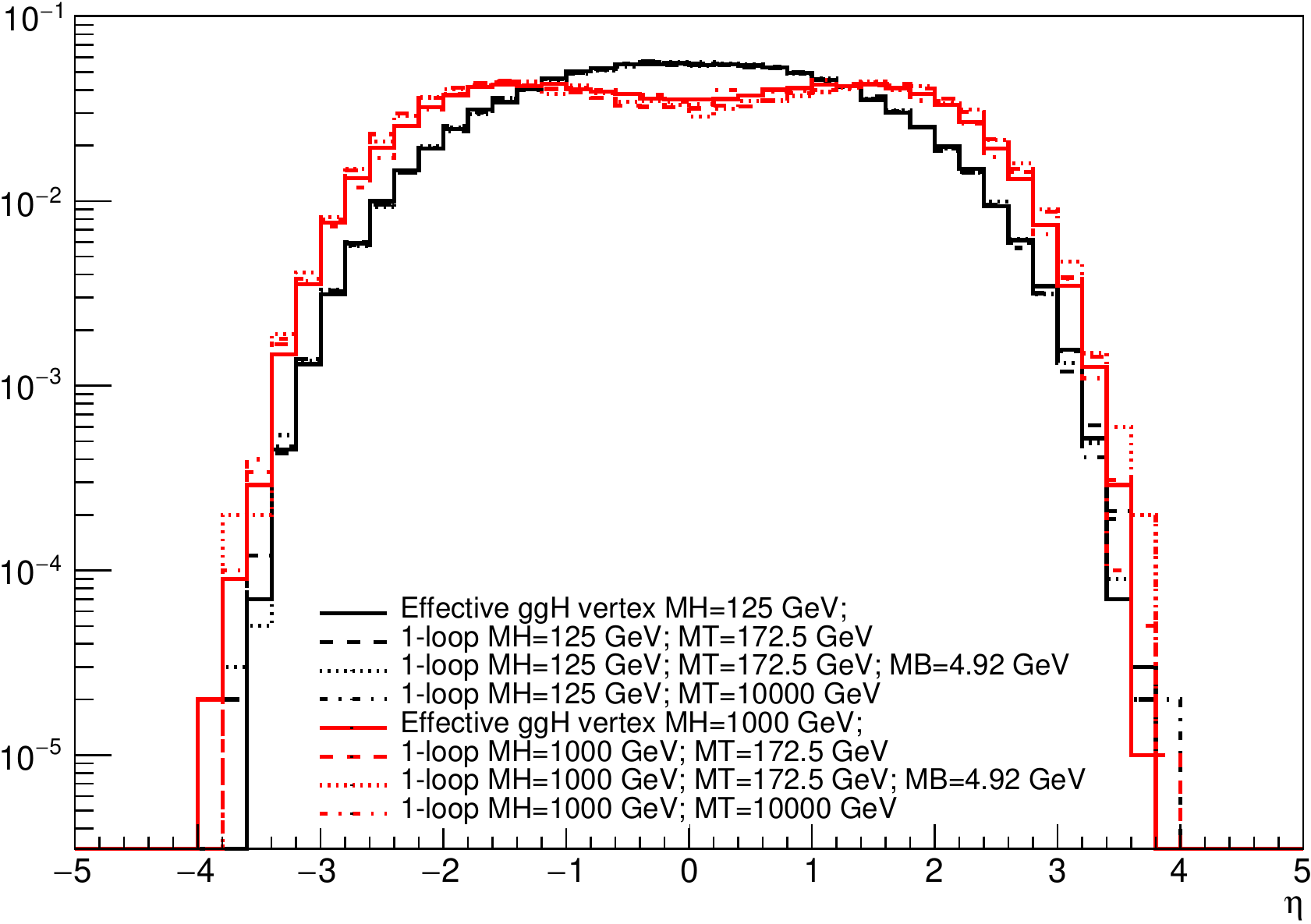}
\caption{\label{fig:EFTNLOdist} Shape of transverse momentum (top panel) and pseudo-rapidity (bottom panel) distributions for a Higgs boson produced in association with one jet. Solid curves correspond to the distributions for the
effective $ggH$ vertex approximation, dashed ones are for the one-loop result with only the top-quark in the loop,
dotted ones are for the one-loop result with  both top- and bottom-quarks in the loop and dot-dashed ones are  for a
very heavy quark in the loop ($m_\mathrm{t}=\unit{10}~{\TeV}$) for the purpose of cross-checking  the effective $ggH$ vertex approximation.
Black and red colours  correspond to $M_{(\rm{DM,DM})}=125$~GeV and 1000 GeV, respectively.}
\end{figure}

\clearpage

\begin{figure}[htb]
\centering
\includegraphics[width=0.8\textwidth]{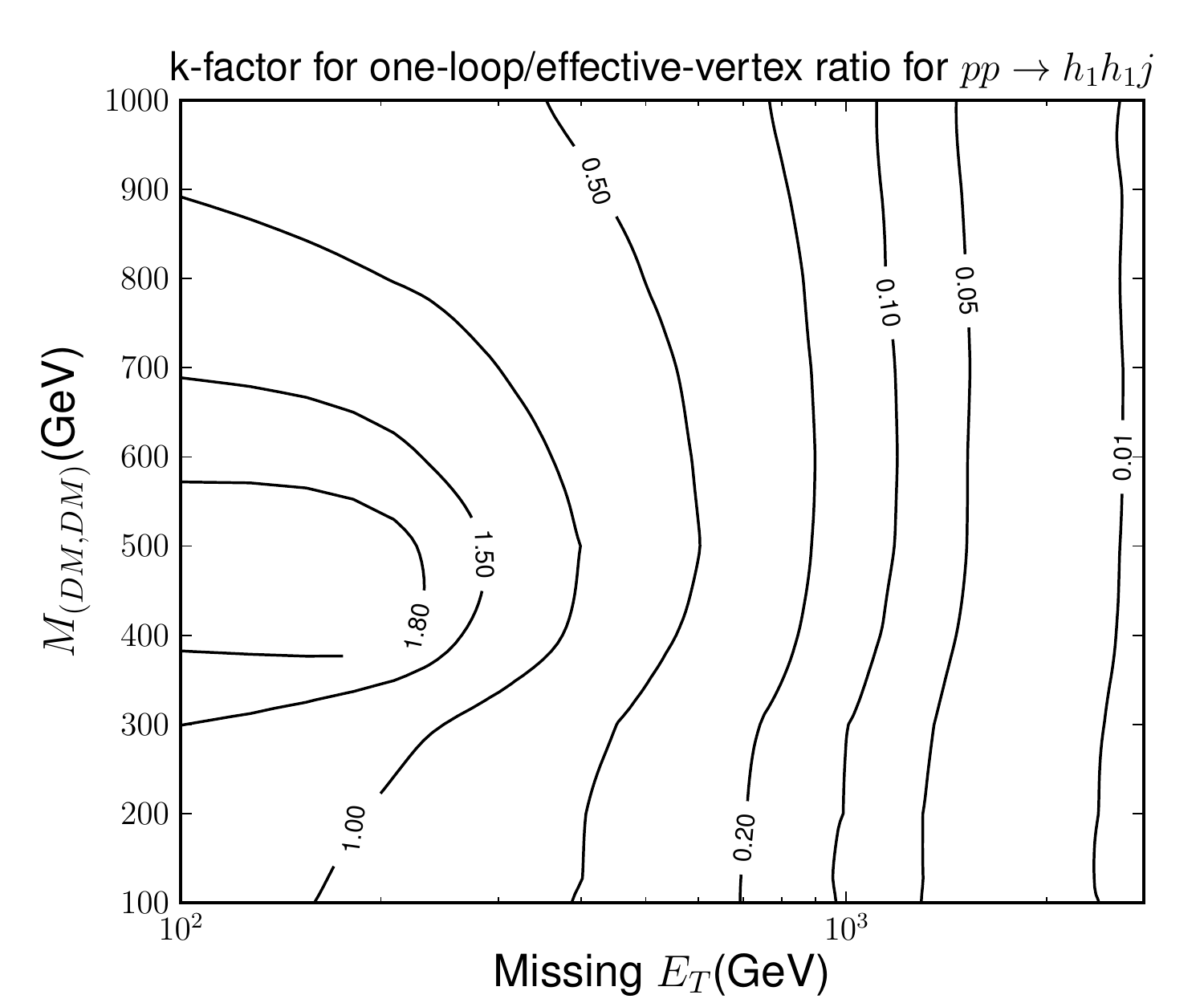}
\caption{\label{fig:kf} The $k$-factor $k_f$ defined in the text  as a function of the \MET{} and  invariant mass of the DM pair.
The $k$-factor values are indicated in the contour lines.}
\end{figure}

For this specific study, our simulations have been performed with {\sc MadGraph5} \cite{Alwall:2011uj,Alwall:2014hca} using the {\sc NNPDF2.3} PDF set~\cite{Ball:2012cx}. We  have compared results for two models:
{\sc MadGraph5} native SM implementation  with the effective $ggH$ vertex and the SM at  one-loop implementation.
Using this setup we have scanned  over a range of  Higgs boson masses  and
compared the Higgs boson $p_T$  and $\eta$ distributions for the effective $ggH$  vertex and one-loop level implementations. The results presented in  Fig.~\ref{fig:EFTNLOdist}
are evaluated  for different benchmarks, corresponding to different Higgs masses (for both effective vertex and one-loop simulations) and with contributions of either top- or bottom-quarks, or both. We have applied an initial cut $\PT^{\rm jet}>\unit{200}\GeV$ for this study. The differences between effective vertex and one-loop distributions are quite large for large transverse momenta and the role of the bottom quark in the loop is -- as one may expect -- rather marginal. The pseudo-rapidity distribution is not affected as much. It is however interesting to notice that larger invariant masses shift the distribution from a central-peaked shape to a more forward-backward behaviour. As a sanity check we have also evaluated the effect of setting the top-quark mass to 10 TeV in the one-loop calculation, so that it can be effectively cross-checked with the effective vertex results
and, as one can see, indeed it agrees with those.

As a result of this comparison, we have defined  a $k$-factor
\begin{equation}
k_f=\frac{\sigma(pp\to h_1h_1j)_{\rm one-loop}}{\sigma(pp\to h_1h_1j)_{\rm EFT}},
\end{equation}
which provides correspondence  between the effective vertex and one-loop results in the  $p_T$ distribution of the Higgs boson as a function of
two variables: \MET\ and $M_{(\rm{DM,DM})}$ (the invariant mass of the DM pair). This $k$-factor is pictorially presented in Fig.~\ref{fig:kf}. One can see that, for large $\MET$ values, the effect of the breakdown of the effective vertex approximation is dramatic. For example, for  $\MET$=1 TeV, the effective vertex approximation overestimates the one-loop result by one order of magnitude, i.e. $k_f\simeq 0.1$.  At the same time, for smaller values of $\MET$,
the effective vertex approximation can even underestimate the one-loop result, which happens for large values of the $M_{(\rm{DM,DM})}$ invariant mass of the  DM pair: for example, for $\MET=300$ GeV and $M_{(\rm{DM,DM})}=500$ GeV, one finds $k_f\simeq 1.5$.

It is finally very important to stress that the $k$-factor which
was found in two recent studies  at Next-to-Leading Order (NLO) in QCD~\cite{Lindert:2018iug,Jones:2018hbb}
is very close (within few percent) to the one we have found here at the LO only (see also \cite{Neumann:2018bsx}
where similar work is presented). Hence,
based on our findings in this section,
for our analysis below we use one-loop results (i.e at LO in QCD) and  take into account the contributions from both top- and bottom-quarks.

\subsection{LHC potential to probe the i2HDM parameter space}

\subsubsection{Benchmarks}
Taking into account all the constraints, and especially the recent XENON1T ones, we suggest a set of six Bench-Mark (BM) points, BM1 to BM6, summarised in Tab. \ref{tab:i2HDMbenchMarks} and described below.
\begin{itemize}
\item {\bf BM1} -- Both $M_{h_{1}}$ and $M_{h_{2}}$ are below $\MH/2$ contributing to about $20\%$ of the invisible Higgs boson decay
and yielding about 800 fb of cross section for the mono-jet signature (which is high enough to be tested at the HL-LHC as we will discuss below) coming from the cumulative sum of the
${h_1 h_1 j}$, ${h_1 h_2 j}$ and ${h_2 h_2 j}$ processes. To measure the XENON1T sensitivity we use the SI DM scattering rate on the proton ($\sigma_{\rm SI}^p$)
accompanied by its ratio to the experimental limit from XENON1T, following re-scaling with the relic density, $R_{\rm SI}^{\rm{XENON1T}} =(\sigma_{\rm SI}^p/\sigma_{\rm SI}^{\rm{XENON1T}})\cdot (\Omega_{\rm DM}/\Omega_{\rm DM}^{\rm Planck})$, which is equal to 0.29 for this benchmark, {i.e.} about a factor of three below the current XENON1T sensitivity. The DM relic density for this point is below the Planck constraints because of the $h_1h_2$ co-annihilation.
\item {\bf BM2} -- Only  $M_{h_{1}}$ is below $\MH/2$ and the value of $\lambda_{345}$ is chosen to be small enough for the DM relic density to match both the upper and  {lower}
Planck constraints. In this case,  the invisible Higgs boson decay to DM is only 2\% and the respective rate of the ${h_1 h_1 j}$ mono-jet signal is only 74.6 fb. This point, with $R_{\rm SI}^{\rm{XENON1T}}=0.75$, is likely to be tested with future DM DD experiments since its value is not far from the present XENON1T limit.
\item {\bf BM3} -- Only  $M_{h_{1}}=60$ GeV is below $\MH/2$, but $M_{h_{2}}=68$~GeV is quite close to it. Because of  the large  invisible Higgs boson decay to DM with $Br(H\to h_1 h_2) = 0.25\% $, the leading signal at the LHC will be mono-jet from ${h_1 h_1 j}$, with a rate above 800 fb, complemented by the ${h_1 h_2 j}$ process with rate 77.4 fb, which are high enough to be tested at the HL-LHC.
\item {\bf BM4} -- $M_{h_{1}}=60$ GeV and  $M_{h_{2}}=68$~GeV as in BM3, but $\lambda_{345}$ is chosen to be low enough such that the DM relic density, governed by $h_1h_2$ co-annihilation, is within the upper and {lower} Planck constraints. This point is unlikely to be tested by DD DM experiments in the near future while the LHC could potentially test it shortly via a combination of $h_1 h_2j$, $h_1 h^{\pm}j$, $h_2 h^\pm j$ and $h^\pm h^\pm j$ signatures, which are  outside the scope of this paper.
\item {\bf BM5} -- With all inert scalars close in mass, $M_{h_{1}}=70$ GeV, $M_{h_{2}}=78$~GeV, $M_{h^\pm}=78$~GeV, so all $h_1 h_2j$, $h_1 h^{\pm}j$, $h_2 h^\pm j$ and $h^\pm h^\pm j$
channels contribute to the mono-jet signature (since both  $h_2$ and
$h^\pm$ promptly decay to $h_1$ and soft leptons escaping detection)
with a total rate of about 250 fb, which is close to the exclusion limit at the HL-LHC as we will see below.
\item {\bf BM6} -- With all inert scalars even more close  in mass in comparison to BM5, since $M_{h_{1}}=80$ GeV, $M_{h_{2}}=81$~GeV and  $M_{h^\pm}=81$~GeV,  as well as $\lambda_{345}=0$ (hence $h_1h_1j$ and $h_2h_2j$ are not possible)
so that all $h_1 h_2j$, $h_1 h^{\pm}j$, $h_2 h^\pm j$ and $h^\pm h^\pm j$ channels contribute to the mono-jet signature  with a total rate of about 210 fb, again close the exclusion limit at the HL-LHC.
\end{itemize}

\begin{table}[!t]
\scriptsize
\centering
\begin{tabular}{ccccccc}
\toprule
 {\bf BM}                       &  {\bf 1}  		& {\bf 2}  	   	& {\bf 3}  		& {\bf 4}  		 & {\bf 5}		 & {\bf 6}		  	\\
  \midrule
  $M_{h_{1}}$ (GeV)     	& 55      		& 55 			& 60 		      	& 60		     	 & 70			 & 80			 	\\
  $M_{h_{2}}$ (GeV)     	& 62      		& 110 			& 68  		      	& 68	       	    	 & 78			 & 81  		 	\\
  $M_{h_{+}}$ (GeV)   		& 120     		& 120 			&100  	  	      	& 100		      	 & 78			 & 81  		 	\\
  $\lambda_{345}$       	& $0.01$  	 	& $0.0065$  		& $0.033$ 		& $0.0001$ 	       	 & $0.01$		 & $0.0$		 	\\
  $\lambda_{2}$         	&  1.0    		& 1.0		        & 1.0	    	      	& 1.0		     	 & 1.0  		 & 1.0  		 	\\
  \midrule
  $\Gamma_{h_{2}}$ (GeV)    & 1.307$\times10^{-8}$ & 2.926$\times10^{-4}$ & 2.564$\times10^{-8}$ & 2.564$\times10^{-8}$ & 2.627$\times10^{-8}$ & 7.314$\times10^{-13}$ \\
  $\Gamma_{h_{+}}$ (GeV)   	& 1.549$\times10^{-3}$ & 9.905$\times10^{-4}$ & 1.137$\times10^{-4}$ & 1.137$\times10^{-4}$ & 3.666$\times10^{-8}$ & 7.587$\times10^{-13}$ \\
  \midrule
  $\Omega_{\rm DM} h^2$         & $1.78 \times 10^{-2}$ & $1.10\times 10^{-1}$  & $1.37 \times 10^{-4}$ & $1.04 \times 10^{-1}$	 & $4.56 \times 10^{-2}$ & $7.52 \times 10^{-3}$  	\\
  \midrule
  $\sigma_{\rm SI}^p$ (pb)    	& $1.75 \times 10^{-10}$& $7.37\times 10^{-11}$ & $1.59 \times 10^{-9}$ & $1.46 \times 10^{-14}$ & $1.07 \times 10^{-10}$& $0.0$ 			\\
  $R_{\rm SI}^{\rm{XENON1T}}$     	& $0.29$ 		&  $0.75$		& $0.020$  		& $1.4 \times 10^{-4}$   & $0.45$		  & $0.0$		 	\\
 \midrule
  $Br(H\to h_1 h_1)$     	& $4.15\times 10^{-2}$ 	&  $0.022$  		& $0.25$  		& $3.1 \times 10^{-6}$   & $0.0$		  & $0.0$		 	\\
  $Br(H\to h_2 h_2)$     	& $1.59\times 10^{-1}$ 	&  $0.0$  		& $0.0$  		& $0.0$  		 & $0.0$		  & $0.0$		 	\\
\midrule
  $\sigma_{{\rm LHC}@13~{\rm TeV}}$ (fb) &&&&\\
  ${h_1 h_1 j}$  		& $1.46\times 10^{2}$ 	&  $74.6$ 		& $857$  		& $1.08\times 10^{-2}$ 	 & $4.96\times 10^{-3}$	 & $0.0$  	\\
  ${h_2 h_2 j}$  		& $5.47\times 10^{2}$ 	&  $3.88\times 10^{-1}$ & $3.06\times 10^{-1}$  & $8.00\times 10^{-2}$ 	 & $5.50\times 10^{-2}$  & $5.34\times 10^{-4}$  	\\
  ${h_1 h_2 j}$  		& $1.04\times 10^{2}$   &  $34.1$  		& $77.4$  		& $77.0$ 		 & $49.6$		 & $39.0$		 	\\
  ${h_1 h^{\pm} j}$  		& $49.2$   		&  $49.0$  		& $65.0$  		& $65.5$ 		 & $83.9$		 & $66.5$		 	\\
  ${h_2 h^{\pm} j}$  		& $44.9$   		&  $24.7$  		& $58.0$  		& $57.9$ 		 & $72.1$		 & $65.4$		 	\\
  ${h^{\pm} h^{\pm} j}$  	& $13.0$   		&  $13.0$  		& $20.9$  		& $16.2$ 		 & $39.2$		 & $35.3$		 	\\
   \bottomrule
\end{tabular}
\caption{BM points from the i2HDM parameter space
together with corresponding observables: DM relic density ($\Omega_{\rm DM} h^2$),
 SI  DM scattering rate on the proton ($\sigma_{\rm SI}^p$)
accompanied by its ratio to the experimental limit from XENON1T following re-scaling with the relic density,
$R_{\rm SI}^{\rm{XENON1T}} =(\sigma_{\rm SI}^p/\sigma_{\rm SI}^{\rm{XENON1T}})\cdot (\Omega_{\rm DM}/\Omega_{\rm DM}^{\rm Planck})$ plus
 the LHC cross sections  at the LHC@13 TeV with a $p_T^{\rm jet} > 100$ GeV cut applied.
\label{tab:i2HDMbenchMarks}}
\end{table}

The  masses of DM for BM1--BM6 were chosen below 100 GeV in anticipation of the LHC sensitivity to the parameter space which we present below.
At the time of  writing, the LHC experimental collaborations ATLAS and CMS do not have specific searches for the
i2HDM, however, the results for generic DM searches in the jet+\MET{} channel can be reinterpreted in the context of such a model.
In order to compare the i2HDM to those limits, the following procedure is followed.
\begin{itemize}
\item The matrix elements that describe the hard interaction are simulated with CalcHEP and event samples for
different values of $M_{h_1}$ are produced. In order to concentrate on a region of  phenomenological interest
and simulate events with  enhanced statistics, a lower threshold on the final state
parton (either $q$ or $g$) is set at $\PT >$ 100~GeV. The event samples are produced in the Les Houches Event format for further processing.
\item In order to accurately describe the $\PT$ distribution each event is weighted with the $k$-factor estimated in the previous section, according to its parton $\PT$ and the invariant mass of the DM-DM system.
\item Each event sample is then passed to PYTHIA~8.2~\cite{Sjostrand:2006za,Sjostrand:2007gs}
for the proper treatment of parton showering, hadronisation and underlying event effects. The aforementioned {\sc NNPDF} set is again deployed through the LHAPDF6 tool~\cite{Buckley:2014ana}.
\item The DELPHES 3 framework for fast simulation of generic collider experiments \cite{deFavereau:2013fsa}
is used to simulate the event reconstruction by the CMS experiment. Specifically, the detector parametrisation for CMS described as standard in the card \texttt{delphes\_card\_CMS.tcl}  from the DELPHES distribution is used.
\item A set of selection criteria is applied to the simulated reconstructed events. In the experimental collaborations, these criteria aim to reduce both the SM backgrounds (mainly composed of inclusive $W/Z$  boson production) and instrumental noise that mimics the appearance of a single, highly energetic jet in the event. We disregard the effect of the latter phenomenon in our analysis, though.
\end{itemize}
After the  fast detector level simulation described above we have performed an  analysis of the missing transverse momentum
distribution \MET{} of the  signal events. 
{We compare the \MET{} distribution predicted by a given signal sample to
the standard model background prediction. That background prediction is a
fundamental ingredient for our analysis: it can either be explicitly given
by the experimental collaborations or estimated by an explicit calculation
of the inclusive $W/Z$ boson production cross section shape.
For each signal sample, we set upper limits on the production cross section of the mono-jet process.
We compute the limits following 
an asymptotic approximation to the modified frequentist prescription known as 
the CL$_{\rm S}$ technique~\cite{Read:2002hq,Cowan:2010js},
in which systematic uncertainties are treated as 
nuisance parameters through use of the profile likelihood ratio.
The only systematic uncertainty we consider in our analysis is
the uncertainty in the background prediction.
Throughout our study, the \texttt{theta} framework~\cite{ref:thetaFramework}
for modelling and inference is used for all statistical analyses and limit-setting procedures.
}

We study the jet+\MET{} signature from two signal processes: $h_1 h_1j$ and $h_1 h_2j$, in presence of a
  small (a few GeV) $M_{h_2}-M_{h_1}$ mass gap making $h_1 h_2j$ to contribute to the mono-jet signature.
In our study, we analyse $h_1 h_1j$ and $h_1 h_2j$ separately because of two reasons:
a) the rate of these  processes depends upon different parameters, so they complement each other as i2HDM parameter space probes;
b) these processes have different  shapes in the \MET{} distribution because of the different nature and mass of the mediators.


\subsubsection{Results from Run 2 data}

At the  beginning of  Run 2   the LHC@13TeV delivered a  total   integrated luminosity of 4.2~\fbinv. The CMS collaboration released a public result where 2.3~\fbinv of data were used to search for DM production in association with jets or hadronically decaying vector bosons~\cite{CMS:2016tns}. {Henceforth, we will refer to this result as the ``CMS Run 2 analysis''.} Supplementary material -- data and Monte Carlo (MC)  background distributions as well as their uncertainties -- were made available by the collaboration and used to set limits on the i2HDM. Tab.~\ref{tab:selectionEXO16013} summarises the experimental selection used for the CMS result while Tab.~\ref{tab:dataFromEXO16013} presents the data used for our study at\unit{13}{\TeV}.

\begin{table}[htbp]
\centering
\begin{tabular}{ll}
\toprule
Quantity & Selection\\
\midrule
Leading jet $\PT$    & $> 100$~GeV \\
Leading jet $|\eta|$ & $<$~2.5     \\
$\MET$               & $>$~200~GeV\\
$\Delta\phi (\MET,\textrm{jet}_{1\ldots4})$ & $> 0.5$\\
\bottomrule
\end{tabular}
\caption{\label{tab:selectionEXO16013}Initial  selection cuts for the CMS Run 2 mono-jet analysis at  $\sqrt{s}$ = \unit{13}{\TeV}~\cite{CMS:2016tns}. Jets considered for the jet multiplicity and angular configuration selections are required to have
$\PT^{\rm jet} > 30$~GeV and $|\eta^{\rm jet}| < 2.5$.}
\end{table}

\begin{table}[htbp]
\scriptsize
\begin{minipage}{.48\textwidth}
\centering
\begin{tabular}{lrclr} 
\toprule
Bin range (GeV) & \multicolumn{3}{c}{SM background} & Observed data\\
\midrule
200	--	230	&		28654	&$\pm$&	171		&	28601	\\
230	--	260	&		14675	&$\pm$&	97		&	14756	\\
260	--	290	&		7666	&$\pm$&	68		&	7770	\\
290	--	320	&		4215	&$\pm$&	48		&	4195	\\
320	--	350	&		2407	&$\pm$&	37		&	2364	\\
350	--	390	&		1826	&$\pm$&	32		&	1875	\\
390	--	430	&		998		&$\pm$&	23		&	1006	\\
430	--	470	&		574		&$\pm$&	17		&	543	\\
470	--	510	&		344		&$\pm$&	12		&	349	\\
510	--	550	&		219		&$\pm$&	9		&	216	\\
550	--	590	&		134		&$\pm$&	7		&	142	\\
\bottomrule
\end{tabular}
\end{minipage}\hfill
\begin{minipage}{.48\textwidth}
\centering
\begin{tabular}{lrclr} 
\toprule
Bin range (GeV) & \multicolumn{3}{c}{SM background} & Observed data\\
\midrule
590	--	640	&		98.5	&$\pm$&	5.8		&	111	\\
640	--	690	&		58.0	&$\pm$&	4.1		&	61	\\
690	--	740	&		35.2	&$\pm$&	2.9		&	32	\\
740	--	790	&		27.7	&$\pm$&	2.7		&	28	\\
790	--	840	&		16.8	&$\pm$&	2.2		&	14	\\
840	--	900	&		12.0	&$\pm$&	1.6		&	13	\\
900	--	960	&		6.9		&$\pm$&	1.2		&	7	\\
960	--	1020 &		4.5		&$\pm$&	1.0		&	3	\\
1020 --	1160 &		3.2		&$\pm$&	0.9		&	1	\\
1160 --	1250 &		2.2		&$\pm$&	0.7		&	2	\\
1250 --	inf	 &		1.6		&$\pm$&	0.6		&	3	\\
\bottomrule
\end{tabular}
\end{minipage}
\caption{\label{tab:dataFromEXO16013}SM background prediction and observed data for the CMS Run 2 mono-jet analysis at $\sqrt{s}$ = \unit{13}{\TeV}~\cite{CMS:2016tns} as function of the \MET{} variable.}
\end{table}

\begin{figure}[htb] 
   \centering
   \includegraphics[width=0.53\textwidth]{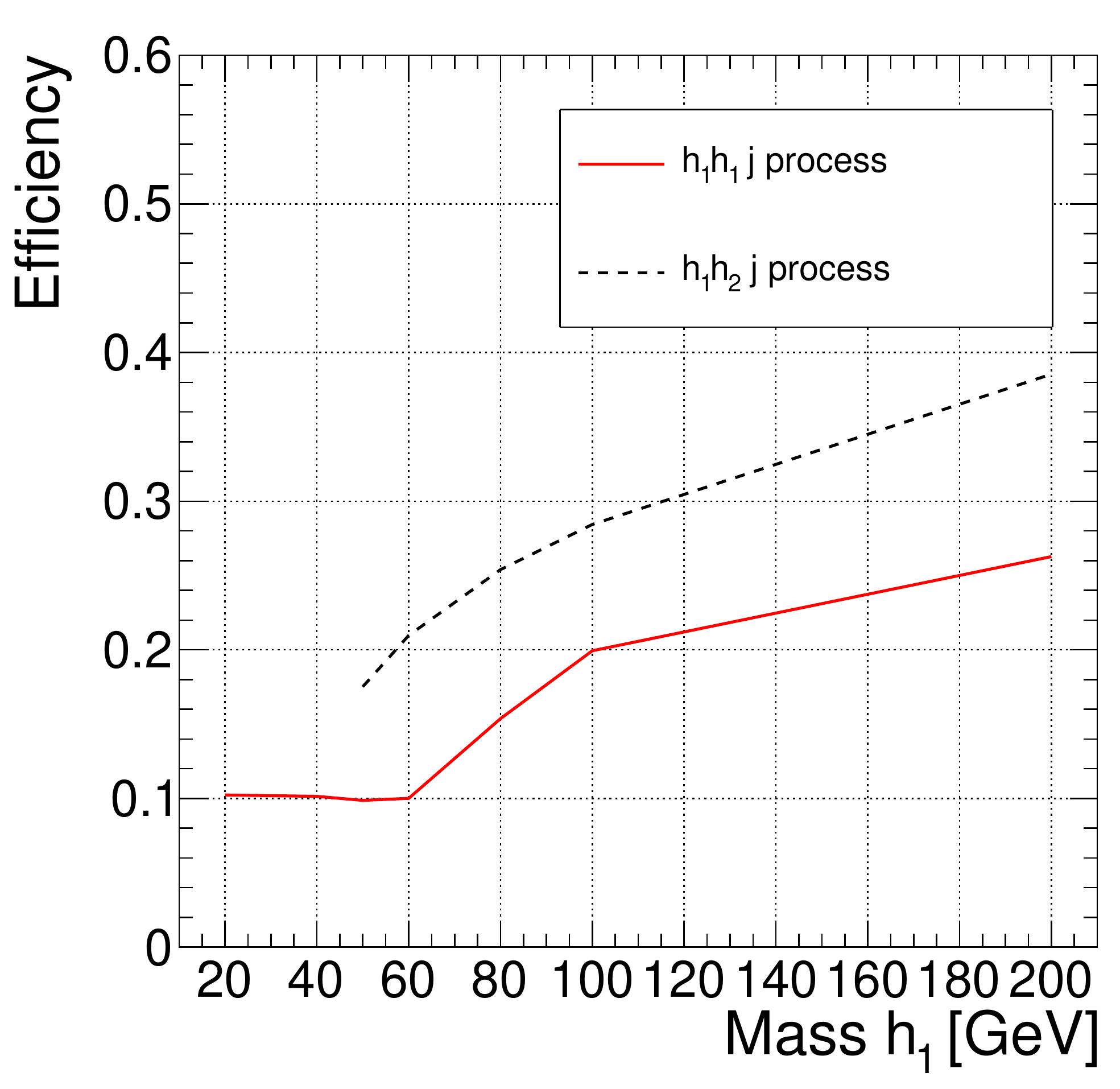}
   \caption{\label{fig:selectionEfficiency}Efficiency of the selection criteria described in Tab.~\ref{tab:selectionEXO16013}, on the simulated reconstructed events, for  $h_1h_1j$  (solid red line) and $h_1h_2j$ (dashed black line). Here, $\sqrt s=13$ TeV}
\end{figure}

\begin{figure}[htb]
   \centering
   \includegraphics[width=0.52\textwidth]{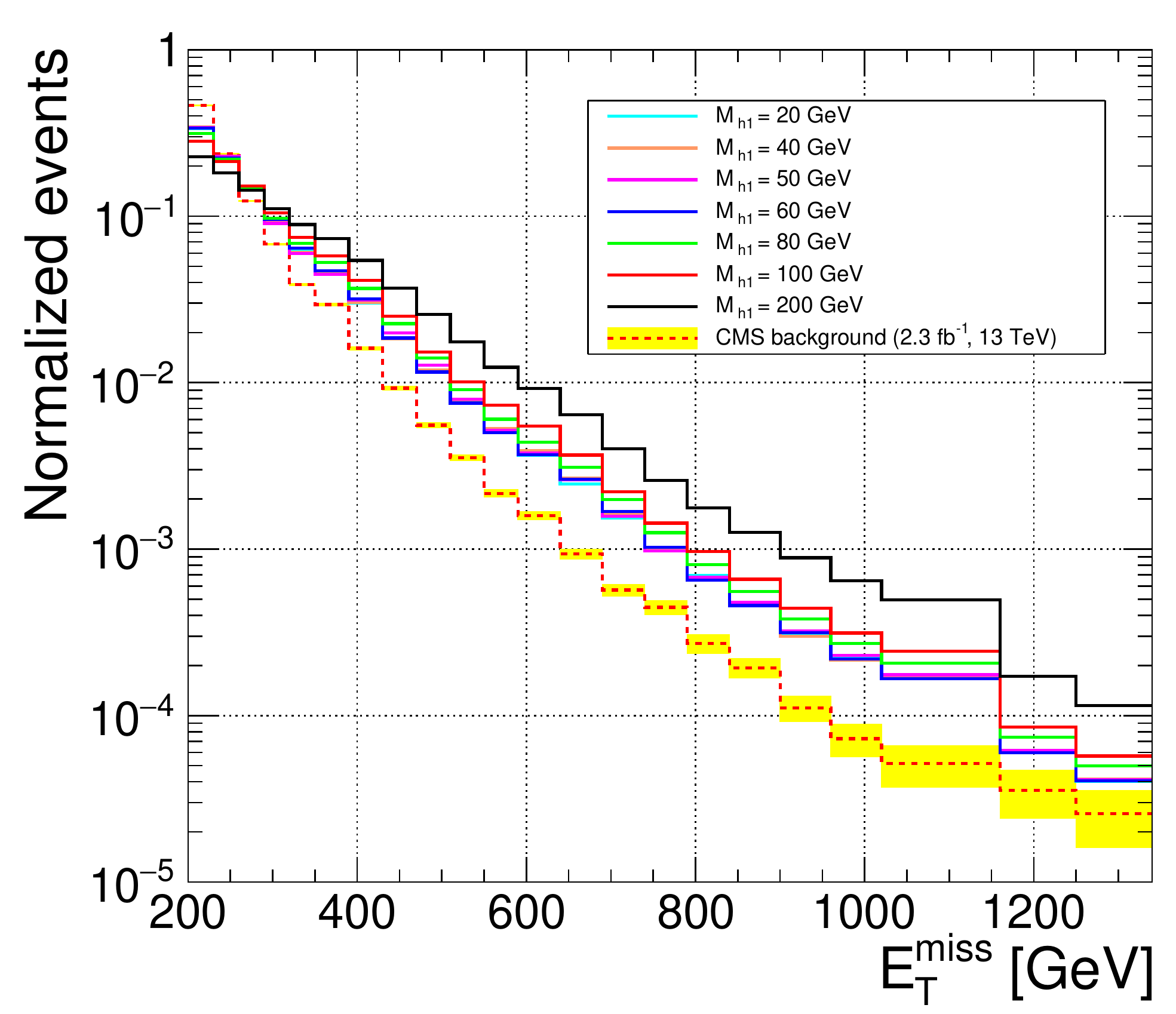}%
   \includegraphics[width=0.52\textwidth]{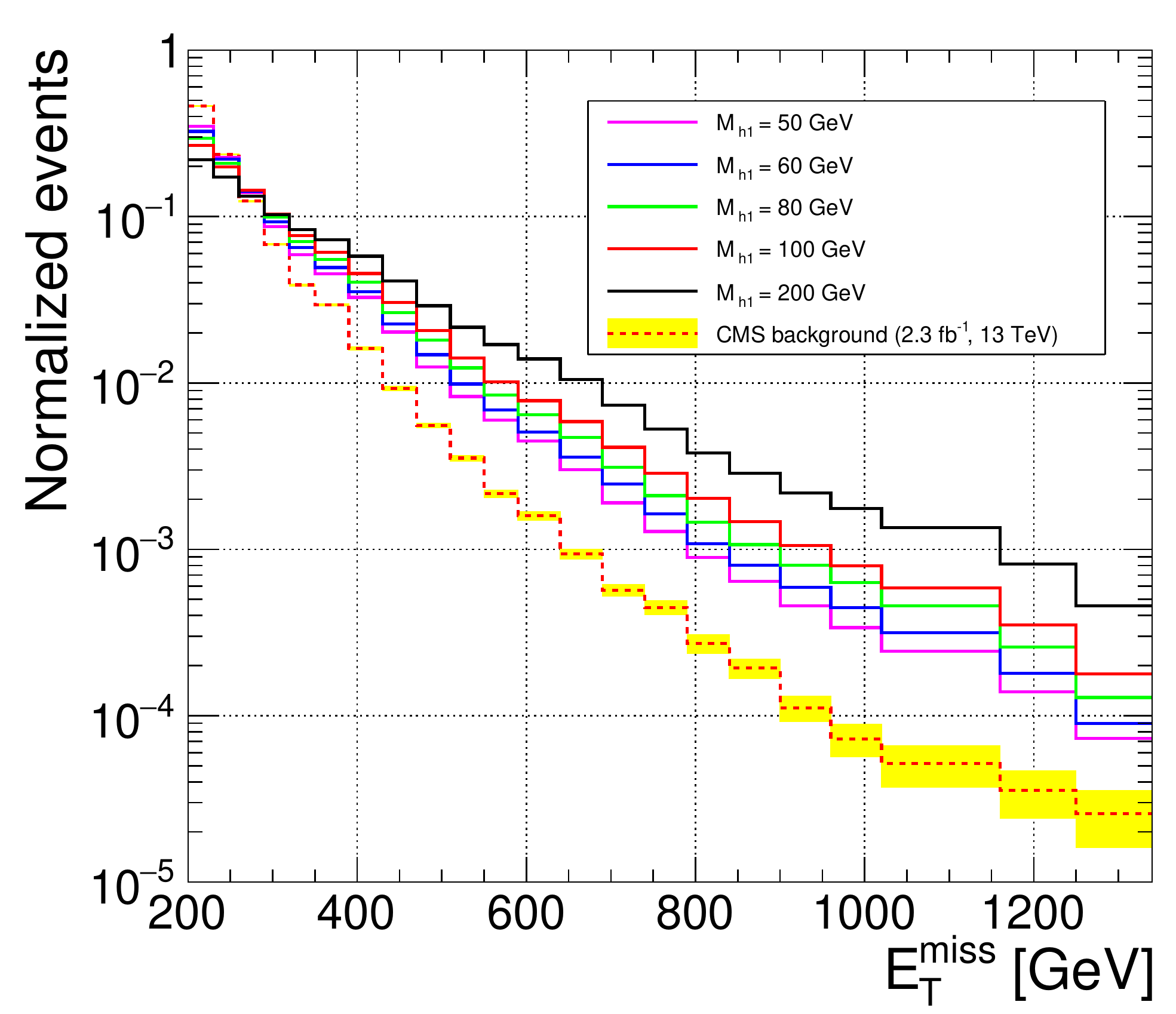}
   \caption{Comparison of \MET{} distributions between signals, for $h_1 h_1j$ (left) and $h_1 h_2j$ (right), for various DM masses, alongside the  estimated (by CMS)  experimental background for $\sqrt s =$ 13 TeV.}
   \label{fig:CMS13TeV_SigBack}
\end{figure}

The main change for the Run 2 selection was the update of the angular discriminant to suppress QCD multijet contributions: whereas in Run 1 a strict requirement was imposed on the jet multiplicity and leading jets azimuthal distance in Run 2 CMS opted instead for an overall requirement of azimuthal separation between the measured \MET{} and the four leading hadronic jets. The selection efficiency for both the $h_1h_1j$ and $h_1h_2j$ processes can be seen in Fig.~\ref{fig:selectionEfficiency} and is around 10--25\% for the former and 18--40\% for the latter. We can understand this difference by noticing that   $h_1h_2j$ production is mediated by a $Z$ boson while $h_1h_1j$ production is mediated by the SM-like Higgs boson, which leads to a different \MET{} spectrum. Fig.~\ref{fig:CMS13TeV_SigBack} presents the comparison of the   \MET{} distributions  for  different DM masses  for the signal from the  $h_1h_1j$ (left panel) and  $h_1h_2j$ (right panel) processes
as well as for  the background. One can  notice that the  \MET{} distribution for the $h_1h_2j$ signal is indeed harder than the one for the $h_1h_1j$ case.
This difference in \MET{} shapes are related to the difference in the
invariant mass of DM pair distributions,  for $h_1h_2j$ and  $h_1h_1j$  signals:
as discussed in \cite{Belyaev:2016pxe}, a scalar mediator defines a
softer invariant mass of the DM pair  than a vector mediator (for similar masses),
while the invariant mass of the DM pair in its turn is correlated with the shape of the  \MET{} distribution.

\begin{figure}[htbp]
   \centering
   \includegraphics[width=0.5\textwidth]{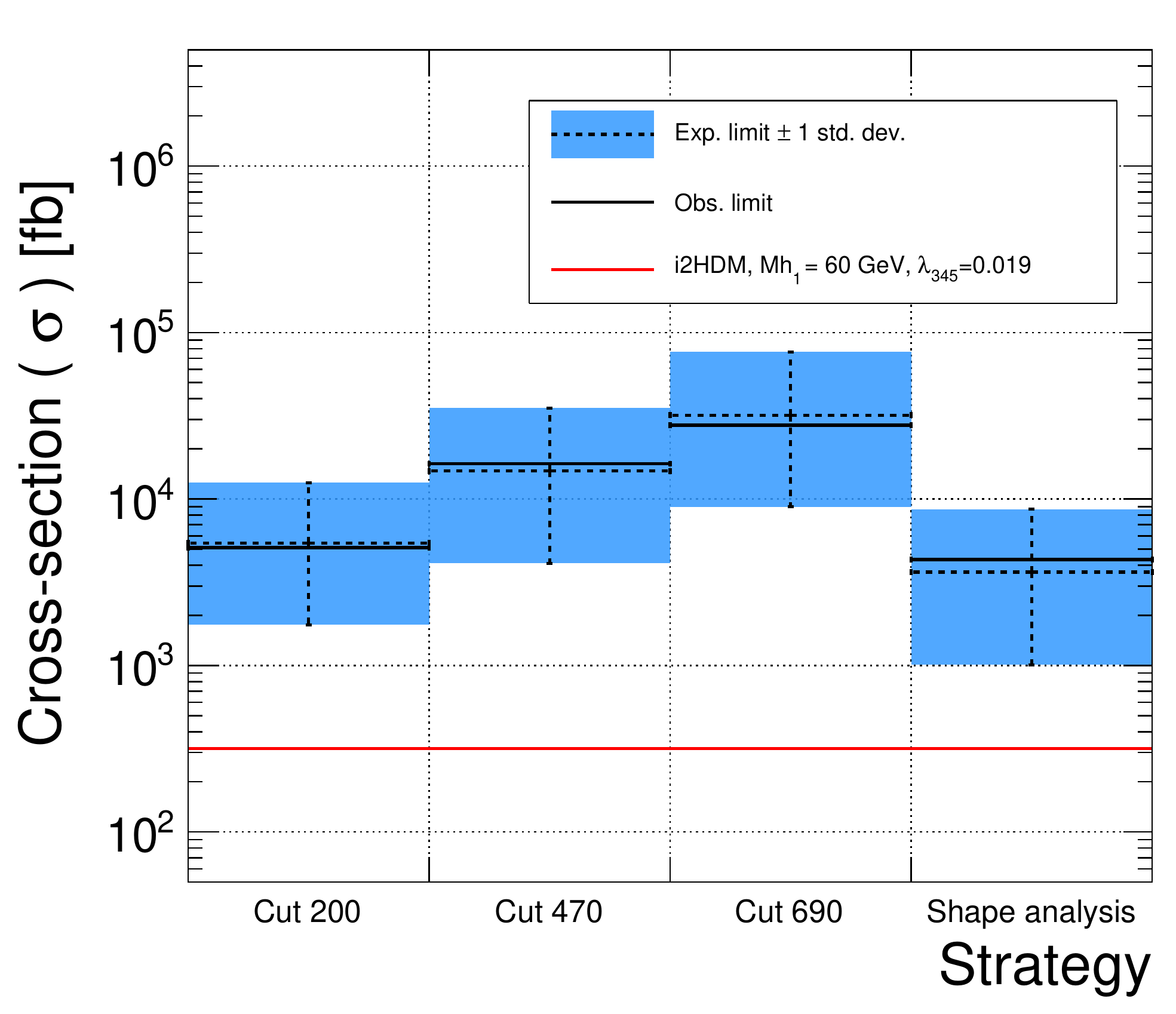}
   \caption{Expected and observed limits for four different analysis strategies: three counting experiments, with lower thresholds of 200, 470 and 690~GeV, and a shape analysis with a lower threshold of 200~GeV. The counting experiments are able to set expected limits of
   \unit{5.44}{\pb},
   \unit{14.7}{\pb},
   and \unit{31.9}{\pb}, respectively, while the shape analysis is able to set an expected limit of
   \unit{3.64}{\pb}, a 30\% improvement over the counting experiment with lowest threshold. The red line is for the $h_1h_1j$ process with $M_{h_1}$ = \unit{60}{\GeV}, $ M_{h_2}$ = \unit{200}{\GeV}, $\lambda_{345} = 0.019$, giving a cross section of ${0.317}$ {pb}.
}
   \label{fig:comparisonCountingShape}
\end{figure}

It can be observed from Fig.~\ref{fig:CMS13TeV_SigBack} that the \MET{} spectrum of the  signal is harder than that of the background for
the whole range of DM masses sampled, especially for the large values, which agrees with the findings of~\cite{Belyaev:2016pxe},
where it was shown that distributions at larger values of $M_{(\rm DM,DM)}$ have a flatter  \MET{} shape.
Eventually, for higher values of $M_{h_1}$,  $M_{(\rm DM,DM)}$ will also be higher.
{This suggests two strategies for the signal and background comparison. A simpler analysis would be a so-called \emph{counting experiment}, where a lower \MET{} threshold (``cut'') is defined and the spectrum is integrated above that value. This procedure produces a single event yield (with uncertainty) for both signal ($N_{\text{sig}}$) and background ($N_{\text{bkg}}$) and, in the case of an observed limit, the total number of observed events ($N_{\text{obs}}$) would also be available. Those are input to the limit-setting technique described in the previous section, through a single likelihood $\mathcal{L}(N_{\text{sig}}, N_{\text{bkg}}, N_{\text{obs}})$. A more sophisticated analysis, in contrast, could take into account the coherent enhancement over all the \MET{} spectrum that the presence of a signal would entail. In this strategy, the binned likelihood is written as the product of the single likelihoods of each bin over the relevant \MET{} range. This will be called the \emph{shape analysis} strategy.
The CMS Run 1 analysis~\cite{Khachatryan:2014rra} used a series of counting experiments with different \MET{} ranges whilst the CMS  Run 2 analysis employs a shape analysis.}
Fig.~\ref{fig:comparisonCountingShape} shows, {for our signal samples and the background estimates from the CMS Run 2 analysis}, the difference amongst four different analysis strategies: three counting experiments, with  respective \MET{} cuts of 200, 470 and 690~GeV, and a shape analysis with a lower threshold of 200~GeV. One can see that higher \MET{} thresholds in the counting experiment make the expected limit become worse while the shape analysis is able to leverage the coherent enhancements in all bins of \MET{} that arises from the signal presence to set a better limit, an order of 30\% improvement. We will therefore adopt the shape analysis strategy for the rest of this study.

\begin{figure}[htbp]
   \centering
   \includegraphics[width=0.5\textwidth]{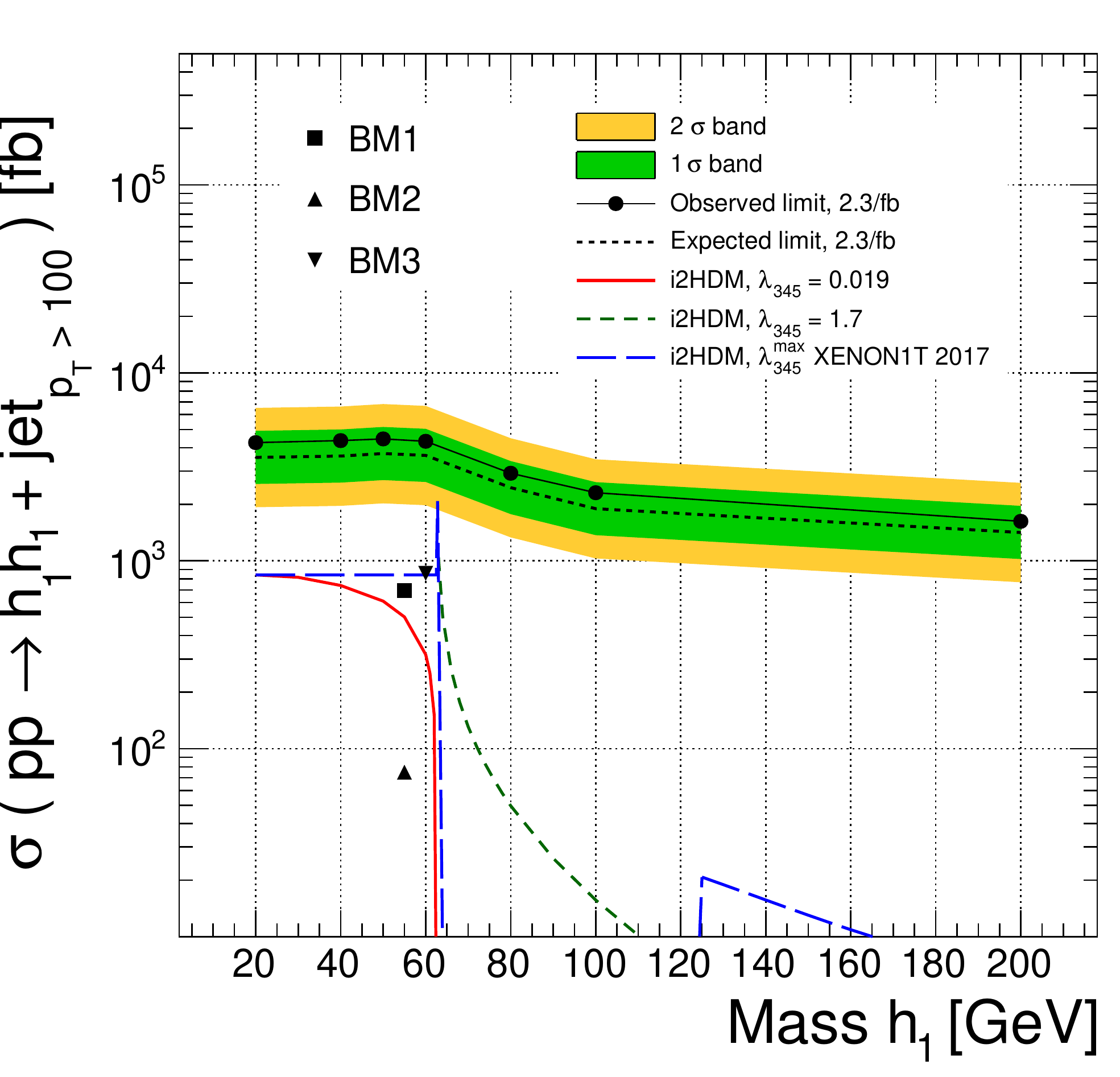}%
   \includegraphics[width=0.5\textwidth]{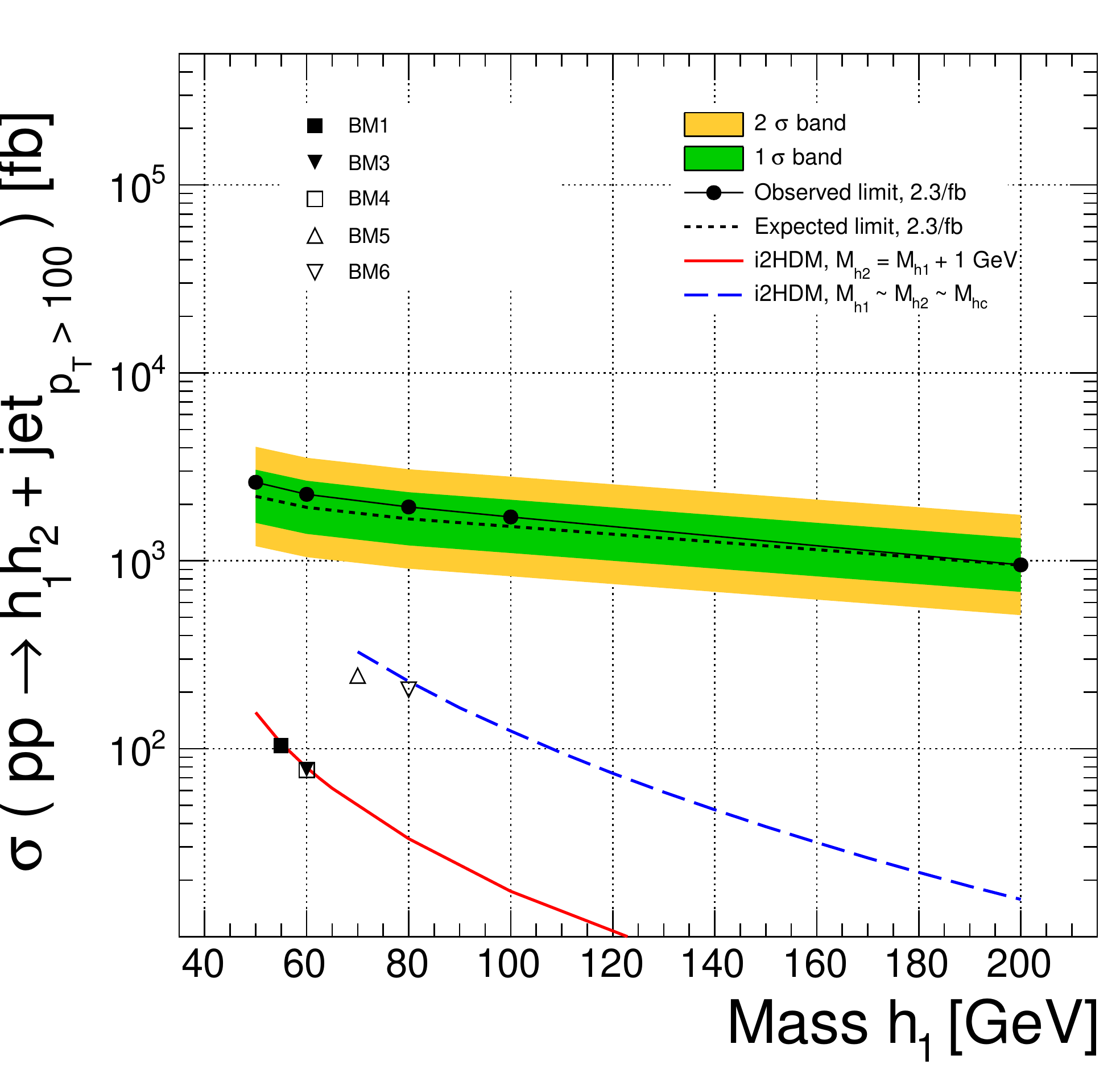}
   \caption{Left: Expected and observed limits on the $h_1 h_1j$ process for 2.3~\fbinv{} of 13 TeV $pp$ collision data.
   The red solid  line is the cross section for the parameter set
   $\lambda_{345} = 0.019$, $M_{h_2}$ = \unit{200}{\GeV}
   while the green short dashed line is the cross section for the parameter set
   $\lambda_{345} = 1.7$, $M_{h_2}$ = \unit{200}{\GeV}.
   The blue dashed line is the combined contribution $h_1h_1j+h_1h_2j$ that is still allowed by XENON1T data (see text).
   Right: Expected and observed limits on the $h_1 h_2j$ process for 2.3~\fbinv{} of 13 TeV $pp$ collision data.
   The red solid  line is the cross section for
   $M_{h_2}=M_{h_1}+1$~GeV.
   The blue short dashed line is the cross section for a full degeneracy between $h_1$, $h_2$ and $h_c$, where additional processes
   involving the charged scalar could mimic the $h_1h_2j$ process. The cross section is plotted for values of $M_{h_1}$ larger than $\sim$ 70 GeV to comply with the LEP bound on the charged Higgs mass~\cite{Belyaev:2016lok}.
   In all cases the isolated symbols represent the benchmark points discussed in Table~\ref{tab:i2HDMbenchMarks}.
   All cross sections are  given for a $\PT^{\rm jet} > \unit{100}{\GeV}$ requirement.
   }
   \label{fig:CMS13TeV_limit}
\end{figure}

 Fig.~\ref{fig:CMS13TeV_limit} shows the {95\% confidence level (CL) expected and observed exclusion limits as a function of  $M_{h_1}$} derived using the CMS 13 TeV background prediction and observed data with the Run 2 selection as described in Tab.~\ref{tab:selectionEXO16013}. On the left panel we show the limit for the $h_1h_1j$  process. In order to compare with the actual signal rate we show two signal lines for different values of $\lambda_{345} $. A red solid line presents the i2HDM cross section for $\lambda_{345} = 0.019$ that is near the maximum allowed by the Higgs invisible decay search, when $M_{h_1} < \MH/2$. In this region the SM-like  Higgs boson is produced on-shell, which enhances substantially the production cross section, and we can further notice a steep drop of the latter for $M_{h_1} > 60$ GeV. In contrast, for $M_{h_1} > \MH/2$, there is no bound on $\lambda_{345} $  from the Higgs invisible decay width and the cross section scales with $\lambda_{345}$ squared.
 We show with a blue dashed line the expected i2HDM cross section for
 maximally allowed $\lambda_{345}$ by present data:
 it reaches $1.6$ times the value
 that is around the maximum   for $M_{h_1}\simeq \MH/2$
 allowed by vacuum stability. Outside of the $M_{h_1}\simeq \MH/2$
 region  $\lambda_{345}$ is strongly   excluded by XENON1T data,
 e.g. in the interval  $65\mbox{\ GeV}< M_{h_1} < 70\mbox{\ GeV}$
the only  values of  $\lambda_{345}\lesssim 0.025$ are allowed.
{We can see that, for the 2.3~\fbinv{} dataset, 
we exclude the $h_1h_1j$ process cross sections in the range of 4.3--1.6~pb for $M_{h_1}$ in the range 20--200~GeV,
which does not exclude the i2HDM even for the highest allowed value of $\lambda_{345}$.
For the $h_1h_2j$ process, we exclude cross sections in the range of 2.6--0.95~pb for $M_{h_1}$ in the range 50--200~GeV,
also not enough to set relevant limits on the i2HDM.}

 When $M_{h_1} \simeq M_{h_2}$, $h_2$ will decay to $h_1$ plus very soft products, thus $h_1h_1j$, $h_2h_2j$ and $h_1h_2j$ production will contribute to the jet+\MET{} signature. The $h_2h_2j$  and  $h_1h_1j$ channels proceed via the same mediator and can be combined since they have the same \MET{} shape (for small values of $\Delta M=  M_{h_2}-M_{h_1}$).
 We indicate the predicted combined  $h_1h_1j$ and   $h_2h_2j$
 cross section by the purple dashed line for $\lambda_{345}^{max}$
 in Fig.~\ref{fig:CMS13TeV_limit} (left panel).
One can see that $2.3$fb$^{-1}$ mono-jet data are not quite sensitive
even to the combined  $h_1h_1j$ and   $h_2h_2j$ signal at $\lambda_{345}^{max}$.

 As mentioned above, the $h_1h_2j$ production is mediated by $Z $ boson exchange (see Fig.~\ref{fig:fd-mono-jet2}) and has
 therefore a  different\MET{} so  we investigate it separately.
 The right panel of Fig.~\ref{fig:CMS13TeV_limit}  presents  the limit for the $h_1h_2j$ production process (for $\Delta M =1$~GeV)
 and indeed demonstrates that the cross section limit for this process is different from the $h_1h_1j$ one because of their different kinematics. This process does not depend on the $\lambda_{345}$ coupling and thus the cross section is determined by the masses of $h_1$ and $h_2$ only, here, the expected signal rate represented by the red line indicates that it is well below the present limit.

\subsubsection{Projections for the HL-LHC }

\begin{figure}[htbp]
   \centering
   \includegraphics[width=0.5\textwidth]{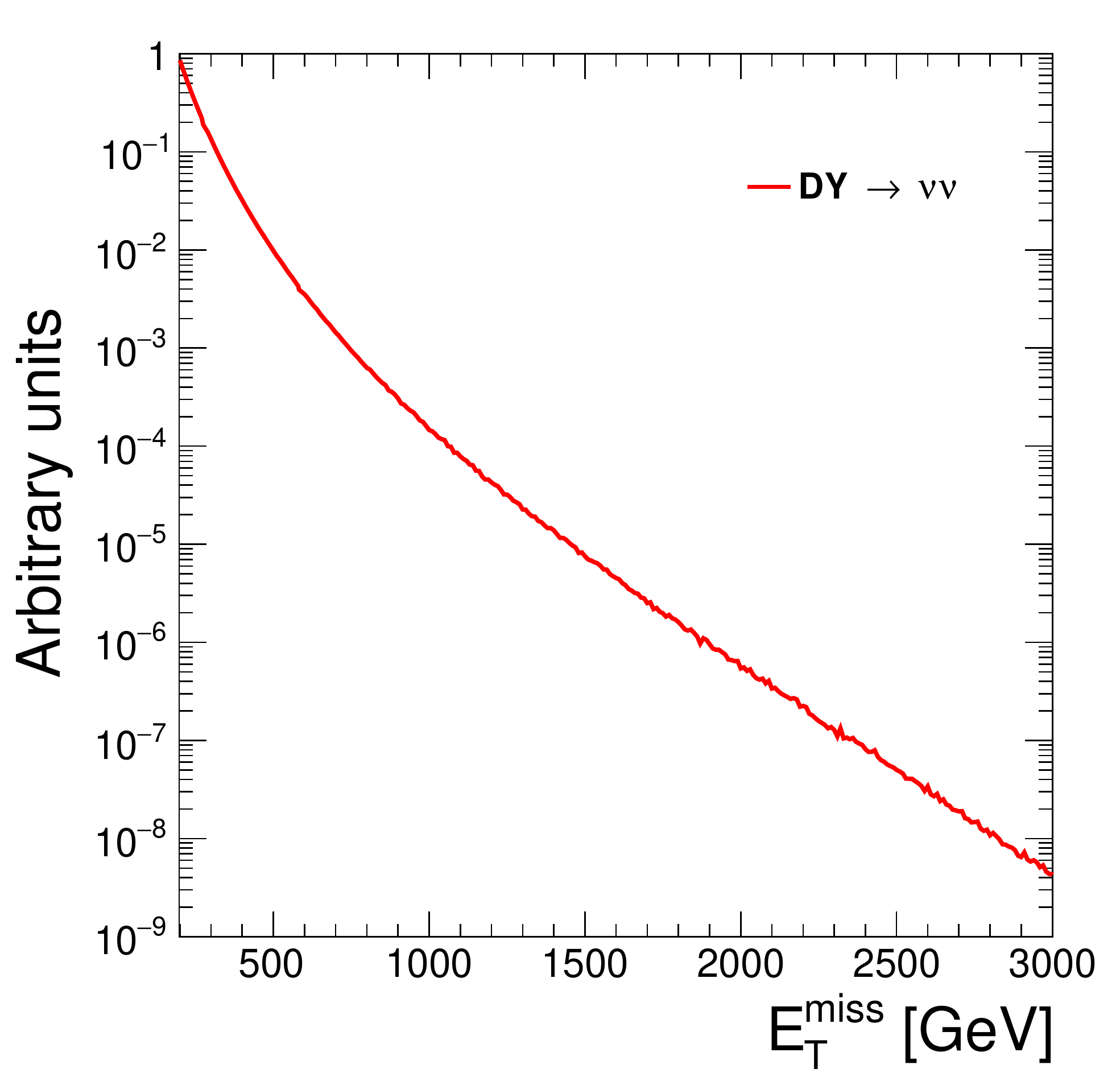}%
   \includegraphics[width=0.5\textwidth]{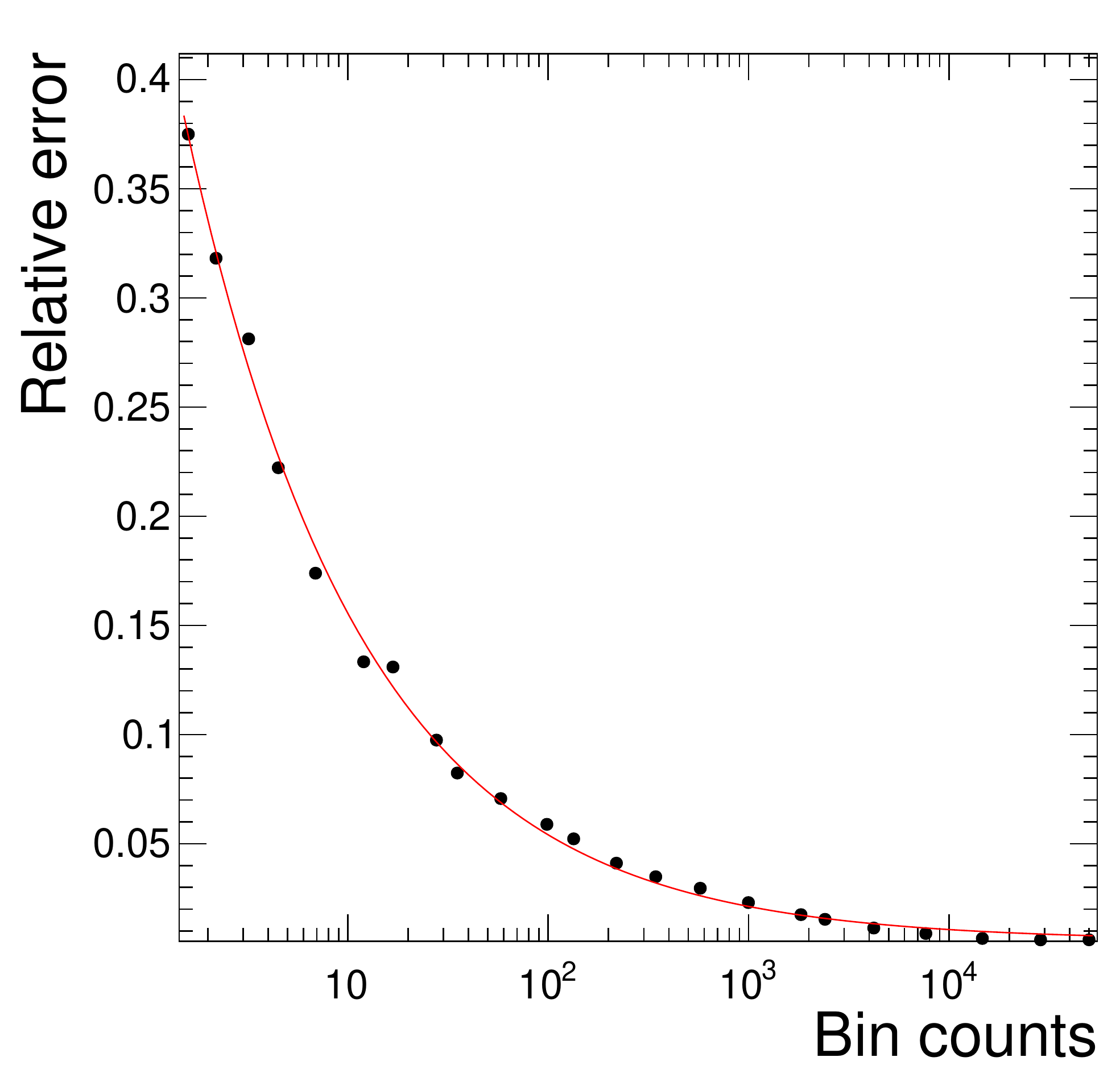}
   \caption{Left: \MET distribution from $pp\to Z \rightarrow \nu\overline{\nu}j$ process  produced with CalcHEP, for $pp$ collisions at $\sqrt s= 13$ TeV. Right: Relative error in the background estimate as function of bin counts, as extracted from Tab.~\ref{tab:dataFromEXO16013}.}
   \label{fig:DYtheoreticalShapeAndBKGError}
\end{figure}

As a next step in our study  we have found the projected LHC potential  at higher integrated luminosities of 30, 300 and 3000~\fbinv
with the last value posited as the ultimate benchmark for the HL-LHC. For this study, we made the following simplifying assumptions.
\begin{itemize}
\item The SM background to the mono-jet searches at the HL-LHC is still going to be dominated by inclusive EW production of $W$ and $Z$ bosons, with strong production of $t \bar t$ pairs being a minor background.
\item The upgraded experiments will be successful in maintaining the physics performance demonstrated during Run 1 and Run 2, even in view of a much higher pileup in the range of $\langle PU\rangle$ = 140--200.
\item The  change from 13 to 14~TeV centre-of-mass energy will not change the kinematic distribution of the reconstructed object in any significant way, neither for the SM background nor for the i2HDM processes.
\item The overall analysis strategy will be kept very similar to that in Tab.~\ref{tab:selectionEXO16013}. As such, shape, yield and uncertainty of both signal and background can be scaled to the desired luminosities.
\end{itemize}
While the extrapolation of the signal distributions to the HL-LHC is a simple rescaling, the estimate of the tails of the $W/Z$ inclusive \PT{} distributions is far from trivial. For the purposes of our study, we estimated the shape of the SM background directly from a simulation of $Z \rightarrow \nu\overline{\nu}j$  produced with CalcHEP, shown in the left panel of Fig.~\ref{fig:DYtheoreticalShapeAndBKGError} while the normalisation is approximated by a rescaling of the CMS results, since the efficiency of the selection is assumed to be the same. Since the background is primarily estimated from data distributions in control regions, we expect that the overall uncertainty in the \MET{} prediction follows approximately a $1/\sqrt{N}$ distribution. The right panel of Fig~\ref{fig:DYtheoreticalShapeAndBKGError} shows the relative errors in each bin from Tab.~\ref{tab:dataFromEXO16013} as function of the bin content. One can see that, indeed, it follows the aforementioned distribution, but in addition it also has a constant term ($\sim 0.6\%$) that can be understood to represent uncertainties that are not statistical in nature. We use the following equation for our bin-by-bin error estimate:
{\begin{equation}
\sigma^{\text{(rel)}}_{\text{bin}} \equiv 
\frac{\sigma_{\text{bin}}}{N_{\text{bin}}} \simeq \frac{0.46}{\sqrt{N_{\text{bin}}}} + 0.6\%,
\label{eq:errorFormula}
\end{equation}
where $N_{\text{bin}}$ and $\sigma_{\text{bin}}$ are the content and uncertainty of the given bin. The numerical values in Eq.~(\ref{eq:errorFormula}) are obtained through a fit to the relative errors from Tab.~\ref{tab:dataFromEXO16013}.}

\begin{figure}[htbp]
   \centering
   \includegraphics[width=0.5\textwidth]{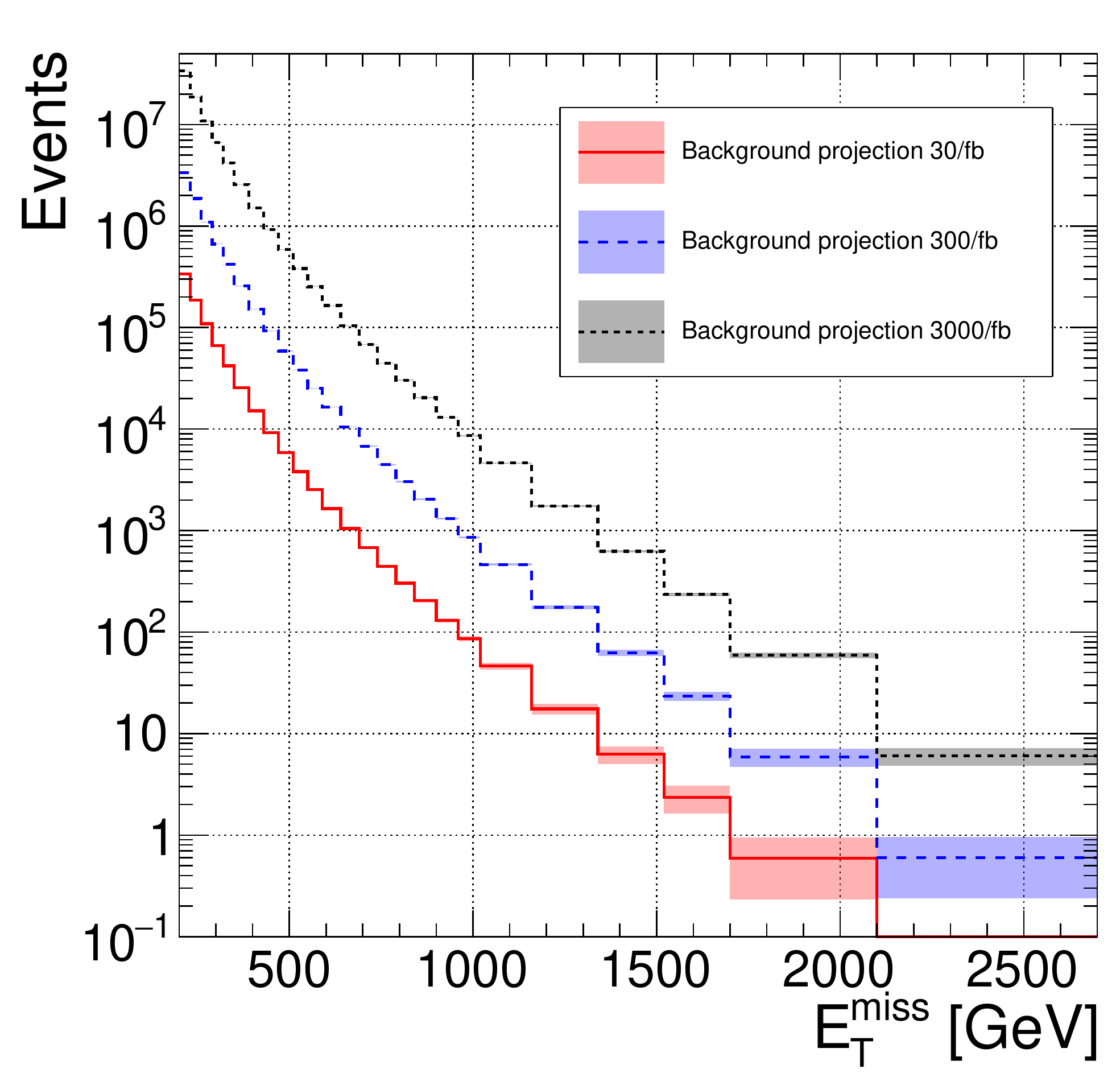}
   \caption{Background extrapolation for
   \unit{30}{\fbinv{}} (red solid line),
   \unit{300}{\fbinv{}} (blue dashed line)
   and
   \unit{3000}{\fbinv{}} (black dotted line).
   The errors (shaded areas) are estimated through the procedure described in the text.}
   \label{fig:CMS13TeV_extrapolation}
\end{figure}

Our final background estimate for the extrapolation to the HL-LHC is therefore done through the following procedure.
\begin{itemize}
\item We find the shape of the \MET distribution from the $pp\to Z \rightarrow \nu\overline{\nu}j$ process (Fig.~\ref{fig:DYtheoreticalShapeAndBKGError}).
\item We normalise the histogram such that the integral $I_L$ in the range 200--1250~GeV is:
\[
I_L = \frac{L_\mathrm{target}}{L_\mathrm{2015}} \cdot N_\textrm{events},
\]
where $L_\mathrm{target}$ is the target luminosity (30, 300 or \unit{3000}{\fbinv{}}), $L_\mathrm{2015} = \unit{2.3}{\fbinv{}}$ is the integrated luminosity of Ref.~\cite{CMS:2016tns} and $N_\textrm{events}$ = 61978.6 is the total number of events in the aforementioned range, from Tab.~\ref{tab:dataFromEXO16013}. This normalisation is produced  to approximate the efficiency of the CMS selection on the real SM background.
\item We find the bin-by-bin errors according to the formula in Eq.~(\ref{eq:errorFormula}).
\end{itemize}
This procedure guarantees that our background estimate has a reasonably correct  shape, normalisation and uncertainty. Fig.~\ref{fig:CMS13TeV_extrapolation} shows the background extrapolation for 30, 300 and 3000~\fbinv{} together with the errors.
The signal shapes are the same as in Fig.~\ref{fig:CMS13TeV_SigBack} and, with these inputs, we then  evaluate  the expected limits for the  values of integrated luminosity under consideration.

\begin{figure}[htb]
   \centering
   \includegraphics[width=0.5\textwidth]{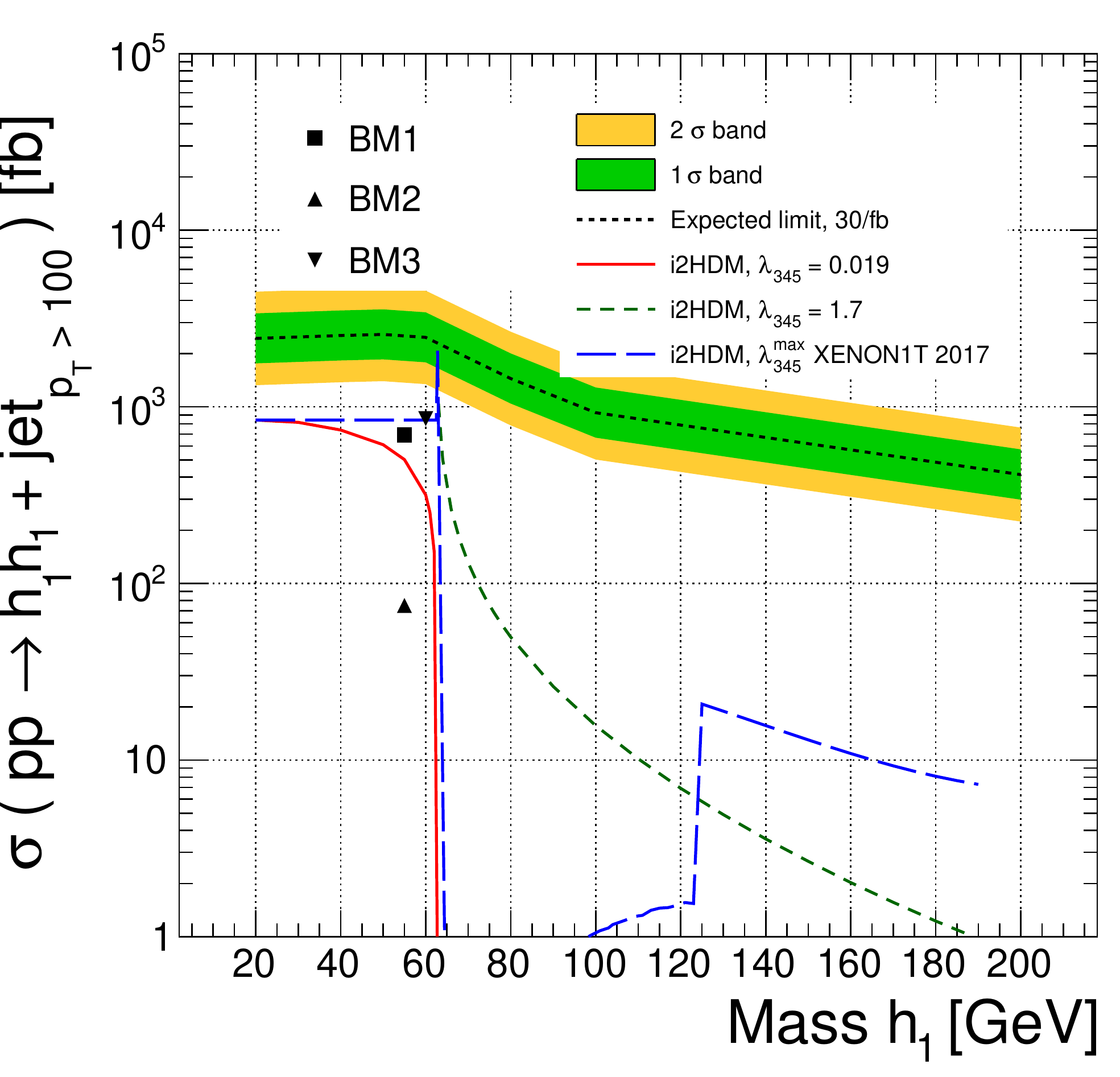}%
   \includegraphics[width=0.5\textwidth]{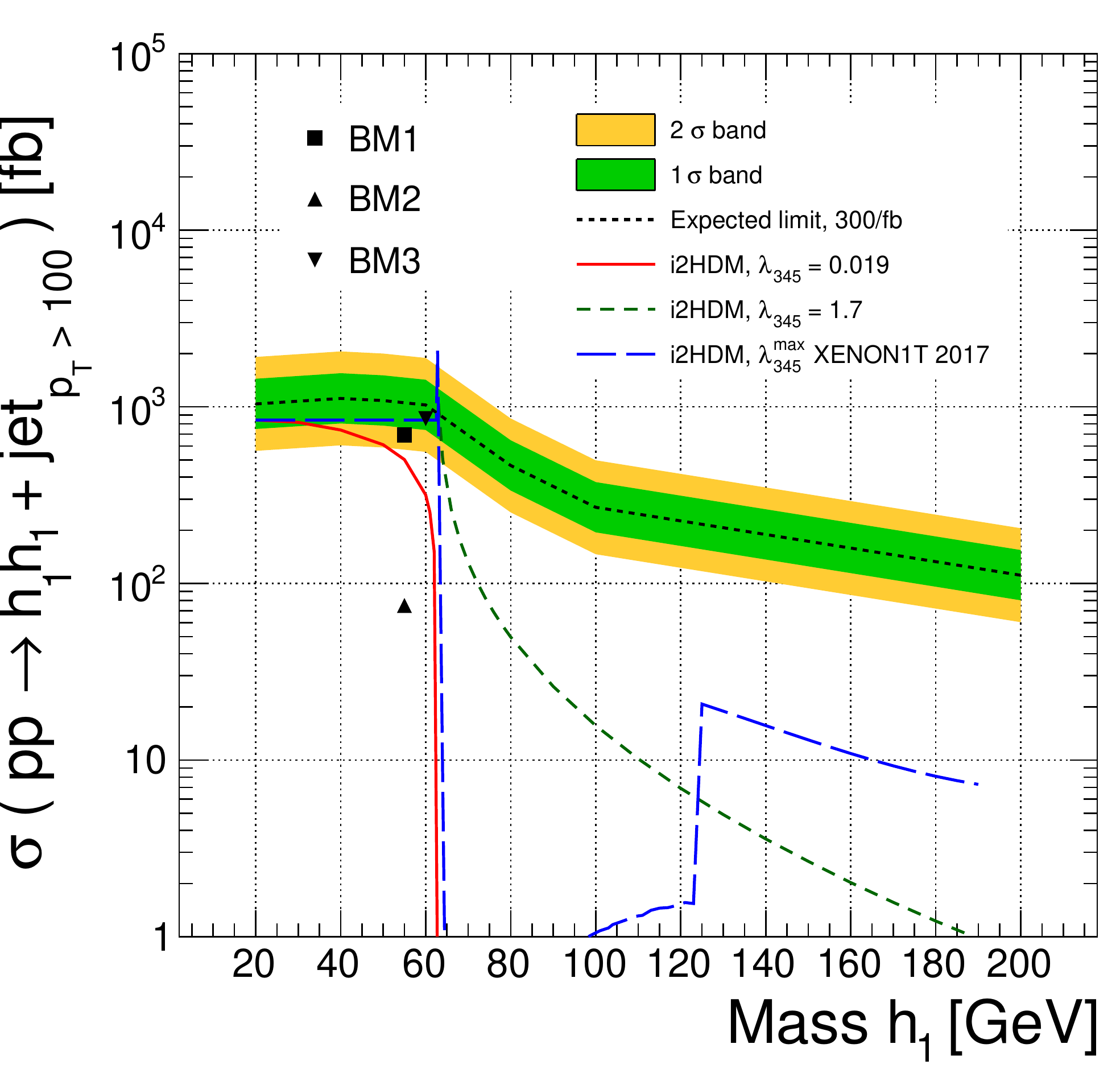}\\
   \includegraphics[width=0.5\textwidth]{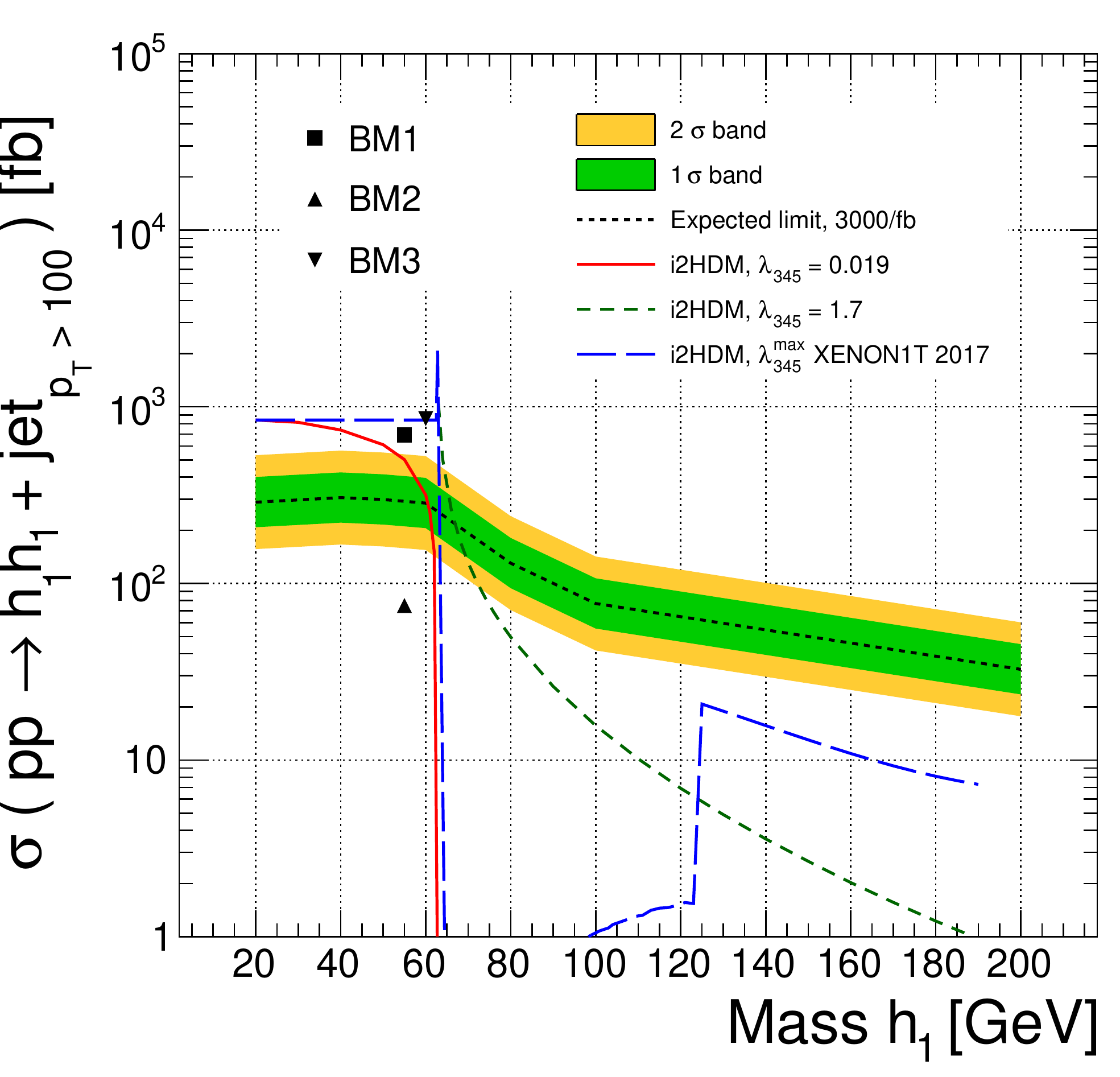}
   \caption{Expected limits for the $h_1 h_1j$ process for \unit{30}{\fbinv{}} (top left), \unit{300}{\fbinv{}} (top right) and \unit{3000}{\fbinv{}} (bottom).
   The red solid  line is the cross section for the parameter set
   $\lambda_{345} = 0.019$, $M_{h_2}$ = \unit{200}{\GeV}
   while the green short dashed line is the cross section for the parameter set
   $\lambda_{345} = 1.7$, $M_{h_2}$ = \unit{200}{\GeV}.
   The blue dashed line is the combined contribution $h_1h_1j+h_1h_2j$
   for $M_{h_2}=M_{h_1}+1\mathrm{\, GeV}$
   and the maximal value of $\lambda_{345}^\mathrm{max}$  allowed  by XENON1T data (see text).
   The isolated symbols represent the benchmark points discussed in Table~\ref{tab:i2HDMbenchMarks}.
   All cross sections are always given for a $\PT^{\rm jet} > \unit{100}{\GeV}$ requirement.}
   \label{fig:limits_h1h1j_extrapolation}
\end{figure}

In Fig.~\ref{fig:limits_h1h1j_extrapolation} we present the {95\% CL limits for the $h_1h_1j$ and $h_2h_2j$ processes as function of $M_{h_1}$}, together with production cross sections for 
$\lambda_{345} = 0.019$ (red solid line) and
$\lambda_{345} = 1.7$ (green short dashed line) with $M_{h_2}$ = \unit{200}{\GeV}, for both values of $\lambda_{345}$.
The blue dashed line is the combined cross section
for $h_1h_1j+h_1h_2j$ production for $M_{h_2}=M_{h_1}+1\mathrm{\, GeV}$
and the maximal value of $\lambda_{345}^\mathrm{max}$  allowed  by XENON1T data.
We find that, with 30~\fbinv{} of integrated luminosity, one can exclude   masses very close to $\MH/2$, for the maximum allowed value of  $\lambda_{345}$. Further, with 3000~\fbinv{} of the HL-LHC, one will be able to exclude all the region of $M_{h_1} < \MH/2$ for $\lambda_{345} = 0.019$.
At the same time one can see that for values of $\lambda_{345}$ allowed by  XENON1T
LHC it will not be possible to probe $M_{h_1} > \MH/2$ with 3000~\fbinv{}
with $h_1h_1j$/$h_2h_2j$ process
even if its  cross section is maximized for $M_{h_1}\simeq M_{h_2}$.
In Fig.~\ref{fig:limits_h1h1j_extrapolation} we also present the relevant benchmark points discussed in Table~\ref{tab:i2HDMbenchMarks}.
One can see that BM1 and BM3 with large (but still experimentally allowed)
$Br(H\to h_2 h_2)$ and $Br(H\to h_1 h_1)$ respectively
can be probed at the LHC at high luminosity. One should note that these benchmarks predict a too low DM relic density, requiring additional source for DM from
somewhere else. At the same time the BM2 scenario
with a DM relic density which is  in agreement  with the upper and lower limits
from Planck collaboration, requires too low values of $\lambda_{345}$
and respectively too low $Br(H\to h_1 h_1) = 0.022$  to be observed
at the LHC even in the high-luminosity stage. We would like to stress, however,
that future DM DD experiments including XENON will be able to probe
this benchmark since, as one  can see from Table~\ref{tab:i2HDMbenchMarks}
the $\sigma^p_\mathrm{SI}$ is already close to the XENON1T exclusion limit.

\begin{figure}[htb]
   \centering
   \includegraphics[width=0.5\textwidth]{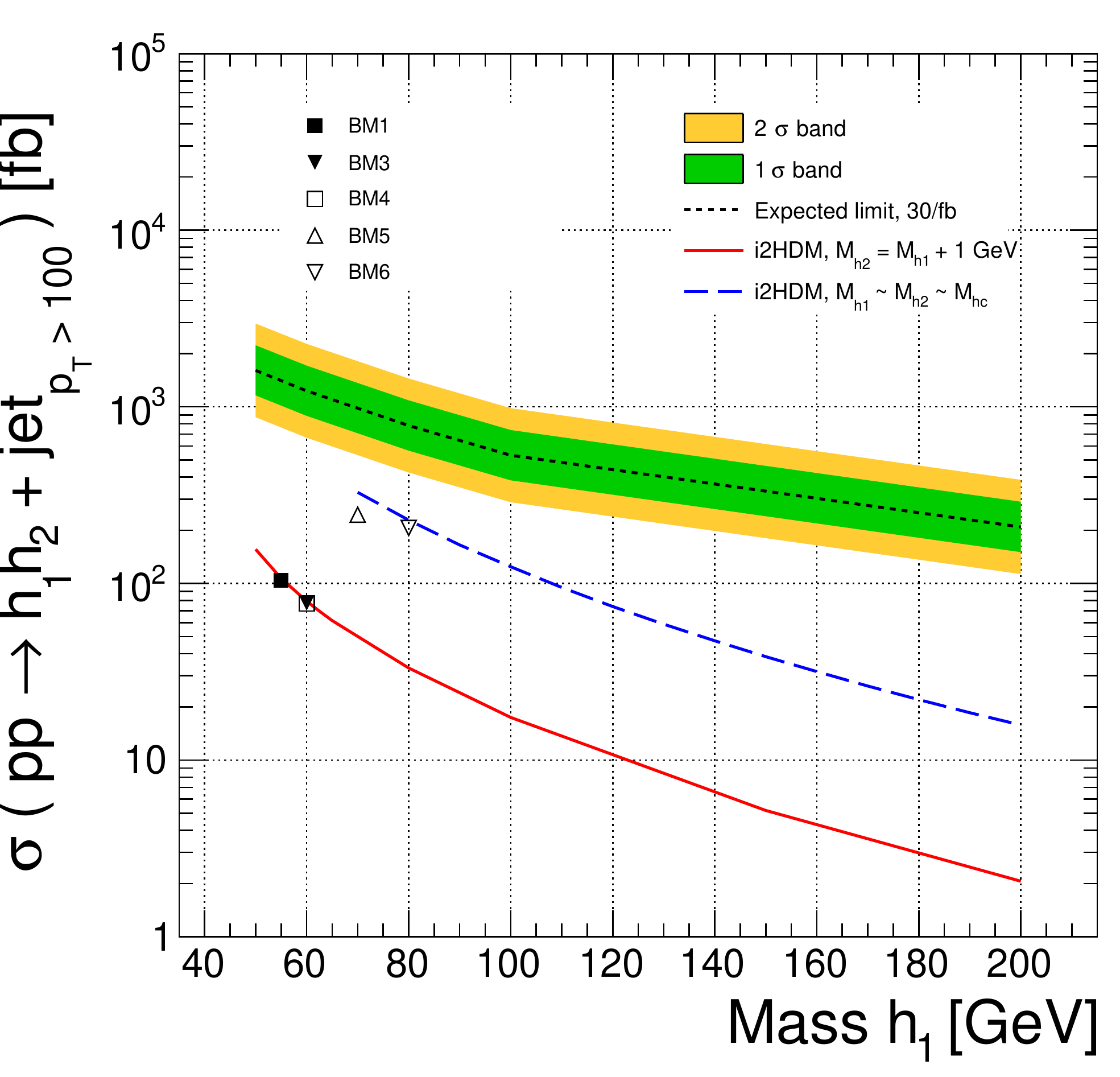}%
   \includegraphics[width=0.5\textwidth]{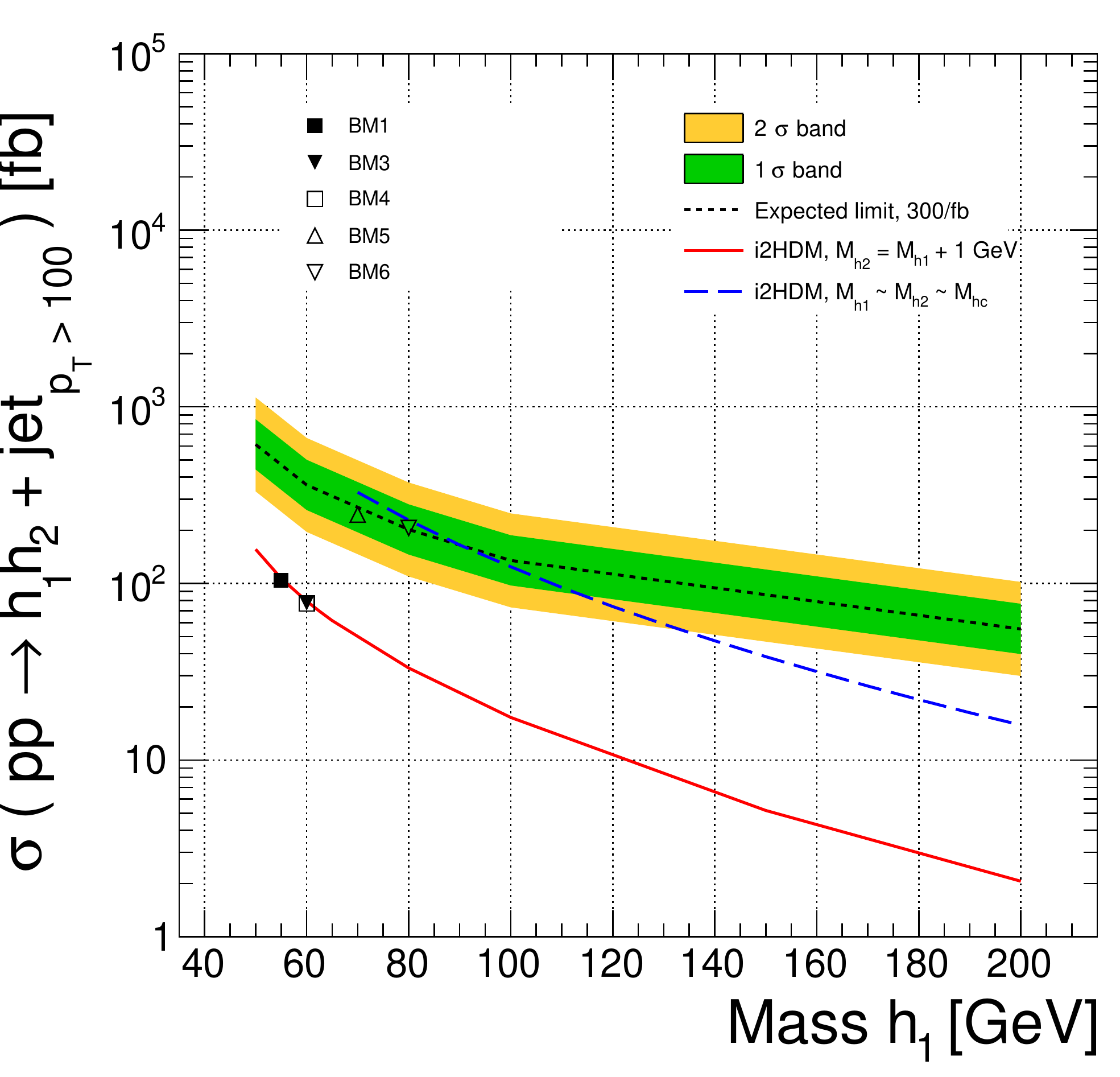}\\
   \includegraphics[width=0.5\textwidth]{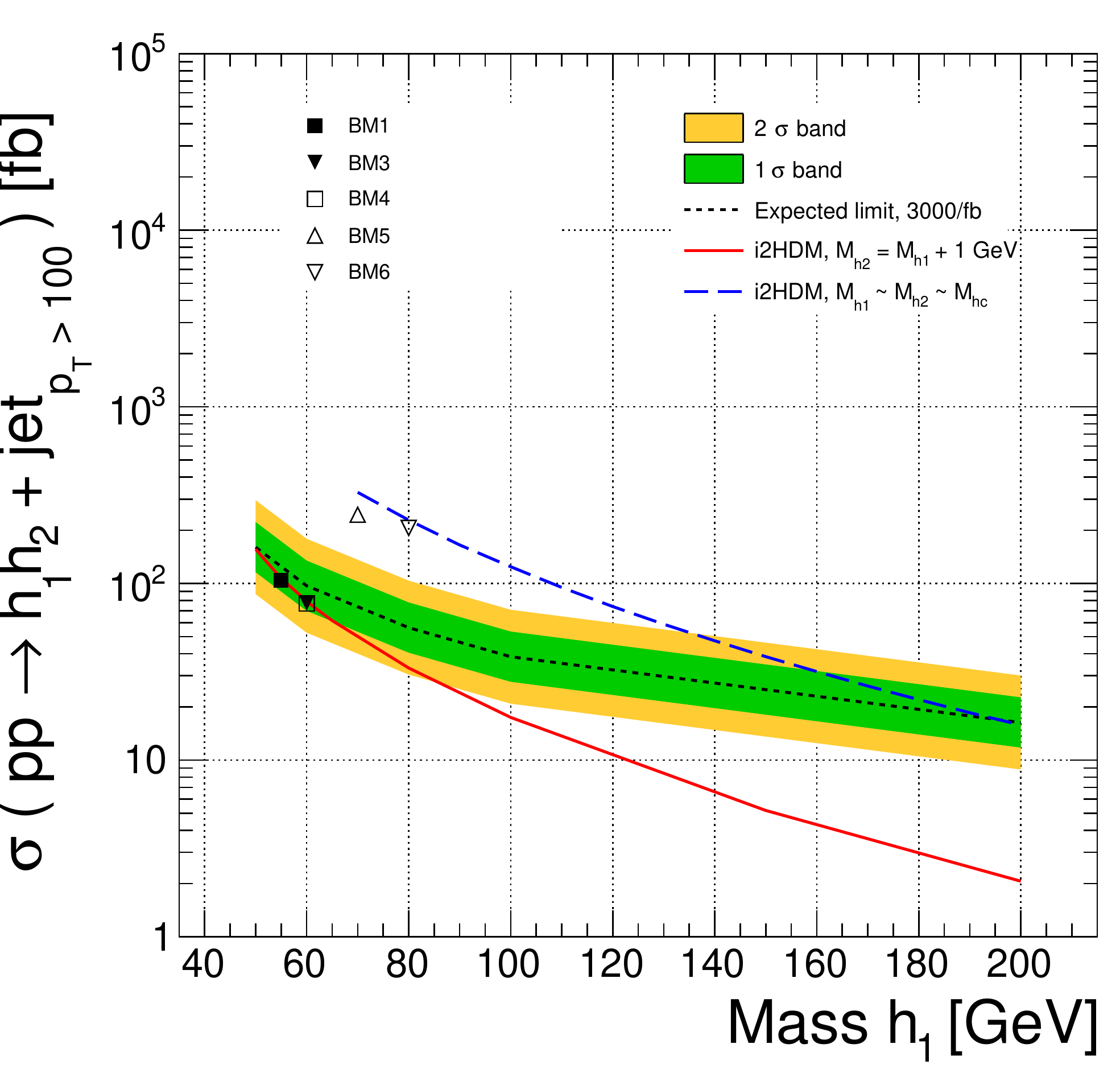}
   \caption{Expected limits for the $h_1 h_2j$ process for \unit{30}{\fbinv{}} (top left), \unit{300}{\fbinv{}} (top right) and \unit{3000}{\fbinv{}} (bottom).
   The red solid line is the cross section for
   $M_{h_2}=M_{h_1}+1$~GeV.
   The  blue short dashed line presents the cross section
   for the case when all inert scalars are close in mass:
   $M_{h_2}=M_{h_c}=M_{h_1}+1\mathrm{\,GeV}$. The cross section is plotted for values of $M_{h_1}$ larger than $\sim$ 70 GeV, corresponding to the LEP bound for the charged Higgs~\cite{Belyaev:2016lok}.
   The isolated symbols represent the benchmark points from Table~\ref{tab:i2HDMbenchMarks}.
   All cross sections are always given for a $\PT^{\rm jet} > \unit{100}{\GeV}$ requirement.}
   \label{fig:limits_h1h2j_extrapolation}
\end{figure}

In Fig.~\ref{fig:limits_h1h2j_extrapolation} we present the {95\% CL limits for the $h_1h_2j$ process as function of $M_{h_1}$.}  Only for  very high luminosity and for lower $M_{h_1} \simeq M_{h_2}$ masses the LHC might be sensitive to this process alone.
It is important to stress one again that this process
does not depend on $\lambda_{345}$
and therefore very complementary to the Higgs boson mediated one.
One should notice that, in the $M_{h_1} \simeq M_{h_2}$  region, the actual limit should be given by a combination of this process with the  $h_1h_1j$ and $h_2h_2j$ ones. The  $h_1h_1j$ and $h_2h_2j$ combination is a trivial one, we just sum both cross sections and the limit is given by Fig.~\ref{fig:limits_h1h1j_extrapolation}. However, the combination with the $h_1h_2j$ process is not trivial since it has a different shape of \MET distribution and the relative weights of $h_1h_1j$/$h_2h_2j$ and
$h_1h_2j$
distributions are eventually  depend on the value of $\lambda_{345}$.
One should also note that the sensitivity of the LHC
to the $h_1h_1j$/$h_2h_2j$ process is very limited
for $M_{h_1}>\MH/2$ as one can see from Fig.~\ref{fig:limits_h1h1j_extrapolation} since XENON1T
puts a very stringent upper limit on the $\lambda_{345}$ coupling.
Therefore, the  $h_1h_2j$ process is likely to be a unique one for the LHC
to probe the i2HDM parameter space beyond for $M_{h_1}>\MH/2$.
If all (pseudo)scalar masses, $M_{h_1}$,
$M_{h_2}$ and $M_{h^+}$, are similar, the LHC will be sensitive to the $M_{h_1}$ up to about 100 GeV with  300 fb$^{-1}$
and up to about 200 GeV with 3000 fb$^{-1}$
as demonstrated in the right and  bottom frames of Fig.~\ref{fig:limits_h1h2j_extrapolation} respectively.
The red solid line in this figure gives the cross section for
   $M_{h_2}=M_{h_1}+1$~GeV while
   the blue short dashed line is the cross section
   for the case when all inert scalars are close in mass
   ($M_{h_2}=M_{h_c}=M_{h_1}+1\mathrm{\,GeV}$) and
  the  processes with  the charged scalar(s) mimics the
  signature from  the $h_1h_2j$ process.
  The Fig.~\ref{fig:limits_h1h2j_extrapolation} shows that
  benchmarks  BM5 and BM6 with all nearly degenerate inert scalars can be tested already  with  300 fb$^{-1}$
integrated luminosity, while BM1, BM3 and BM4 with
nearly degenerate $h_1$ and $h_2$ can be excluded with  3000 fb$^{-1}$.

\begin{figure}[htbp]
   \centering
   \includegraphics[width=0.8\textwidth]{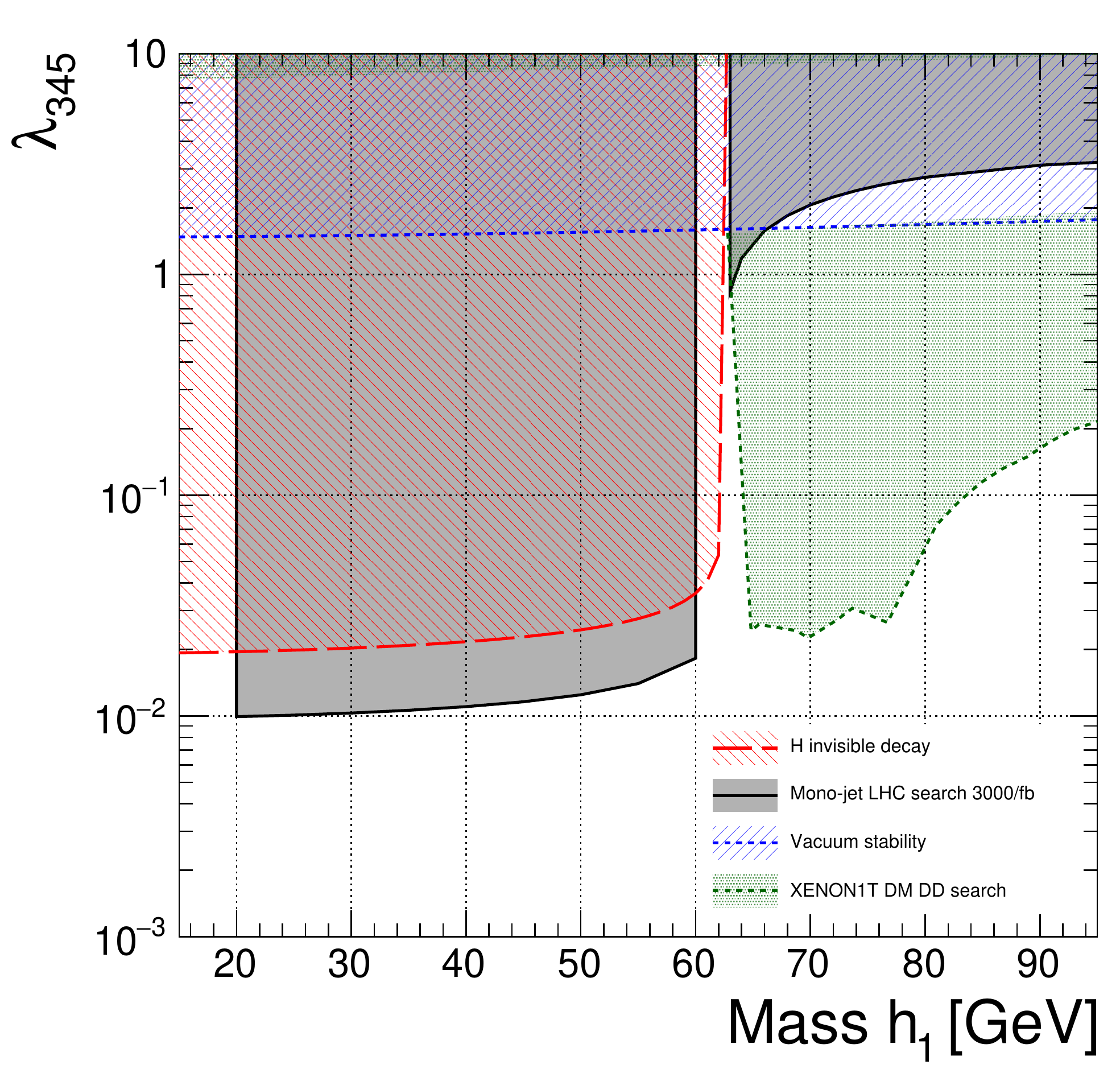}
   \caption{Expected exclusion region on the $(M_{h_1} , \lambda_{345})$ plane for 3000~\fbinv{}. The curve corresponding to Eq.~(\ref{l345limit}) is given by the red dashed contour whilst the expected result for 3000~\fbinv{} is given by the black solid contour. Also shown are the limits from vacuum stability (hashed blue region, dotted contour) and the XENON1T direct detection search (shaded green region, dotted contour).}
   \label{fig:limit_ld345XMh1}
\end{figure}

One can finally use the dependence of the cross section upon $\lambda_{345}$ to calculate an exclusion region on the $(M_{h_1} , \lambda_{345})$ plane. Fig.~\ref{fig:limit_ld345XMh1} shows the excluded values of $\lambda_{345}$ as function of $M_{h_1}$ for 3000~\fbinv{}. A mono-jet search at the HL-LHC will therefore exclude values of $\lambda_{345}$ larger than 0.011--0.02, for the range of masses $M_{h_1} < \MH/2$. For higher values of $M_{h_1}$, one would instead need a coupling value as large as $\lambda_{345} = 4.9$ in order to exclude $M_{h_1} < $~100~GeV. Also shown are the experimentally excluded regions from the invisible Higgs decay constraints as well the theoretically allowed maximum of  $\lambda_{345} $ from  vacuum stability.

\section{Conclusions}
In this paper, we have assessed the scope of the LHC in accessing a mono-jet signal stemming from  the i2HDM wherein the lightest inert Higgs state $h_1$  is a DM candidate, produced in pair from gluon-gluon fusion into the SM-like Higgs $H$ and accompanied by (at least)  a hard jet with transverse momentum above 100 GeV, i.e. a $h_1h_1j$ final state. The second-lightest inert Higgs boson $h_2$ can also contribute to a mono-jet signature, whenever it is degenerate enough with  the lightest one so that its decay products produced alongside the $h_1$ state are too soft to be detected. This can happen in $h_2h_2 j$ (again produced by gluon-gluon fusion into the SM-like Higgs) as well as $h_1h_2j$ (induced by $Z$ mediation) final states.

Before proceeding to such an assessment, we have established the viable parameter space of the i2HDM following both theoretical and experimental constraints. The former are dominated by vacuum stability requirements whereas the latter  are extracted from LEP, EWPT, LHC, relic density as well as LUX and, especially, XENON1T data, which greatly reduce the accessible volume of i2HDM parameter space. The established impact of XENON1T results  is in fact one of the main results of our analysis.

Over the surviving i2HDM parameter space, we have defined a several benchmark  points, wherein $M_{h_1}$ varies from 55 to 80 GeV and $M_{h_2}$ is between 1 and 55 GeV apart,  and tested them against a CMS inspired selection. However, in relation to the latter, we have adopted a somewhat orthogonal approach, as we have exploited the shape of the \MET distribution (as opposed to a standard {counting experiment} analysis). {We have indeed shown that the shape analysis is able to obtain a better sensitivity than a counting experiment.}
Furthermore, we have extrapolated such sensitivity to much higher luminosities, typical of the end of Run 2, Run 3 and high luminosity LHC.

By adopting an improved version of standard analysis tools (i.e matrix element, parton shower and hadronisation generators as well as detector software)   which further accounts for a $k$-factor enabling to correct the EFT approach for the emulation of the explicit loop entering the $gg\to H$ process  in the signal and a sophisticated  background treatment, we have been able to establish that the advocated shape analysis has significant scope in constraining mono-jet signals induced by i2HDM dynamics.

We have found that $h_1h_1j$ (plus $h_2h_2j$) and $h_1h_2j$ processes are very complementary to each other in probing the i2HDM parameter space. The former covers the $M_{h_1} < \MH/2$ region and will allow to put constraints on (in case of void searches) or else extract (in case of discovery)  two fundamental parameters of the i2HDM entering the leading mono-jet process. These are the $h_1$ mass and the trilinear self-coupling $\lambda_{345}$ connecting the SM-like Higgs to the DM candidate pair.   For example, for $M_{h_1}<\MH/2$, no values for $\lambda_{345}$ above 0.01-0.03 would be allowed in the case of no discovery. At the same time this process is not sensitive to  $M_{h_1} > \MH/2$ for values of $\lambda_{345}$ allowed by DM DD constraints. On the other hand  $\lambda_{345}$-independent $h_1h_2j$ process can be used to probe the  $M_{h_1} > \MH/2$ region of the parameter space via mono-jet signature in case $M_{h_1}\simeq M_{h_2}$. Moreover, the $h_1h_2j$ process has a slightly less steeply falling \MET distribution than the $h_1h_1j$ one because of different mediator (Z boson instead of Higgs boson) and respectively  slightly better LHC limit. If all the (pseudo)scalar masses, $M_{h_1}$, $M_{h_2}$ and $M_{h^+}$, are similar, the LHC will be sensitive to $M_{h_1}$ up to about 100 GeV with  300 fb$^{-1}$ and up to about 200 GeV with 3000 fb$^{-1}$.


\begin{acknowledgments}
AB acknowledges partial  support from the STFC grant ST/L000296/1, Royal Society Leverhulme Trust Senior Research Fellowship LT140094, and
Soton-FAPESP grant.
AB also thanks the NExT Institute and   Royal Society International Exchange grant IE150682,
partial support from the InvisiblesPlus RISE from the European
Union Horizon 2020 research and innovation programme under the Marie Sklodowska-Curie grant
agreement No 690575.
The work of CSM has been supported by the National Research Foundation of Korea(NRF) grant funded by the Korea government(MSIT)(No. 2018R1A6A1A06024970, 2018R1C1B5045624).
SM is supported in part through the NExT Institute and the STFC CG ST/L000296/1. SM and LP acknowledge funding via the H2020-MSCA-RISE-2014 grant no. 645722 (NonMinimalHiggs).
SN, TT and PM would like to thank FAPESP for support through the project grant 2013/01907-0 and by  the Cooperation Agreement
(SPRINT Program) between FAPESP and the University of Southampton (UK -- FAPESP  2013/50905--0).  TT would additionally like to thank FAPESP for support through grant 2016/15897-4.
\end{acknowledgments}

\bibliography{dm}

\begin{thebibliography}{73}%
\makeatletter
\providecommand \@ifxundefined [1]{%
 \@ifx{#1\undefined}
}%
\providecommand \@ifnum [1]{%
 \ifnum #1\expandafter \@firstoftwo
 \else \expandafter \@secondoftwo
 \fi
}%
\providecommand \@ifx [1]{%
 \ifx #1\expandafter \@firstoftwo
 \else \expandafter \@secondoftwo
 \fi
}%
\providecommand \natexlab [1]{#1}%
\providecommand \enquote  [1]{``#1''}%
\providecommand \bibnamefont  [1]{#1}%
\providecommand \bibfnamefont [1]{#1}%
\providecommand \citenamefont [1]{#1}%
\providecommand \href@noop [0]{\@secondoftwo}%
\providecommand \href [0]{\begingroup \@sanitize@url \@href}%
\providecommand \@href[1]{\@@startlink{#1}\@@href}%
\providecommand \@@href[1]{\endgroup#1\@@endlink}%
\providecommand \@sanitize@url [0]{\catcode `\\12\catcode `\$12\catcode
  `\&12\catcode `\#12\catcode `\^12\catcode `\_12\catcode `\%12\relax}%
\providecommand \@@startlink[1]{}%
\providecommand \@@endlink[0]{}%
\providecommand \url  [0]{\begingroup\@sanitize@url \@url }%
\providecommand \@url [1]{\endgroup\@href {#1}{\urlprefix }}%
\providecommand \urlprefix  [0]{URL }%
\providecommand \Eprint [0]{\href }%
\providecommand \doibase [0]{http://dx.doi.org/}%
\providecommand \selectlanguage [0]{\@gobble}%
\providecommand \bibinfo  [0]{\@secondoftwo}%
\providecommand \bibfield  [0]{\@secondoftwo}%
\providecommand \translation [1]{[#1]}%
\providecommand \BibitemOpen [0]{}%
\providecommand \bibitemStop [0]{}%
\providecommand \bibitemNoStop [0]{.\EOS\space}%
\providecommand \EOS [0]{\spacefactor3000\relax}%
\providecommand \BibitemShut  [1]{\csname bibitem#1\endcsname}%
\let\auto@bib@innerbib\@empty
\bibitem [{\citenamefont {Fox}\ \emph {et~al.}(2012)\citenamefont {Fox},
  \citenamefont {Harnik}, \citenamefont {Kopp},\ and\ \citenamefont
  {Tsai}}]{Fox:2011pm}%
  \BibitemOpen
  \bibfield  {author} {\bibinfo {author} {\bibfnamefont {P.~J.}\ \bibnamefont
  {Fox}}, \bibinfo {author} {\bibfnamefont {R.}~\bibnamefont {Harnik}},
  \bibinfo {author} {\bibfnamefont {J.}~\bibnamefont {Kopp}}, \ and\ \bibinfo
  {author} {\bibfnamefont {Y.}~\bibnamefont {Tsai}},\ }\href {\doibase
  10.1103/PhysRevD.85.056011} {\bibfield  {journal} {\bibinfo  {journal} {Phys.
  Rev.}\ }\textbf {\bibinfo {volume} {D85}},\ \bibinfo {pages} {056011}
  (\bibinfo {year} {2012})},\ \Eprint {http://arxiv.org/abs/1109.4398}
  {arXiv:1109.4398 [hep-ph]} \BibitemShut {NoStop}%
\bibitem [{\citenamefont {Rajaraman}\ \emph {et~al.}(2011)\citenamefont
  {Rajaraman}, \citenamefont {Shepherd}, \citenamefont {Tait},\ and\
  \citenamefont {Wijangco}}]{Rajaraman:2011wf}%
  \BibitemOpen
  \bibfield  {author} {\bibinfo {author} {\bibfnamefont {A.}~\bibnamefont
  {Rajaraman}}, \bibinfo {author} {\bibfnamefont {W.}~\bibnamefont {Shepherd}},
  \bibinfo {author} {\bibfnamefont {T.~M.~P.}\ \bibnamefont {Tait}}, \ and\
  \bibinfo {author} {\bibfnamefont {A.~M.}\ \bibnamefont {Wijangco}},\ }\href
  {\doibase 10.1103/PhysRevD.84.095013} {\bibfield  {journal} {\bibinfo
  {journal} {Phys. Rev.}\ }\textbf {\bibinfo {volume} {D84}},\ \bibinfo {pages}
  {095013} (\bibinfo {year} {2011})},\ \Eprint {http://arxiv.org/abs/1108.1196}
  {arXiv:1108.1196 [hep-ph]} \BibitemShut {NoStop}%
\bibitem [{\citenamefont {Goodman}\ \emph {et~al.}(2010)\citenamefont
  {Goodman}, \citenamefont {Ibe}, \citenamefont {Rajaraman}, \citenamefont
  {Shepherd}, \citenamefont {Tait} \emph {et~al.}}]{Goodman:2010ku}%
  \BibitemOpen
  \bibfield  {author} {\bibinfo {author} {\bibfnamefont {J.}~\bibnamefont
  {Goodman}}, \bibinfo {author} {\bibfnamefont {M.}~\bibnamefont {Ibe}},
  \bibinfo {author} {\bibfnamefont {A.}~\bibnamefont {Rajaraman}}, \bibinfo
  {author} {\bibfnamefont {W.}~\bibnamefont {Shepherd}}, \bibinfo {author}
  {\bibfnamefont {T.~M.}\ \bibnamefont {Tait}},  \emph {et~al.},\ }\href
  {\doibase 10.1103/PhysRevD.82.116010} {\bibfield  {journal} {\bibinfo
  {journal} {Phys.Rev.}\ }\textbf {\bibinfo {volume} {D82}},\ \bibinfo {pages}
  {116010} (\bibinfo {year} {2010})},\ \Eprint {http://arxiv.org/abs/1008.1783}
  {arXiv:1008.1783 [hep-ph]} \BibitemShut {NoStop}%
\bibitem [{\citenamefont {Bai}\ \emph {et~al.}(2010)\citenamefont {Bai},
  \citenamefont {Fox},\ and\ \citenamefont {Harnik}}]{Bai:2010hh}%
  \BibitemOpen
  \bibfield  {author} {\bibinfo {author} {\bibfnamefont {Y.}~\bibnamefont
  {Bai}}, \bibinfo {author} {\bibfnamefont {P.~J.}\ \bibnamefont {Fox}}, \ and\
  \bibinfo {author} {\bibfnamefont {R.}~\bibnamefont {Harnik}},\ }\href
  {\doibase 10.1007/JHEP12(2010)048} {\bibfield  {journal} {\bibinfo  {journal}
  {JHEP}\ }\textbf {\bibinfo {volume} {12}},\ \bibinfo {pages} {048} (\bibinfo
  {year} {2010})},\ \Eprint {http://arxiv.org/abs/1005.3797} {arXiv:1005.3797
  [hep-ph]} \BibitemShut {NoStop}%
\bibitem [{\citenamefont {Beltran}\ \emph {et~al.}(2010)\citenamefont
  {Beltran}, \citenamefont {Hooper}, \citenamefont {Kolb}, \citenamefont
  {Krusberg},\ and\ \citenamefont {Tait}}]{Beltran:2010ww}%
  \BibitemOpen
  \bibfield  {author} {\bibinfo {author} {\bibfnamefont {M.}~\bibnamefont
  {Beltran}}, \bibinfo {author} {\bibfnamefont {D.}~\bibnamefont {Hooper}},
  \bibinfo {author} {\bibfnamefont {E.~W.}\ \bibnamefont {Kolb}}, \bibinfo
  {author} {\bibfnamefont {Z.~A.~C.}\ \bibnamefont {Krusberg}}, \ and\ \bibinfo
  {author} {\bibfnamefont {T.~M.~P.}\ \bibnamefont {Tait}},\ }\href {\doibase
  10.1007/JHEP09(2010)037} {\bibfield  {journal} {\bibinfo  {journal} {JHEP}\
  }\textbf {\bibinfo {volume} {09}},\ \bibinfo {pages} {037} (\bibinfo {year}
  {2010})},\ \Eprint {http://arxiv.org/abs/1002.4137} {arXiv:1002.4137
  [hep-ph]} \BibitemShut {NoStop}%
\bibitem [{\citenamefont {Goodman}\ \emph {et~al.}(2011)\citenamefont
  {Goodman}, \citenamefont {Ibe}, \citenamefont {Rajaraman}, \citenamefont
  {Shepherd}, \citenamefont {Tait},\ and\ \citenamefont {Yu}}]{Goodman:2010yf}%
  \BibitemOpen
  \bibfield  {author} {\bibinfo {author} {\bibfnamefont {J.}~\bibnamefont
  {Goodman}}, \bibinfo {author} {\bibfnamefont {M.}~\bibnamefont {Ibe}},
  \bibinfo {author} {\bibfnamefont {A.}~\bibnamefont {Rajaraman}}, \bibinfo
  {author} {\bibfnamefont {W.}~\bibnamefont {Shepherd}}, \bibinfo {author}
  {\bibfnamefont {T.~M.~P.}\ \bibnamefont {Tait}}, \ and\ \bibinfo {author}
  {\bibfnamefont {H.-B.}\ \bibnamefont {Yu}},\ }\href {\doibase
  10.1016/j.physletb.2010.11.009} {\bibfield  {journal} {\bibinfo  {journal}
  {Phys. Lett.}\ }\textbf {\bibinfo {volume} {B695}},\ \bibinfo {pages} {185}
  (\bibinfo {year} {2011})},\ \Eprint {http://arxiv.org/abs/1005.1286}
  {arXiv:1005.1286 [hep-ph]} \BibitemShut {NoStop}%
\bibitem [{\citenamefont {Fox}\ \emph {et~al.}(2011)\citenamefont {Fox},
  \citenamefont {Harnik}, \citenamefont {Kopp},\ and\ \citenamefont
  {Tsai}}]{Fox:2011fx}%
  \BibitemOpen
  \bibfield  {author} {\bibinfo {author} {\bibfnamefont {P.~J.}\ \bibnamefont
  {Fox}}, \bibinfo {author} {\bibfnamefont {R.}~\bibnamefont {Harnik}},
  \bibinfo {author} {\bibfnamefont {J.}~\bibnamefont {Kopp}}, \ and\ \bibinfo
  {author} {\bibfnamefont {Y.}~\bibnamefont {Tsai}},\ }\href {\doibase
  10.1103/PhysRevD.84.014028} {\bibfield  {journal} {\bibinfo  {journal} {Phys.
  Rev.}\ }\textbf {\bibinfo {volume} {D84}},\ \bibinfo {pages} {014028}
  (\bibinfo {year} {2011})},\ \Eprint {http://arxiv.org/abs/1103.0240}
  {arXiv:1103.0240 [hep-ph]} \BibitemShut {NoStop}%
\bibitem [{\citenamefont {Shoemaker}\ and\ \citenamefont
  {Vecchi}(2012)}]{Shoemaker:2011vi}%
  \BibitemOpen
  \bibfield  {author} {\bibinfo {author} {\bibfnamefont {I.~M.}\ \bibnamefont
  {Shoemaker}}\ and\ \bibinfo {author} {\bibfnamefont {L.}~\bibnamefont
  {Vecchi}},\ }\href {\doibase 10.1103/PhysRevD.86.015023} {\bibfield
  {journal} {\bibinfo  {journal} {Phys. Rev.}\ }\textbf {\bibinfo {volume}
  {D86}},\ \bibinfo {pages} {015023} (\bibinfo {year} {2012})},\ \Eprint
  {http://arxiv.org/abs/1112.5457} {arXiv:1112.5457 [hep-ph]} \BibitemShut
  {NoStop}%
\bibitem [{\citenamefont {Fox}\ and\ \citenamefont
  {Williams}(2013)}]{Fox:2012ru}%
  \BibitemOpen
  \bibfield  {author} {\bibinfo {author} {\bibfnamefont {P.~J.}\ \bibnamefont
  {Fox}}\ and\ \bibinfo {author} {\bibfnamefont {C.}~\bibnamefont {Williams}},\
  }\href {\doibase 10.1103/PhysRevD.87.054030} {\bibfield  {journal} {\bibinfo
  {journal} {Phys. Rev.}\ }\textbf {\bibinfo {volume} {D87}},\ \bibinfo {pages}
  {054030} (\bibinfo {year} {2013})},\ \Eprint {http://arxiv.org/abs/1211.6390}
  {arXiv:1211.6390 [hep-ph]} \BibitemShut {NoStop}%
\bibitem [{\citenamefont {Haisch}\ \emph {et~al.}(2013)\citenamefont {Haisch},
  \citenamefont {Kahlhoefer},\ and\ \citenamefont {Unwin}}]{Haisch:2012kf}%
  \BibitemOpen
  \bibfield  {author} {\bibinfo {author} {\bibfnamefont {U.}~\bibnamefont
  {Haisch}}, \bibinfo {author} {\bibfnamefont {F.}~\bibnamefont {Kahlhoefer}},
  \ and\ \bibinfo {author} {\bibfnamefont {J.}~\bibnamefont {Unwin}},\ }\href
  {\doibase 10.1007/JHEP07(2013)125} {\bibfield  {journal} {\bibinfo  {journal}
  {JHEP}\ }\textbf {\bibinfo {volume} {07}},\ \bibinfo {pages} {125} (\bibinfo
  {year} {2013})},\ \Eprint {http://arxiv.org/abs/1208.4605} {arXiv:1208.4605
  [hep-ph]} \BibitemShut {NoStop}%
\bibitem [{\citenamefont {Busoni}\ \emph
  {et~al.}(2014{\natexlab{a}})\citenamefont {Busoni}, \citenamefont
  {De~Simone}, \citenamefont {Morgante},\ and\ \citenamefont
  {Riotto}}]{Busoni:2013lha}%
  \BibitemOpen
  \bibfield  {author} {\bibinfo {author} {\bibfnamefont {G.}~\bibnamefont
  {Busoni}}, \bibinfo {author} {\bibfnamefont {A.}~\bibnamefont {De~Simone}},
  \bibinfo {author} {\bibfnamefont {E.}~\bibnamefont {Morgante}}, \ and\
  \bibinfo {author} {\bibfnamefont {A.}~\bibnamefont {Riotto}},\ }\href
  {\doibase 10.1016/j.physletb.2013.11.069} {\bibfield  {journal} {\bibinfo
  {journal} {Phys. Lett.}\ }\textbf {\bibinfo {volume} {B728}},\ \bibinfo
  {pages} {412} (\bibinfo {year} {2014}{\natexlab{a}})},\ \Eprint
  {http://arxiv.org/abs/1307.2253} {arXiv:1307.2253 [hep-ph]} \BibitemShut
  {NoStop}%
\bibitem [{\citenamefont {Busoni}\ \emph
  {et~al.}(2014{\natexlab{b}})\citenamefont {Busoni}, \citenamefont
  {De~Simone}, \citenamefont {Gramling}, \citenamefont {Morgante},\ and\
  \citenamefont {Riotto}}]{Busoni2014a}%
  \BibitemOpen
  \bibfield  {author} {\bibinfo {author} {\bibfnamefont {G.}~\bibnamefont
  {Busoni}}, \bibinfo {author} {\bibfnamefont {A.}~\bibnamefont {De~Simone}},
  \bibinfo {author} {\bibfnamefont {J.}~\bibnamefont {Gramling}}, \bibinfo
  {author} {\bibfnamefont {E.}~\bibnamefont {Morgante}}, \ and\ \bibinfo
  {author} {\bibfnamefont {A.}~\bibnamefont {Riotto}},\ }\href {\doibase
  10.1088/1475-7516/2014/06/060} {\bibfield  {journal} {\bibinfo  {journal}
  {JCAP}\ }\textbf {\bibinfo {volume} {1406}},\ \bibinfo {pages} {060}
  (\bibinfo {year} {2014}{\natexlab{b}})},\ \Eprint
  {http://arxiv.org/abs/1402.1275} {arXiv:1402.1275 [hep-ph]} \BibitemShut
  {NoStop}%
\bibitem [{\citenamefont {Belyaev}\ \emph {et~al.}(2017)\citenamefont
  {Belyaev}, \citenamefont {Panizzi}, \citenamefont {Pukhov},\ and\
  \citenamefont {Thomas}}]{Belyaev:2016pxe}%
  \BibitemOpen
  \bibfield  {author} {\bibinfo {author} {\bibfnamefont {A.}~\bibnamefont
  {Belyaev}}, \bibinfo {author} {\bibfnamefont {L.}~\bibnamefont {Panizzi}},
  \bibinfo {author} {\bibfnamefont {A.}~\bibnamefont {Pukhov}}, \ and\ \bibinfo
  {author} {\bibfnamefont {M.}~\bibnamefont {Thomas}},\ }\href {\doibase
  10.1007/JHEP04(2017)110} {\bibfield  {journal} {\bibinfo  {journal} {JHEP}\
  }\textbf {\bibinfo {volume} {04}},\ \bibinfo {pages} {110} (\bibinfo {year}
  {2017})},\ \Eprint {http://arxiv.org/abs/1610.07545} {arXiv:1610.07545
  [hep-ph]} \BibitemShut {NoStop}%
\bibitem [{\citenamefont {Buchmueller}\ \emph {et~al.}(2014)\citenamefont
  {Buchmueller}, \citenamefont {Dolan},\ and\ \citenamefont
  {McCabe}}]{Buchmueller:2013dya}%
  \BibitemOpen
  \bibfield  {author} {\bibinfo {author} {\bibfnamefont {O.}~\bibnamefont
  {Buchmueller}}, \bibinfo {author} {\bibfnamefont {M.~J.}\ \bibnamefont
  {Dolan}}, \ and\ \bibinfo {author} {\bibfnamefont {C.}~\bibnamefont
  {McCabe}},\ }\href {\doibase 10.1007/JHEP01(2014)025} {\bibfield  {journal}
  {\bibinfo  {journal} {JHEP}\ }\textbf {\bibinfo {volume} {1401}},\ \bibinfo
  {pages} {025} (\bibinfo {year} {2014})},\ \Eprint
  {http://arxiv.org/abs/1308.6799} {arXiv:1308.6799 [hep-ph]} \BibitemShut
  {NoStop}%
\bibitem [{\citenamefont {Busoni}\ \emph
  {et~al.}(2014{\natexlab{c}})\citenamefont {Busoni}, \citenamefont
  {De~Simone}, \citenamefont {Gramling}, \citenamefont {Morgante},\ and\
  \citenamefont {Riotto}}]{Busoni:2014sya}%
  \BibitemOpen
  \bibfield  {author} {\bibinfo {author} {\bibfnamefont {G.}~\bibnamefont
  {Busoni}}, \bibinfo {author} {\bibfnamefont {A.}~\bibnamefont {De~Simone}},
  \bibinfo {author} {\bibfnamefont {J.}~\bibnamefont {Gramling}}, \bibinfo
  {author} {\bibfnamefont {E.}~\bibnamefont {Morgante}}, \ and\ \bibinfo
  {author} {\bibfnamefont {A.}~\bibnamefont {Riotto}},\ }\href {\doibase
  10.1088/1475-7516/2014/06/060} {\bibfield  {journal} {\bibinfo  {journal}
  {JCAP}\ }\textbf {\bibinfo {volume} {1406}},\ \bibinfo {pages} {060}
  (\bibinfo {year} {2014}{\natexlab{c}})},\ \Eprint
  {http://arxiv.org/abs/1402.1275} {arXiv:1402.1275 [hep-ph]} \BibitemShut
  {NoStop}%
\bibitem [{\citenamefont {Busoni}\ \emph
  {et~al.}(2014{\natexlab{d}})\citenamefont {Busoni}, \citenamefont
  {De~Simone}, \citenamefont {Jacques}, \citenamefont {Morgante},\ and\
  \citenamefont {Riotto}}]{Busoni:2014haa}%
  \BibitemOpen
  \bibfield  {author} {\bibinfo {author} {\bibfnamefont {G.}~\bibnamefont
  {Busoni}}, \bibinfo {author} {\bibfnamefont {A.}~\bibnamefont {De~Simone}},
  \bibinfo {author} {\bibfnamefont {T.}~\bibnamefont {Jacques}}, \bibinfo
  {author} {\bibfnamefont {E.}~\bibnamefont {Morgante}}, \ and\ \bibinfo
  {author} {\bibfnamefont {A.}~\bibnamefont {Riotto}},\ }\href {\doibase
  10.1088/1475-7516/2014/09/022} {\bibfield  {journal} {\bibinfo  {journal}
  {JCAP}\ }\textbf {\bibinfo {volume} {1409}},\ \bibinfo {pages} {022}
  (\bibinfo {year} {2014}{\natexlab{d}})},\ \Eprint
  {http://arxiv.org/abs/1405.3101} {arXiv:1405.3101 [hep-ph]} \BibitemShut
  {NoStop}%
\bibitem [{\citenamefont {Buchmueller}\ \emph {et~al.}(2015)\citenamefont
  {Buchmueller}, \citenamefont {Dolan}, \citenamefont {Malik},\ and\
  \citenamefont {McCabe}}]{Buchmueller:2014yoa}%
  \BibitemOpen
  \bibfield  {author} {\bibinfo {author} {\bibfnamefont {O.}~\bibnamefont
  {Buchmueller}}, \bibinfo {author} {\bibfnamefont {M.~J.}\ \bibnamefont
  {Dolan}}, \bibinfo {author} {\bibfnamefont {S.~A.}\ \bibnamefont {Malik}}, \
  and\ \bibinfo {author} {\bibfnamefont {C.}~\bibnamefont {McCabe}},\ }\href
  {\doibase 10.1007/JHEP01(2015)037} {\bibfield  {journal} {\bibinfo  {journal}
  {JHEP}\ }\textbf {\bibinfo {volume} {1501}},\ \bibinfo {pages} {037}
  (\bibinfo {year} {2015})},\ \Eprint {http://arxiv.org/abs/1407.8257}
  {arXiv:1407.8257 [hep-ph]} \BibitemShut {NoStop}%
\bibitem [{\citenamefont {Buckley}\ \emph
  {et~al.}(2015{\natexlab{a}})\citenamefont {Buckley}, \citenamefont {Feld},\
  and\ \citenamefont {Goncalves}}]{Buckley:2014fba}%
  \BibitemOpen
  \bibfield  {author} {\bibinfo {author} {\bibfnamefont {M.~R.}\ \bibnamefont
  {Buckley}}, \bibinfo {author} {\bibfnamefont {D.}~\bibnamefont {Feld}}, \
  and\ \bibinfo {author} {\bibfnamefont {D.}~\bibnamefont {Goncalves}},\ }\href
  {\doibase 10.1103/PhysRevD.91.015017} {\bibfield  {journal} {\bibinfo
  {journal} {Phys. Rev.}\ }\textbf {\bibinfo {volume} {D91}},\ \bibinfo {pages}
  {015017} (\bibinfo {year} {2015}{\natexlab{a}})},\ \Eprint
  {http://arxiv.org/abs/1410.6497} {arXiv:1410.6497 [hep-ph]} \BibitemShut
  {NoStop}%
\bibitem [{\citenamefont {Abdallah}\ \emph {et~al.}(2015)\citenamefont
  {Abdallah} \emph {et~al.}}]{Abdallah:2015ter}%
  \BibitemOpen
  \bibfield  {author} {\bibinfo {author} {\bibfnamefont {J.}~\bibnamefont
  {Abdallah}} \emph {et~al.},\ }\href {\doibase 10.1016/j.dark.2015.08.001}
  {\bibfield  {journal} {\bibinfo  {journal} {Phys. Dark Univ.}\ }\textbf
  {\bibinfo {volume} {9-10}},\ \bibinfo {pages} {8} (\bibinfo {year} {2015})},\
  \Eprint {http://arxiv.org/abs/1506.03116} {arXiv:1506.03116 [hep-ph]}
  \BibitemShut {NoStop}%
\bibitem [{\citenamefont {Abdallah}\ \emph {et~al.}(2014)\citenamefont
  {Abdallah}, \citenamefont {Ashkenazi}, \citenamefont {Boveia}, \citenamefont
  {Busoni}, \citenamefont {De~Simone} \emph {et~al.}}]{Abdallah:2014hon}%
  \BibitemOpen
  \bibfield  {author} {\bibinfo {author} {\bibfnamefont {J.}~\bibnamefont
  {Abdallah}}, \bibinfo {author} {\bibfnamefont {A.}~\bibnamefont {Ashkenazi}},
  \bibinfo {author} {\bibfnamefont {A.}~\bibnamefont {Boveia}}, \bibinfo
  {author} {\bibfnamefont {G.}~\bibnamefont {Busoni}}, \bibinfo {author}
  {\bibfnamefont {A.}~\bibnamefont {De~Simone}},  \emph {et~al.},\ }\href@noop
  {} {\  (\bibinfo {year} {2014})},\ \Eprint {http://arxiv.org/abs/1409.2893}
  {arXiv:1409.2893 [hep-ph]} \BibitemShut {NoStop}%
\bibitem [{\citenamefont {Abercrombie}\ \emph {et~al.}(2015)\citenamefont
  {Abercrombie} \emph {et~al.}}]{Abercrombie:2015wmb}%
  \BibitemOpen
  \bibfield  {author} {\bibinfo {author} {\bibfnamefont {D.}~\bibnamefont
  {Abercrombie}} \emph {et~al.},\ }\href@noop {} {\  (\bibinfo {year}
  {2015})},\ \Eprint {http://arxiv.org/abs/1507.00966} {arXiv:1507.00966
  [hep-ex]} \BibitemShut {NoStop}%
\bibitem [{\citenamefont {Deshpande}\ and\ \citenamefont
  {Ma}(1978)}]{Deshpande:1977rw}%
  \BibitemOpen
  \bibfield  {author} {\bibinfo {author} {\bibfnamefont {N.~G.}\ \bibnamefont
  {Deshpande}}\ and\ \bibinfo {author} {\bibfnamefont {E.}~\bibnamefont {Ma}},\
  }\href {\doibase 10.1103/PhysRevD.18.2574} {\bibfield  {journal} {\bibinfo
  {journal} {Phys.Rev.}\ }\textbf {\bibinfo {volume} {D18}},\ \bibinfo {pages}
  {2574} (\bibinfo {year} {1978})}\BibitemShut {NoStop}%
\bibitem [{\citenamefont {Ma}(2006)}]{Ma:2006km}%
  \BibitemOpen
  \bibfield  {author} {\bibinfo {author} {\bibfnamefont {E.}~\bibnamefont
  {Ma}},\ }\href {\doibase 10.1103/PhysRevD.73.077301} {\bibfield  {journal}
  {\bibinfo  {journal} {Phys.Rev.}\ }\textbf {\bibinfo {volume} {D73}},\
  \bibinfo {pages} {077301} (\bibinfo {year} {2006})},\ \Eprint
  {http://arxiv.org/abs/hep-ph/0601225} {arXiv:hep-ph/0601225 [hep-ph]}
  \BibitemShut {NoStop}%
\bibitem [{\citenamefont {Barbieri}\ \emph {et~al.}(2006)\citenamefont
  {Barbieri}, \citenamefont {Hall},\ and\ \citenamefont
  {Rychkov}}]{Barbieri:2006dq}%
  \BibitemOpen
  \bibfield  {author} {\bibinfo {author} {\bibfnamefont {R.}~\bibnamefont
  {Barbieri}}, \bibinfo {author} {\bibfnamefont {L.~J.}\ \bibnamefont {Hall}},
  \ and\ \bibinfo {author} {\bibfnamefont {V.~S.}\ \bibnamefont {Rychkov}},\
  }\href {\doibase 10.1103/PhysRevD.74.015007} {\bibfield  {journal} {\bibinfo
  {journal} {Phys.Rev.}\ }\textbf {\bibinfo {volume} {D74}},\ \bibinfo {pages}
  {015007} (\bibinfo {year} {2006})},\ \Eprint
  {http://arxiv.org/abs/hep-ph/0603188} {arXiv:hep-ph/0603188 [hep-ph]}
  \BibitemShut {NoStop}%
\bibitem [{\citenamefont {Lopez~Honorez}\ \emph {et~al.}(2007)\citenamefont
  {Lopez~Honorez}, \citenamefont {Nezri}, \citenamefont {Oliver},\ and\
  \citenamefont {Tytgat}}]{LopezHonorez:2006gr}%
  \BibitemOpen
  \bibfield  {author} {\bibinfo {author} {\bibfnamefont {L.}~\bibnamefont
  {Lopez~Honorez}}, \bibinfo {author} {\bibfnamefont {E.}~\bibnamefont
  {Nezri}}, \bibinfo {author} {\bibfnamefont {J.~F.}\ \bibnamefont {Oliver}}, \
  and\ \bibinfo {author} {\bibfnamefont {M.~H.}\ \bibnamefont {Tytgat}},\
  }\href {\doibase 10.1088/1475-7516/2007/02/028} {\bibfield  {journal}
  {\bibinfo  {journal} {JCAP}\ }\textbf {\bibinfo {volume} {0702}},\ \bibinfo
  {pages} {028} (\bibinfo {year} {2007})},\ \Eprint
  {http://arxiv.org/abs/hep-ph/0612275} {arXiv:hep-ph/0612275 [hep-ph]}
  \BibitemShut {NoStop}%
\bibitem [{\citenamefont {Arina}\ \emph {et~al.}(2009)\citenamefont {Arina},
  \citenamefont {Ling},\ and\ \citenamefont {Tytgat}}]{Arina:2009um}%
  \BibitemOpen
  \bibfield  {author} {\bibinfo {author} {\bibfnamefont {C.}~\bibnamefont
  {Arina}}, \bibinfo {author} {\bibfnamefont {F.-S.}\ \bibnamefont {Ling}}, \
  and\ \bibinfo {author} {\bibfnamefont {M.~H.~G.}\ \bibnamefont {Tytgat}},\
  }\href {\doibase 10.1088/1475-7516/2009/10/018} {\bibfield  {journal}
  {\bibinfo  {journal} {JCAP}\ }\textbf {\bibinfo {volume} {0910}},\ \bibinfo
  {pages} {018} (\bibinfo {year} {2009})},\ \Eprint
  {http://arxiv.org/abs/0907.0430} {arXiv:0907.0430 [hep-ph]} \BibitemShut
  {NoStop}%
\bibitem [{\citenamefont {Nezri}\ \emph {et~al.}(2009)\citenamefont {Nezri},
  \citenamefont {Tytgat},\ and\ \citenamefont {Vertongen}}]{Nezri:2009jd}%
  \BibitemOpen
  \bibfield  {author} {\bibinfo {author} {\bibfnamefont {E.}~\bibnamefont
  {Nezri}}, \bibinfo {author} {\bibfnamefont {M.~H.~G.}\ \bibnamefont
  {Tytgat}}, \ and\ \bibinfo {author} {\bibfnamefont {G.}~\bibnamefont
  {Vertongen}},\ }\href {\doibase 10.1088/1475-7516/2009/04/014} {\bibfield
  {journal} {\bibinfo  {journal} {JCAP}\ }\textbf {\bibinfo {volume} {0904}},\
  \bibinfo {pages} {014} (\bibinfo {year} {2009})},\ \Eprint
  {http://arxiv.org/abs/0901.2556} {arXiv:0901.2556 [hep-ph]} \BibitemShut
  {NoStop}%
\bibitem [{\citenamefont {Miao}\ \emph {et~al.}(2010)\citenamefont {Miao},
  \citenamefont {Su},\ and\ \citenamefont {Thomas}}]{Miao:2010rg}%
  \BibitemOpen
  \bibfield  {author} {\bibinfo {author} {\bibfnamefont {X.}~\bibnamefont
  {Miao}}, \bibinfo {author} {\bibfnamefont {S.}~\bibnamefont {Su}}, \ and\
  \bibinfo {author} {\bibfnamefont {B.}~\bibnamefont {Thomas}},\ }\href
  {\doibase 10.1103/PhysRevD.82.035009} {\bibfield  {journal} {\bibinfo
  {journal} {Phys. Rev.}\ }\textbf {\bibinfo {volume} {D82}},\ \bibinfo {pages}
  {035009} (\bibinfo {year} {2010})},\ \Eprint {http://arxiv.org/abs/1005.0090}
  {arXiv:1005.0090 [hep-ph]} \BibitemShut {NoStop}%
\bibitem [{\citenamefont {Gustafsson}\ \emph {et~al.}(2012)\citenamefont
  {Gustafsson}, \citenamefont {Rydbeck}, \citenamefont {Lopez-Honorez},\ and\
  \citenamefont {Lundstrom}}]{Gustafsson:2012aj}%
  \BibitemOpen
  \bibfield  {author} {\bibinfo {author} {\bibfnamefont {M.}~\bibnamefont
  {Gustafsson}}, \bibinfo {author} {\bibfnamefont {S.}~\bibnamefont {Rydbeck}},
  \bibinfo {author} {\bibfnamefont {L.}~\bibnamefont {Lopez-Honorez}}, \ and\
  \bibinfo {author} {\bibfnamefont {E.}~\bibnamefont {Lundstrom}},\ }\href
  {\doibase 10.1103/PhysRevD.86.075019} {\bibfield  {journal} {\bibinfo
  {journal} {Phys. Rev.}\ }\textbf {\bibinfo {volume} {D86}},\ \bibinfo {pages}
  {075019} (\bibinfo {year} {2012})},\ \Eprint {http://arxiv.org/abs/1206.6316}
  {arXiv:1206.6316 [hep-ph]} \BibitemShut {NoStop}%
\bibitem [{\citenamefont {Arhrib}\ \emph {et~al.}(2012)\citenamefont {Arhrib},
  \citenamefont {Benbrik},\ and\ \citenamefont {Gaur}}]{Arhrib:2012ia}%
  \BibitemOpen
  \bibfield  {author} {\bibinfo {author} {\bibfnamefont {A.}~\bibnamefont
  {Arhrib}}, \bibinfo {author} {\bibfnamefont {R.}~\bibnamefont {Benbrik}}, \
  and\ \bibinfo {author} {\bibfnamefont {N.}~\bibnamefont {Gaur}},\ }\href
  {\doibase 10.1103/PhysRevD.85.095021} {\bibfield  {journal} {\bibinfo
  {journal} {Phys. Rev.}\ }\textbf {\bibinfo {volume} {D85}},\ \bibinfo {pages}
  {095021} (\bibinfo {year} {2012})},\ \Eprint {http://arxiv.org/abs/1201.2644}
  {arXiv:1201.2644 [hep-ph]} \BibitemShut {NoStop}%
\bibitem [{\citenamefont {Swiezewska}\ and\ \citenamefont
  {Krawczyk}(2013)}]{Swiezewska:2012eh}%
  \BibitemOpen
  \bibfield  {author} {\bibinfo {author} {\bibfnamefont {B.}~\bibnamefont
  {Swiezewska}}\ and\ \bibinfo {author} {\bibfnamefont {M.}~\bibnamefont
  {Krawczyk}},\ }\href {\doibase 10.1103/PhysRevD.88.035019} {\bibfield
  {journal} {\bibinfo  {journal} {Phys. Rev.}\ }\textbf {\bibinfo {volume}
  {D88}},\ \bibinfo {pages} {035019} (\bibinfo {year} {2013})},\ \Eprint
  {http://arxiv.org/abs/1212.4100} {arXiv:1212.4100 [hep-ph]} \BibitemShut
  {NoStop}%
\bibitem [{\citenamefont {Goudelis}\ \emph {et~al.}(2013)\citenamefont
  {Goudelis}, \citenamefont {Herrmann},\ and\ \citenamefont
  {Stål}}]{Goudelis:2013uca}%
  \BibitemOpen
  \bibfield  {author} {\bibinfo {author} {\bibfnamefont {A.}~\bibnamefont
  {Goudelis}}, \bibinfo {author} {\bibfnamefont {B.}~\bibnamefont {Herrmann}},
  \ and\ \bibinfo {author} {\bibfnamefont {O.}~\bibnamefont {Stål}},\ }\href
  {\doibase 10.1007/JHEP09(2013)106} {\bibfield  {journal} {\bibinfo  {journal}
  {JHEP}\ }\textbf {\bibinfo {volume} {1309}},\ \bibinfo {pages} {106}
  (\bibinfo {year} {2013})},\ \Eprint {http://arxiv.org/abs/1303.3010}
  {arXiv:1303.3010 [hep-ph]} \BibitemShut {NoStop}%
\bibitem [{\citenamefont {Arhrib}\ \emph {et~al.}(2014)\citenamefont {Arhrib},
  \citenamefont {Tsai}, \citenamefont {Yuan},\ and\ \citenamefont
  {Yuan}}]{Arhrib:2013ela}%
  \BibitemOpen
  \bibfield  {author} {\bibinfo {author} {\bibfnamefont {A.}~\bibnamefont
  {Arhrib}}, \bibinfo {author} {\bibfnamefont {Y.-L.~S.}\ \bibnamefont {Tsai}},
  \bibinfo {author} {\bibfnamefont {Q.}~\bibnamefont {Yuan}}, \ and\ \bibinfo
  {author} {\bibfnamefont {T.-C.}\ \bibnamefont {Yuan}},\ }\href {\doibase
  10.1088/1475-7516/2014/06/030} {\bibfield  {journal} {\bibinfo  {journal}
  {JCAP}\ }\textbf {\bibinfo {volume} {1406}},\ \bibinfo {pages} {030}
  (\bibinfo {year} {2014})},\ \Eprint {http://arxiv.org/abs/1310.0358}
  {arXiv:1310.0358 [hep-ph]} \BibitemShut {NoStop}%
\bibitem [{\citenamefont {Krawczyk}\ \emph
  {et~al.}(2013{\natexlab{a}})\citenamefont {Krawczyk}, \citenamefont
  {Sokolowska}, \citenamefont {Swaczyna},\ and\ \citenamefont
  {Swiezewska}}]{Krawczyk:2013jta}%
  \BibitemOpen
  \bibfield  {author} {\bibinfo {author} {\bibfnamefont {M.}~\bibnamefont
  {Krawczyk}}, \bibinfo {author} {\bibfnamefont {D.}~\bibnamefont
  {Sokolowska}}, \bibinfo {author} {\bibfnamefont {P.}~\bibnamefont
  {Swaczyna}}, \ and\ \bibinfo {author} {\bibfnamefont {B.}~\bibnamefont
  {Swiezewska}},\ }\href {\doibase 10.1007/JHEP09(2013)055} {\bibfield
  {journal} {\bibinfo  {journal} {JHEP}\ }\textbf {\bibinfo {volume} {09}},\
  \bibinfo {pages} {055} (\bibinfo {year} {2013}{\natexlab{a}})},\ \Eprint
  {http://arxiv.org/abs/1305.6266} {arXiv:1305.6266 [hep-ph]} \BibitemShut
  {NoStop}%
\bibitem [{\citenamefont {Krawczyk}\ \emph
  {et~al.}(2013{\natexlab{b}})\citenamefont {Krawczyk}, \citenamefont
  {Sokołowska}, \citenamefont {Swaczyna},\ and\ \citenamefont
  {Świeżewska}}]{Krawczyk:2013pea}%
  \BibitemOpen
  \bibfield  {author} {\bibinfo {author} {\bibfnamefont {M.}~\bibnamefont
  {Krawczyk}}, \bibinfo {author} {\bibfnamefont {D.}~\bibnamefont
  {Sokołowska}}, \bibinfo {author} {\bibfnamefont {P.}~\bibnamefont
  {Swaczyna}}, \ and\ \bibinfo {author} {\bibfnamefont {B.}~\bibnamefont
  {Świeżewska}},\ }\bibfield  {booktitle} {\emph {\bibinfo {booktitle}
  {{Proceedings, 37th International Conference of Theoretical Physics on Matter
  to the Deepest: Recent Developments in Physics of Fundamental
  Interactions}}},\ }\href {\doibase 10.5506/APhysPolB.44.2163} {\bibfield
  {journal} {\bibinfo  {journal} {Acta Phys. Polon.}\ }\textbf {\bibinfo
  {volume} {B44}},\ \bibinfo {pages} {2163} (\bibinfo {year}
  {2013}{\natexlab{b}})},\ \Eprint {http://arxiv.org/abs/1309.7880}
  {arXiv:1309.7880 [hep-ph]} \BibitemShut {NoStop}%
\bibitem [{\citenamefont {Ilnicka}\ \emph {et~al.}(2016)\citenamefont
  {Ilnicka}, \citenamefont {Krawczyk},\ and\ \citenamefont
  {Robens}}]{Ilnicka:2015jba}%
  \BibitemOpen
  \bibfield  {author} {\bibinfo {author} {\bibfnamefont {A.}~\bibnamefont
  {Ilnicka}}, \bibinfo {author} {\bibfnamefont {M.}~\bibnamefont {Krawczyk}}, \
  and\ \bibinfo {author} {\bibfnamefont {T.}~\bibnamefont {Robens}},\ }\href
  {\doibase 10.1103/PhysRevD.93.055026} {\bibfield  {journal} {\bibinfo
  {journal} {Phys. Rev.}\ }\textbf {\bibinfo {volume} {D93}},\ \bibinfo {pages}
  {055026} (\bibinfo {year} {2016})},\ \Eprint
  {http://arxiv.org/abs/1508.01671} {arXiv:1508.01671 [hep-ph]} \BibitemShut
  {NoStop}%
\bibitem [{\citenamefont {Díaz}\ \emph {et~al.}(2016)\citenamefont {Díaz},
  \citenamefont {Koch},\ and\ \citenamefont {Urrutia-Quiroga}}]{Diaz:2015pyv}%
  \BibitemOpen
  \bibfield  {author} {\bibinfo {author} {\bibfnamefont {M.~A.}\ \bibnamefont
  {Díaz}}, \bibinfo {author} {\bibfnamefont {B.}~\bibnamefont {Koch}}, \ and\
  \bibinfo {author} {\bibfnamefont {S.}~\bibnamefont {Urrutia-Quiroga}},\
  }\href {\doibase 10.1155/2016/8278375} {\bibfield  {journal} {\bibinfo
  {journal} {Adv. High Energy Phys.}\ }\textbf {\bibinfo {volume} {2016}},\
  \bibinfo {pages} {8278375} (\bibinfo {year} {2016})},\ \Eprint
  {http://arxiv.org/abs/1511.04429} {arXiv:1511.04429 [hep-ph]} \BibitemShut
  {NoStop}%
\bibitem [{\citenamefont {Modak}\ and\ \citenamefont
  {Majumdar}(2015)}]{Modak:2015uda}%
  \BibitemOpen
  \bibfield  {author} {\bibinfo {author} {\bibfnamefont {K.~P.}\ \bibnamefont
  {Modak}}\ and\ \bibinfo {author} {\bibfnamefont {D.}~\bibnamefont
  {Majumdar}},\ }\href {\doibase 10.1088/0067-0049/219/2/37} {\bibfield
  {journal} {\bibinfo  {journal} {Astrophys. J. Suppl.}\ }\textbf {\bibinfo
  {volume} {219}},\ \bibinfo {pages} {37} (\bibinfo {year} {2015})},\ \Eprint
  {http://arxiv.org/abs/1502.05682} {arXiv:1502.05682 [hep-ph]} \BibitemShut
  {NoStop}%
\bibitem [{\citenamefont {Queiroz}\ and\ \citenamefont
  {Yaguna}(2016)}]{Queiroz:2015utg}%
  \BibitemOpen
  \bibfield  {author} {\bibinfo {author} {\bibfnamefont {F.~S.}\ \bibnamefont
  {Queiroz}}\ and\ \bibinfo {author} {\bibfnamefont {C.~E.}\ \bibnamefont
  {Yaguna}},\ }\href {\doibase 10.1088/1475-7516/2016/02/038} {\bibfield
  {journal} {\bibinfo  {journal} {JCAP}\ }\textbf {\bibinfo {volume} {1602}},\
  \bibinfo {pages} {038} (\bibinfo {year} {2016})},\ \Eprint
  {http://arxiv.org/abs/1511.05967} {arXiv:1511.05967 [hep-ph]} \BibitemShut
  {NoStop}%
\bibitem [{\citenamefont {Garcia-Cely}\ \emph {et~al.}(2016)\citenamefont
  {Garcia-Cely}, \citenamefont {Gustafsson},\ and\ \citenamefont
  {Ibarra}}]{Garcia-Cely:2015khw}%
  \BibitemOpen
  \bibfield  {author} {\bibinfo {author} {\bibfnamefont {C.}~\bibnamefont
  {Garcia-Cely}}, \bibinfo {author} {\bibfnamefont {M.}~\bibnamefont
  {Gustafsson}}, \ and\ \bibinfo {author} {\bibfnamefont {A.}~\bibnamefont
  {Ibarra}},\ }\href {\doibase 10.1088/1475-7516/2016/02/043} {\bibfield
  {journal} {\bibinfo  {journal} {JCAP}\ }\textbf {\bibinfo {volume} {1602}},\
  \bibinfo {pages} {043} (\bibinfo {year} {2016})},\ \Eprint
  {http://arxiv.org/abs/1512.02801} {arXiv:1512.02801 [hep-ph]} \BibitemShut
  {NoStop}%
\bibitem [{\citenamefont {Hashemi}\ and\ \citenamefont
  {Najjari}(2017)}]{Hashemi:2016wup}%
  \BibitemOpen
  \bibfield  {author} {\bibinfo {author} {\bibfnamefont {M.}~\bibnamefont
  {Hashemi}}\ and\ \bibinfo {author} {\bibfnamefont {S.}~\bibnamefont
  {Najjari}},\ }\href {\doibase 10.1140/epjc/s10052-017-5159-0} {\bibfield
  {journal} {\bibinfo  {journal} {Eur. Phys. J.}\ }\textbf {\bibinfo {volume}
  {C77}},\ \bibinfo {pages} {592} (\bibinfo {year} {2017})},\ \Eprint
  {http://arxiv.org/abs/1611.07827} {arXiv:1611.07827 [hep-ph]} \BibitemShut
  {NoStop}%
\bibitem [{\citenamefont {Poulose}\ \emph {et~al.}(2017)\citenamefont
  {Poulose}, \citenamefont {Sahoo},\ and\ \citenamefont
  {Sridhar}}]{Poulose:2016lvz}%
  \BibitemOpen
  \bibfield  {author} {\bibinfo {author} {\bibfnamefont {P.}~\bibnamefont
  {Poulose}}, \bibinfo {author} {\bibfnamefont {S.}~\bibnamefont {Sahoo}}, \
  and\ \bibinfo {author} {\bibfnamefont {K.}~\bibnamefont {Sridhar}},\ }\href
  {\doibase 10.1016/j.physletb.2016.12.022} {\bibfield  {journal} {\bibinfo
  {journal} {Phys. Lett.}\ }\textbf {\bibinfo {volume} {B765}},\ \bibinfo
  {pages} {300} (\bibinfo {year} {2017})},\ \Eprint
  {http://arxiv.org/abs/1604.03045} {arXiv:1604.03045 [hep-ph]} \BibitemShut
  {NoStop}%
\bibitem [{\citenamefont {Alves}\ \emph {et~al.}(2016)\citenamefont {Alves},
  \citenamefont {Camargo}, \citenamefont {Dias}, \citenamefont {Longas},
  \citenamefont {Nishi},\ and\ \citenamefont {Queiroz}}]{Alves:2016bib}%
  \BibitemOpen
  \bibfield  {author} {\bibinfo {author} {\bibfnamefont {A.}~\bibnamefont
  {Alves}}, \bibinfo {author} {\bibfnamefont {D.~A.}\ \bibnamefont {Camargo}},
  \bibinfo {author} {\bibfnamefont {A.~G.}\ \bibnamefont {Dias}}, \bibinfo
  {author} {\bibfnamefont {R.}~\bibnamefont {Longas}}, \bibinfo {author}
  {\bibfnamefont {C.~C.}\ \bibnamefont {Nishi}}, \ and\ \bibinfo {author}
  {\bibfnamefont {F.~S.}\ \bibnamefont {Queiroz}},\ }\href {\doibase
  10.1007/JHEP10(2016)015} {\bibfield  {journal} {\bibinfo  {journal} {JHEP}\
  }\textbf {\bibinfo {volume} {10}},\ \bibinfo {pages} {015} (\bibinfo {year}
  {2016})},\ \Eprint {http://arxiv.org/abs/1606.07086} {arXiv:1606.07086
  [hep-ph]} \BibitemShut {NoStop}%
\bibitem [{\citenamefont {Datta}\ \emph {et~al.}(2017)\citenamefont {Datta},
  \citenamefont {Ganguly}, \citenamefont {Khan},\ and\ \citenamefont
  {Rakshit}}]{Datta:2016nfz}%
  \BibitemOpen
  \bibfield  {author} {\bibinfo {author} {\bibfnamefont {A.}~\bibnamefont
  {Datta}}, \bibinfo {author} {\bibfnamefont {N.}~\bibnamefont {Ganguly}},
  \bibinfo {author} {\bibfnamefont {N.}~\bibnamefont {Khan}}, \ and\ \bibinfo
  {author} {\bibfnamefont {S.}~\bibnamefont {Rakshit}},\ }\href {\doibase
  10.1103/PhysRevD.95.015017} {\bibfield  {journal} {\bibinfo  {journal} {Phys.
  Rev.}\ }\textbf {\bibinfo {volume} {D95}},\ \bibinfo {pages} {015017}
  (\bibinfo {year} {2017})},\ \Eprint {http://arxiv.org/abs/1610.00648}
  {arXiv:1610.00648 [hep-ph]} \BibitemShut {NoStop}%
\bibitem [{\citenamefont {Belyaev}\ \emph {et~al.}(2018)\citenamefont
  {Belyaev}, \citenamefont {Cacciapaglia}, \citenamefont {Ivanov},
  \citenamefont {Rojas-Abatte},\ and\ \citenamefont
  {Thomas}}]{Belyaev:2016lok}%
  \BibitemOpen
  \bibfield  {author} {\bibinfo {author} {\bibfnamefont {A.}~\bibnamefont
  {Belyaev}}, \bibinfo {author} {\bibfnamefont {G.}~\bibnamefont
  {Cacciapaglia}}, \bibinfo {author} {\bibfnamefont {I.~P.}\ \bibnamefont
  {Ivanov}}, \bibinfo {author} {\bibfnamefont {F.}~\bibnamefont
  {Rojas-Abatte}}, \ and\ \bibinfo {author} {\bibfnamefont {M.}~\bibnamefont
  {Thomas}},\ }\href {\doibase 10.1103/PhysRevD.97.035011} {\bibfield
  {journal} {\bibinfo  {journal} {Phys. Rev.}\ }\textbf {\bibinfo {volume}
  {D97}},\ \bibinfo {pages} {035011} (\bibinfo {year} {2018})},\ \Eprint
  {http://arxiv.org/abs/1612.00511} {arXiv:1612.00511 [hep-ph]} \BibitemShut
  {NoStop}%
\bibitem [{\citenamefont {Aad}\ \emph {et~al.}(2016)\citenamefont {Aad} \emph
  {et~al.}}]{Aad:2015txa}%
  \BibitemOpen
  \bibfield  {author} {\bibinfo {author} {\bibfnamefont {G.}~\bibnamefont
  {Aad}} \emph {et~al.} (\bibinfo {collaboration} {ATLAS}),\ }\href {\doibase
  10.1007/JHEP01(2016)172} {\bibfield  {journal} {\bibinfo  {journal} {JHEP}\
  }\textbf {\bibinfo {volume} {01}},\ \bibinfo {pages} {172} (\bibinfo {year}
  {2016})},\ \Eprint {http://arxiv.org/abs/1508.07869} {arXiv:1508.07869
  [hep-ex]} \BibitemShut {NoStop}%
\bibitem [{\citenamefont {Khachatryan}\ \emph {et~al.}(2017)\citenamefont
  {Khachatryan} \emph {et~al.}}]{Khachatryan:2016whc}%
  \BibitemOpen
  \bibfield  {author} {\bibinfo {author} {\bibfnamefont {V.}~\bibnamefont
  {Khachatryan}} \emph {et~al.} (\bibinfo {collaboration} {CMS}),\ }\href
  {\doibase 10.1007/JHEP02(2017)135} {\bibfield  {journal} {\bibinfo  {journal}
  {JHEP}\ }\textbf {\bibinfo {volume} {02}},\ \bibinfo {pages} {135} (\bibinfo
  {year} {2017})},\ \Eprint {http://arxiv.org/abs/1610.09218} {arXiv:1610.09218
  [hep-ex]} \BibitemShut {NoStop}%
\bibitem [{\citenamefont {Belyaev}\ \emph {et~al.}(2013)\citenamefont
  {Belyaev}, \citenamefont {D.},\ and\ \citenamefont {Pukhov}}]{CALCHEP}%
  \BibitemOpen
  \bibfield  {author} {\bibinfo {author} {\bibfnamefont {A.}~\bibnamefont
  {Belyaev}}, \bibinfo {author} {\bibfnamefont {C.~N.}\ \bibnamefont {D.}}, \
  and\ \bibinfo {author} {\bibfnamefont {A.}~\bibnamefont {Pukhov}},\ }\href
  {\doibase 10.1016/j.cpc.2013.01.014} {\bibfield  {journal} {\bibinfo
  {journal} {Comput. Phys. Commun.}\ }\textbf {\bibinfo {volume} {184}},\
  \bibinfo {pages} {1729} (\bibinfo {year} {2013})},\ \Eprint
  {http://arxiv.org/abs/1207.6082} {arXiv:1207.6082 [hep-ph]} \BibitemShut
  {NoStop}%
\bibitem [{\citenamefont {Belanger}\ \emph {et~al.}(2002)\citenamefont
  {Belanger}, \citenamefont {Boudjema}, \citenamefont {Pukhov},\ and\
  \citenamefont {Semenov}}]{Belanger:2001fz}%
  \BibitemOpen
  \bibfield  {author} {\bibinfo {author} {\bibfnamefont {G.}~\bibnamefont
  {Belanger}}, \bibinfo {author} {\bibfnamefont {F.}~\bibnamefont {Boudjema}},
  \bibinfo {author} {\bibfnamefont {A.}~\bibnamefont {Pukhov}}, \ and\ \bibinfo
  {author} {\bibfnamefont {A.}~\bibnamefont {Semenov}},\ }\href {\doibase
  10.1016/S0010-4655(02)00596-9} {\bibfield  {journal} {\bibinfo  {journal}
  {Comput. Phys. Commun.}\ }\textbf {\bibinfo {volume} {149}},\ \bibinfo
  {pages} {103} (\bibinfo {year} {2002})},\ \Eprint
  {http://arxiv.org/abs/hep-ph/0112278} {arXiv:hep-ph/0112278 [hep-ph]}
  \BibitemShut {NoStop}%
\bibitem [{\citenamefont {Belanger}\ \emph {et~al.}(2006)\citenamefont
  {Belanger}, \citenamefont {Boudjema}, \citenamefont {Pukhov},\ and\
  \citenamefont {Semenov}}]{Belanger:2004yn}%
  \BibitemOpen
  \bibfield  {author} {\bibinfo {author} {\bibfnamefont {G.}~\bibnamefont
  {Belanger}}, \bibinfo {author} {\bibfnamefont {F.}~\bibnamefont {Boudjema}},
  \bibinfo {author} {\bibfnamefont {A.}~\bibnamefont {Pukhov}}, \ and\ \bibinfo
  {author} {\bibfnamefont {A.}~\bibnamefont {Semenov}},\ }\href {\doibase
  10.1016/j.cpc.2005.12.005} {\bibfield  {journal} {\bibinfo  {journal}
  {Comput. Phys. Commun.}\ }\textbf {\bibinfo {volume} {174}},\ \bibinfo
  {pages} {577} (\bibinfo {year} {2006})},\ \Eprint
  {http://arxiv.org/abs/hep-ph/0405253} {arXiv:hep-ph/0405253 [hep-ph]}
  \BibitemShut {NoStop}%
\bibitem [{\citenamefont {Ade}\ \emph {et~al.}(2014)\citenamefont {Ade} \emph
  {et~al.}}]{Ade:2013zuv}%
  \BibitemOpen
  \bibfield  {author} {\bibinfo {author} {\bibfnamefont {P.~A.~R.}\
  \bibnamefont {Ade}} \emph {et~al.} (\bibinfo {collaboration} {Planck}),\
  }\href {\doibase 10.1051/0004-6361/201321591} {\bibfield  {journal} {\bibinfo
   {journal} {Astron. Astrophys.}\ }\textbf {\bibinfo {volume} {571}},\
  \bibinfo {pages} {A16} (\bibinfo {year} {2014})},\ \Eprint
  {http://arxiv.org/abs/1303.5076} {arXiv:1303.5076 [astro-ph.CO]} \BibitemShut
  {NoStop}%
\bibitem [{\citenamefont {Ade}\ \emph {et~al.}(2016)\citenamefont {Ade} \emph
  {et~al.}}]{Ade:2015xua}%
  \BibitemOpen
  \bibfield  {author} {\bibinfo {author} {\bibfnamefont {P.~A.~R.}\
  \bibnamefont {Ade}} \emph {et~al.} (\bibinfo {collaboration} {Planck}),\
  }\href {\doibase 10.1051/0004-6361/201525830} {\bibfield  {journal} {\bibinfo
   {journal} {Astron. Astrophys.}\ }\textbf {\bibinfo {volume} {594}},\
  \bibinfo {pages} {A13} (\bibinfo {year} {2016})},\ \Eprint
  {http://arxiv.org/abs/1502.01589} {arXiv:1502.01589 [astro-ph.CO]}
  \BibitemShut {NoStop}%
\bibitem [{\citenamefont {Akerib}\ \emph {et~al.}(2014)\citenamefont {Akerib}
  \emph {et~al.}}]{Akerib:2013tjd}%
  \BibitemOpen
  \bibfield  {author} {\bibinfo {author} {\bibfnamefont {D.~S.}\ \bibnamefont
  {Akerib}} \emph {et~al.} (\bibinfo {collaboration} {LUX}),\ }\href {\doibase
  10.1103/PhysRevLett.112.091303} {\bibfield  {journal} {\bibinfo  {journal}
  {Phys. Rev. Lett.}\ }\textbf {\bibinfo {volume} {112}},\ \bibinfo {pages}
  {091303} (\bibinfo {year} {2014})},\ \Eprint {http://arxiv.org/abs/1310.8214}
  {arXiv:1310.8214 [astro-ph.CO]} \BibitemShut {NoStop}%
\bibitem [{\citenamefont {Aprile}\ \emph {et~al.}(2017)\citenamefont {Aprile}
  \emph {et~al.}}]{Aprile:2017iyp}%
  \BibitemOpen
  \bibfield  {author} {\bibinfo {author} {\bibfnamefont {E.}~\bibnamefont
  {Aprile}} \emph {et~al.} (\bibinfo {collaboration} {XENON}),\ }\href
  {\doibase 10.1103/PhysRevLett.119.181301} {\bibfield  {journal} {\bibinfo
  {journal} {Phys. Rev. Lett.}\ }\textbf {\bibinfo {volume} {119}},\ \bibinfo
  {pages} {181301} (\bibinfo {year} {2017})},\ \Eprint
  {http://arxiv.org/abs/1705.06655} {arXiv:1705.06655 [astro-ph.CO]}
  \BibitemShut {NoStop}%
\bibitem [{phe()}]{phenodata-xenon1t}%
  \BibitemOpen
  \href@noop {} {}\bibinfo {howpublished}
  {\url{https://hepmdb.soton.ac.uk/phenodata/view.php?id=595e239abb817586383e929d
  }}\BibitemShut {NoStop}%
\bibitem [{\citenamefont {Ilnicka}\ \emph {et~al.}(2018)\citenamefont
  {Ilnicka}, \citenamefont {Robens},\ and\ \citenamefont
  {Stefaniak}}]{Ilnicka:2018def}%
  \BibitemOpen
  \bibfield  {author} {\bibinfo {author} {\bibfnamefont {A.}~\bibnamefont
  {Ilnicka}}, \bibinfo {author} {\bibfnamefont {T.}~\bibnamefont {Robens}}, \
  and\ \bibinfo {author} {\bibfnamefont {T.}~\bibnamefont {Stefaniak}},\ }\href
  {\doibase 10.1142/S0217732318300070} {\bibfield  {journal} {\bibinfo
  {journal} {Mod. Phys. Lett.}\ }\textbf {\bibinfo {volume} {A33}},\ \bibinfo
  {pages} {1830007} (\bibinfo {year} {2018})},\ \Eprint
  {http://arxiv.org/abs/1803.03594} {arXiv:1803.03594 [hep-ph]} \BibitemShut
  {NoStop}%
\bibitem [{\citenamefont {\texttt{https://hepmdb.soton.ac.uk}}(2015)}]{HEPMDB}%
  \BibitemOpen
  \bibfield  {author} {\bibinfo {author} {\bibnamefont
  {\texttt{https://hepmdb.soton.ac.uk}}},\ }\href
  {\url{https://hepmdb.soton.ac.uk}} {\emph {\bibinfo {title} {High Energy
  Physics Models DataBase}}},\ \bibinfo {type} {Webpage}\ (\bibinfo {year}
  {2015})\BibitemShut {NoStop}%
\bibitem [{\citenamefont {Ball}\ \emph {et~al.}(2013)\citenamefont {Ball} \emph
  {et~al.}}]{Ball:2012cx}%
  \BibitemOpen
  \bibfield  {author} {\bibinfo {author} {\bibfnamefont {R.~D.}\ \bibnamefont
  {Ball}} \emph {et~al.},\ }\href {\doibase 10.1016/j.nuclphysb.2012.10.003}
  {\bibfield  {journal} {\bibinfo  {journal} {Nucl. Phys.}\ }\textbf {\bibinfo
  {volume} {B867}},\ \bibinfo {pages} {244} (\bibinfo {year} {2013})},\ \Eprint
  {http://arxiv.org/abs/1207.1303} {arXiv:1207.1303 [hep-ph]} \BibitemShut
  {NoStop}%
\bibitem [{\citenamefont {de~Florian}\ \emph {et~al.}(2016)\citenamefont
  {de~Florian} \emph {et~al.}}]{deFlorian:2016spz}%
  \BibitemOpen
  \bibfield  {author} {\bibinfo {author} {\bibfnamefont {D.}~\bibnamefont
  {de~Florian}} \emph {et~al.} (\bibinfo {collaboration} {LHC Higgs Cross
  Section Working Group}),\ }\href {\doibase 10.23731/CYRM-2017-002} {\
  (\bibinfo {year} {2016}),\ 10.23731/CYRM-2017-002},\ \Eprint
  {http://arxiv.org/abs/1610.07922} {arXiv:1610.07922 [hep-ph]} \BibitemShut
  {NoStop}%
\bibitem [{\citenamefont {Alwall}\ \emph {et~al.}(2011)\citenamefont {Alwall},
  \citenamefont {Herquet}, \citenamefont {Maltoni}, \citenamefont {Mattelaer},\
  and\ \citenamefont {Stelzer}}]{Alwall:2011uj}%
  \BibitemOpen
  \bibfield  {author} {\bibinfo {author} {\bibfnamefont {J.}~\bibnamefont
  {Alwall}}, \bibinfo {author} {\bibfnamefont {M.}~\bibnamefont {Herquet}},
  \bibinfo {author} {\bibfnamefont {F.}~\bibnamefont {Maltoni}}, \bibinfo
  {author} {\bibfnamefont {O.}~\bibnamefont {Mattelaer}}, \ and\ \bibinfo
  {author} {\bibfnamefont {T.}~\bibnamefont {Stelzer}},\ }\href {\doibase
  10.1007/JHEP06(2011)128} {\bibfield  {journal} {\bibinfo  {journal} {JHEP}\
  }\textbf {\bibinfo {volume} {1106}},\ \bibinfo {pages} {128} (\bibinfo {year}
  {2011})},\ \Eprint {http://arxiv.org/abs/1106.0522} {arXiv:1106.0522
  [hep-ph]} \BibitemShut {NoStop}%
\bibitem [{\citenamefont {Alwall}\ \emph {et~al.}(2014)\citenamefont {Alwall},
  \citenamefont {Frederix}, \citenamefont {Frixione}, \citenamefont {Hirschi},
  \citenamefont {Maltoni}, \citenamefont {Mattelaer}, \citenamefont {Shao},
  \citenamefont {Stelzer}, \citenamefont {Torrielli},\ and\ \citenamefont
  {Zaro}}]{Alwall:2014hca}%
  \BibitemOpen
  \bibfield  {author} {\bibinfo {author} {\bibfnamefont {J.}~\bibnamefont
  {Alwall}}, \bibinfo {author} {\bibfnamefont {R.}~\bibnamefont {Frederix}},
  \bibinfo {author} {\bibfnamefont {S.}~\bibnamefont {Frixione}}, \bibinfo
  {author} {\bibfnamefont {V.}~\bibnamefont {Hirschi}}, \bibinfo {author}
  {\bibfnamefont {F.}~\bibnamefont {Maltoni}}, \bibinfo {author} {\bibfnamefont
  {O.}~\bibnamefont {Mattelaer}}, \bibinfo {author} {\bibfnamefont {H.~S.}\
  \bibnamefont {Shao}}, \bibinfo {author} {\bibfnamefont {T.}~\bibnamefont
  {Stelzer}}, \bibinfo {author} {\bibfnamefont {P.}~\bibnamefont {Torrielli}},
  \ and\ \bibinfo {author} {\bibfnamefont {M.}~\bibnamefont {Zaro}},\ }\href
  {\doibase 10.1007/JHEP07(2014)079} {\bibfield  {journal} {\bibinfo  {journal}
  {JHEP}\ }\textbf {\bibinfo {volume} {07}},\ \bibinfo {pages} {079} (\bibinfo
  {year} {2014})},\ \Eprint {http://arxiv.org/abs/1405.0301} {arXiv:1405.0301
  [hep-ph]} \BibitemShut {NoStop}%
\bibitem [{\citenamefont {Lindert}\ \emph {et~al.}(2018)\citenamefont
  {Lindert}, \citenamefont {Kudashkin}, \citenamefont {Melnikov},\ and\
  \citenamefont {Wever}}]{Lindert:2018iug}%
  \BibitemOpen
  \bibfield  {author} {\bibinfo {author} {\bibfnamefont {J.~M.}\ \bibnamefont
  {Lindert}}, \bibinfo {author} {\bibfnamefont {K.}~\bibnamefont {Kudashkin}},
  \bibinfo {author} {\bibfnamefont {K.}~\bibnamefont {Melnikov}}, \ and\
  \bibinfo {author} {\bibfnamefont {C.}~\bibnamefont {Wever}},\ }\href
  {\doibase 10.1016/j.physletb.2018.05.009} {\bibfield  {journal} {\bibinfo
  {journal} {Phys. Lett.}\ }\textbf {\bibinfo {volume} {B782}},\ \bibinfo
  {pages} {210} (\bibinfo {year} {2018})},\ \Eprint
  {http://arxiv.org/abs/1801.08226} {arXiv:1801.08226 [hep-ph]} \BibitemShut
  {NoStop}%
\bibitem [{\citenamefont {Jones}\ \emph {et~al.}(2018)\citenamefont {Jones},
  \citenamefont {Kerner},\ and\ \citenamefont {Luisoni}}]{Jones:2018hbb}%
  \BibitemOpen
  \bibfield  {author} {\bibinfo {author} {\bibfnamefont {S.~P.}\ \bibnamefont
  {Jones}}, \bibinfo {author} {\bibfnamefont {M.}~\bibnamefont {Kerner}}, \
  and\ \bibinfo {author} {\bibfnamefont {G.}~\bibnamefont {Luisoni}},\ }\href
  {\doibase 10.1103/PhysRevLett.120.162001} {\bibfield  {journal} {\bibinfo
  {journal} {Phys. Rev. Lett.}\ }\textbf {\bibinfo {volume} {120}},\ \bibinfo
  {pages} {162001} (\bibinfo {year} {2018})},\ \Eprint
  {http://arxiv.org/abs/1802.00349} {arXiv:1802.00349 [hep-ph]} \BibitemShut
  {NoStop}%
\bibitem [{\citenamefont {Neumann}(2018)}]{Neumann:2018bsx}%
  \BibitemOpen
  \bibfield  {author} {\bibinfo {author} {\bibfnamefont {T.}~\bibnamefont
  {Neumann}},\ }\href@noop {} {\  (\bibinfo {year} {2018})},\ \Eprint
  {http://arxiv.org/abs/1802.02981} {arXiv:1802.02981 [hep-ph]} \BibitemShut
  {NoStop}%
\bibitem [{\citenamefont {Sjostrand}\ \emph {et~al.}(2006)\citenamefont
  {Sjostrand}, \citenamefont {Mrenna},\ and\ \citenamefont
  {Skands}}]{Sjostrand:2006za}%
  \BibitemOpen
  \bibfield  {author} {\bibinfo {author} {\bibfnamefont {T.}~\bibnamefont
  {Sjostrand}}, \bibinfo {author} {\bibfnamefont {S.}~\bibnamefont {Mrenna}}, \
  and\ \bibinfo {author} {\bibfnamefont {P.~Z.}\ \bibnamefont {Skands}},\
  }\href {\doibase 10.1088/1126-6708/2006/05/026} {\bibfield  {journal}
  {\bibinfo  {journal} {JHEP}\ }\textbf {\bibinfo {volume} {05}},\ \bibinfo
  {pages} {026} (\bibinfo {year} {2006})},\ \Eprint
  {http://arxiv.org/abs/hep-ph/0603175} {arXiv:hep-ph/0603175 [hep-ph]}
  \BibitemShut {NoStop}%
\bibitem [{\citenamefont {Sjostrand}\ \emph {et~al.}(2008)\citenamefont
  {Sjostrand}, \citenamefont {Mrenna},\ and\ \citenamefont
  {Skands}}]{Sjostrand:2007gs}%
  \BibitemOpen
  \bibfield  {author} {\bibinfo {author} {\bibfnamefont {T.}~\bibnamefont
  {Sjostrand}}, \bibinfo {author} {\bibfnamefont {S.}~\bibnamefont {Mrenna}}, \
  and\ \bibinfo {author} {\bibfnamefont {P.~Z.}\ \bibnamefont {Skands}},\
  }\href {\doibase 10.1016/j.cpc.2008.01.036} {\bibfield  {journal} {\bibinfo
  {journal} {Comput. Phys. Commun.}\ }\textbf {\bibinfo {volume} {178}},\
  \bibinfo {pages} {852} (\bibinfo {year} {2008})},\ \Eprint
  {http://arxiv.org/abs/0710.3820} {arXiv:0710.3820 [hep-ph]} \BibitemShut
  {NoStop}%
\bibitem [{\citenamefont {Buckley}\ \emph
  {et~al.}(2015{\natexlab{b}})\citenamefont {Buckley}, \citenamefont
  {Ferrando}, \citenamefont {Lloyd}, \citenamefont {Nordström}, \citenamefont
  {Page}, \citenamefont {Rüfenacht}, \citenamefont {Schönherr},\ and\
  \citenamefont {Watt}}]{Buckley:2014ana}%
  \BibitemOpen
  \bibfield  {author} {\bibinfo {author} {\bibfnamefont {A.}~\bibnamefont
  {Buckley}}, \bibinfo {author} {\bibfnamefont {J.}~\bibnamefont {Ferrando}},
  \bibinfo {author} {\bibfnamefont {S.}~\bibnamefont {Lloyd}}, \bibinfo
  {author} {\bibfnamefont {K.}~\bibnamefont {Nordström}}, \bibinfo {author}
  {\bibfnamefont {B.}~\bibnamefont {Page}}, \bibinfo {author} {\bibfnamefont
  {M.}~\bibnamefont {Rüfenacht}}, \bibinfo {author} {\bibfnamefont
  {M.}~\bibnamefont {Schönherr}}, \ and\ \bibinfo {author} {\bibfnamefont
  {G.}~\bibnamefont {Watt}},\ }\href {\doibase 10.1140/epjc/s10052-015-3318-8}
  {\bibfield  {journal} {\bibinfo  {journal} {Eur. Phys. J.}\ }\textbf
  {\bibinfo {volume} {C75}},\ \bibinfo {pages} {132} (\bibinfo {year}
  {2015}{\natexlab{b}})},\ \Eprint {http://arxiv.org/abs/1412.7420}
  {arXiv:1412.7420 [hep-ph]} \BibitemShut {NoStop}%
\bibitem [{\citenamefont {de~Favereau}\ \emph {et~al.}(2014)\citenamefont
  {de~Favereau} \emph {et~al.}}]{deFavereau:2013fsa}%
  \BibitemOpen
  \bibfield  {author} {\bibinfo {author} {\bibfnamefont {J.}~\bibnamefont
  {de~Favereau}} \emph {et~al.} (\bibinfo {collaboration} {DELPHES 3}),\ }\href
  {\doibase 10.1007/JHEP02(2014)057} {\bibfield  {journal} {\bibinfo  {journal}
  {JHEP}\ }\textbf {\bibinfo {volume} {1402}},\ \bibinfo {pages} {057}
  (\bibinfo {year} {2014})},\ \Eprint {http://arxiv.org/abs/1307.6346}
  {arXiv:1307.6346 [hep-ex]} \BibitemShut {NoStop}%
\bibitem [{\citenamefont {Read}(2002)}]{Read:2002hq}%
  \BibitemOpen
  \bibfield  {author} {\bibinfo {author} {\bibfnamefont {A.~L.}\ \bibnamefont
  {Read}},\ }\bibfield  {booktitle} {\emph {\bibinfo {booktitle} {{Advanced
  statistical techniques in particle physics. Proceedings, Conference, Durham,
  UK, March 18-22, 2002}}},\ }\href {\doibase 10.1088/0954-3899/28/10/313}
  {\bibfield  {journal} {\bibinfo  {journal} {J. Phys.}\ }\textbf {\bibinfo
  {volume} {G28}},\ \bibinfo {pages} {2693} (\bibinfo {year} {2002})},\
  \bibinfo {note} {[,11(2002)]}\BibitemShut {NoStop}%
\bibitem [{\citenamefont {Cowan}\ \emph {et~al.}(2011)\citenamefont {Cowan},
  \citenamefont {Cranmer}, \citenamefont {Gross},\ and\ \citenamefont
  {Vitells}}]{Cowan:2010js}%
  \BibitemOpen
  \bibfield  {author} {\bibinfo {author} {\bibfnamefont {G.}~\bibnamefont
  {Cowan}}, \bibinfo {author} {\bibfnamefont {K.}~\bibnamefont {Cranmer}},
  \bibinfo {author} {\bibfnamefont {E.}~\bibnamefont {Gross}}, \ and\ \bibinfo
  {author} {\bibfnamefont {O.}~\bibnamefont {Vitells}},\ }\href {\doibase
  10.1140/epjc/s10052-011-1554-0, 10.1140/epjc/s10052-013-2501-z} {\bibfield
  {journal} {\bibinfo  {journal} {Eur. Phys. J.}\ }\textbf {\bibinfo {volume}
  {C71}},\ \bibinfo {pages} {1554} (\bibinfo {year} {2011})},\ \bibinfo {note}
  {[Erratum: Eur. Phys. J.C73,2501(2013)]},\ \Eprint
  {http://arxiv.org/abs/1007.1727} {arXiv:1007.1727 [physics.data-an]}
  \BibitemShut {NoStop}%
\bibitem [{\citenamefont {M{\"u}ller}\ \emph {et~al.}()\citenamefont
  {M{\"u}ller}, \citenamefont {Ott},\ and\ \citenamefont
  {Wagner-Kuhr}}]{ref:thetaFramework}%
  \BibitemOpen
  \bibfield  {author} {\bibinfo {author} {\bibfnamefont {T.}~\bibnamefont
  {M{\"u}ller}}, \bibinfo {author} {\bibfnamefont {J.}~\bibnamefont {Ott}}, \
  and\ \bibinfo {author} {\bibfnamefont {J.}~\bibnamefont {Wagner-Kuhr}},\
  }\href@noop {} {\enquote {\bibinfo {title} {The theta framework for
  template-based statistical modeling and inference},}\ }\bibinfo
  {howpublished}
  {\url{http://www-ekp.physik.uni-karlsruhe.de/~ott/theta/theta-auto/index.html}}\BibitemShut
  {NoStop}%
\bibitem [{\citenamefont {Collaboration}(2016)}]{CMS:2016tns}%
  \BibitemOpen
  \bibfield  {author} {\bibinfo {author} {\bibfnamefont {C.}~\bibnamefont
  {Collaboration}} (\bibinfo {collaboration} {CMS}),\ }\href@noop {} {\ ,\
  \bibinfo {pages} {CMS} (\bibinfo {year} {2016})}\BibitemShut {NoStop}%
\bibitem [{\citenamefont {Khachatryan}\ \emph {et~al.}(2015)\citenamefont
  {Khachatryan} \emph {et~al.}}]{Khachatryan:2014rra}%
  \BibitemOpen
  \bibfield  {author} {\bibinfo {author} {\bibfnamefont {V.}~\bibnamefont
  {Khachatryan}} \emph {et~al.} (\bibinfo {collaboration} {CMS}),\ }\href
  {\doibase 10.1140/epjc/s10052-015-3451-4} {\bibfield  {journal} {\bibinfo
  {journal} {Eur. Phys. J.}\ }\textbf {\bibinfo {volume} {C75}},\ \bibinfo
  {pages} {235} (\bibinfo {year} {2015})},\ \Eprint
  {http://arxiv.org/abs/1408.3583} {arXiv:1408.3583 [hep-ex]} \BibitemShut
  {NoStop}%
\end{thebibliography}%

\end{document}